\documentclass[a4paper]{article}

\usepackage{amsmath}
\usepackage{amsthm}
\usepackage{amssymb}
\usepackage{mathrsfs}

\usepackage{verbatim}

\usepackage{epsfig}
\usepackage{color}

\author{Volker Schlue\footnote{Department of Pure Mathematics and Mathematical Statistics, University of Cambridge, Cambridge, CB3 0WB, United Kingdom}}
\title{Decay of linear waves\\on higher dimensional Schwarzschild black holes}

\theoremstyle{plain}
\newtheorem{thm}{Theorem}
\newtheorem{prop}{Proposition}[section]
\newtheorem{lemma}[prop]{Lemma}
\newtheorem{cor}[prop]{Corollary}

\theoremstyle{remark}
\newtheorem*{note}{Note}
\newtheorem{remark}[prop]{Remark}
\newtheorem*{notation}{Notation}
\newtheorem{rmkthm}[thm]{Remark}

\newcommand{\M}{$\mathcal{M}$}
\newcommand{\Q}{$\mathcal{Q}$}
\newcommand{\Hp}{$\mathcal{H}^+$}

\newcommand{\rs}{{r^\ast}}
\newcommand{\us}{{u^\ast}}
\newcommand{\vs}{{v^\ast}}
\newcommand{\dus}{\frac{\partial}{\partial\us}}
\newcommand{\dvs}{\frac{\partial}{\partial\vs}}
\newcommand{\dt}{\frac{\partial}{\partial t}}
\newcommand{\drs}{\frac{\partial}{\partial\rs}}
\newcommand{\ddus}[1]{\frac{\partial #1}{\partial\us}}
\newcommand{\ddvs}[1]{\frac{\partial #1}{\partial\vs}}
\newcommand{\dddvs}[1]{\frac{\partial^2 #1}{\partial\vs^2}}
\newcommand{\ddt}[1]{\frac{\partial #1}{\partial t}}
\newcommand{\ddrs}[1]{\frac{\partial #1}{\partial\rs}}
\newcommand{\pvs}{\partial_{\vs}}

\newcommand{\Sn}{{\mathbb{S}^{n-1}}}
\newcommand{\wn}{\omega_n}
\newcommand{\rh}{\sqrt[n-2]{2m}}
\newcommand{\rhp}{(2m)^\frac{1}{n-2}}
\newcommand{\rph}{\sqrt[n-2]{nm}}
\newcommand{\rphp}{(nm)^{\frac{1}{n-2}}}
\newcommand{\rphn}{(\frac{n}{2})^\frac{1}{n-2}}
\newcommand{\rn}{r^\frac{n-1}{2}}
\newcommand{\sg}{\kappa_n}

\newcommand{\gn}{\stackrel{\circ}{\gamma}_{n-1}}
\newcommand{\dm}[1]{\ud \mu_{#1}}
\newcommand{\nablab}{{\nabla\!\!\!\!/\,}}
\newcommand{\nablassph}{\stackrel{\circ}{\nabla\!\!\!\!/\,}_{n-1}}
\newcommand{\laplacesph}{\triangle\!\!\!\!/\,_{r^2\gn}}
\newcommand{\laplacesphs}{\triangle\!\!\!\!/\,}
\newcommand{\laplacessph}{\stackrel{\circ}{\triangle\!\!\!\!/\,}_{n-1}}
\newcommand{\normsph}[1]{\bigl\vert{#1}\bigr\vert_{r^2\gn}}
\newcommand{\deformt}[1]{{}^{(#1)}\pi}
\newcommand{\onh}{\vert_{\mathcal{H}^+}}
\newcommand{\Onh}[1]{\left. #1 \right\vert_{\mathcal{H}^+}}
\newcommand{\cf}{1-\frac{2m}{r^{n-2}}}
\newcommand{\mrn}{\frac{2m}{r^{n-2}}}
\newcommand{\sumn}{\sum_{i=1}^\frac{n(n-1)}{2}}

\newcommand{\St}[1]{{\Sigma_{#1}}}
\newcommand{\Stp}[1]{{\Sigma^\prime_{#1}}}
\newcommand{\Stt}[2]{{\Sigma_{#1}^{#2}}}

\newcommand{\sprime}[1]{{{#1}^\prime}}
\newcommand{\dprime}[1]{{#1}^{\prime\prime}}
\newcommand{\tprime}[1]{{#1}^{\prime\prime\prime}}
\newcommand{\taup}{\sprime{\tau}}
\newcommand{\taub}{{\overline{\tau}}}
\newcommand{\ipr}[2]{{{#1}^\prime_{#2}}}
\newcommand{\idpr}[2]{{{#1}^{\prime\prime}_{#2}}}
\newcommand{\itpr}[2]{{{#1}^{\prime\prime\prime}_{#2}}}
\newcommand{\iqpr}[2]{{{#1}^{\prime\prime\prime\prime}_{#2}}}
\newcommand{\taujp}[1]{\ipr{\tau}{#1}}
\newcommand{\taujdp}[1]{\idpr{\tau}{#1}}

\newcommand{\dP}[3]{{}^{#1}\mathcal{P}_{#2}^{#3}}
\newcommand{\dD}[3]{{}^{#1}\mathcal{D}_{#2}^{#3}}
\newcommand{\dDb}[3]{{}^{#1}\mathcal{D}\!\!\!\backslash\:{}_{#2}^{#3}}

\newcommand{\intD}[1]{\int_{{}^{#1}\mathcal{D}_{\tau_1}^{\tau_2}}}
\newcommand{\intpD}[1]{\int_{\partial{}^{#1}\mathcal{D}_{\tau_1}^{\tau_2}}}
\newcommand{\intDloc}[3]{\int_{#1}^{#2}\!\!\!\!\ud\us\int_{\us+{#3}^\ast}^\infty\!\!\!\!\ud\vs\int_\Sn\dm{\gn}\times}
\newcommand{\intU}[3]{\int_{#2+#1^\ast}^\infty\!\!\!\!\ud\vs\int_\Sn\!\!\dm{\gn}\times\biggl\{#3\biggr\}\Bigr\vert_{\us=#2}}
\newcommand{\intT}[4]{\int_{2 #1+#3^\ast}^{2 #2+#3^\ast}\!\!\!\!\!\!\!\!\!\ud t\int_\Sn\dm{\gn}\,\times\biggl\{ #4 \biggr\}\Bigr\vert_{r=#3}}

\newcommand{\ud}{\,\mathrm{d}}

\newcommand{\Jr}{\stackrel{\scriptscriptstyle{r}}{J}}
\newcommand{\Kr}{\stackrel{\scriptscriptstyle{r}}{K}}
\newcommand{\Jv}{\stackrel{\scriptscriptstyle{v}}{J}}
\newcommand{\Kv}[2]{\stackrel{\scriptscriptstyle{v}}{K_{#2}}(#1)}
\newcommand{\Krp}[2]{\stackrel{\scriptscriptstyle{r}}{K}_{#2}(#1)}

\newcommand{\ned}[2]{\Bigl(J^{#1}(#2),n\Bigr)}
\newcommand{\sned}{(J^N,n)}
\newcommand{\sted}{(J^T,n)}
\newcommand{\nedCT}[1]{\Bigl(\sum_{k=0}^{#1}J^N(T^k\cdot\phi),n\Bigr)}

\newcommand{\sqb}[1]{\Bigl(#1\Bigr)^2}
\newcommand{\sqv}[1]{\bigl\vert #1\bigr\vert^2}
\newcommand{\cdotc}[2]{#1\cdot #2}
\newcommand{\tphi}{\cdotc{T}{\phi}}
\newcommand{\tpsi}{\cdotc{T}{\psi}}

\newcommand{\pll}{\pi_{<L}}
\newcommand{\phl}{\pi_{\geq L}}

\numberwithin{equation}{section}

\begin{document}

\maketitle

\begin{abstract}
In this paper we consider solutions to the linear wave equation on higher dimensional Schwarzschild black hole spacetimes and prove robust nondegenerate energy decay estimates that are in principle required in a nonlinear stability problem.
More precisely, it is shown that for solutions to the wave equation $\Box_g\phi=0$ on the domain of outer communications of the Schwarzschild spacetime manifold $(\mathcal{M}_m^n,g)$ (where $n\geq 3$ is the spatial dimension, and $m>0$ is the mass of the black hole) the associated energy flux $E[\phi](\Sigma_\tau)$ through a foliation of hypersurfaces $\Sigma_\tau$ (terminating at future null infinity and to the future of the bifurcation sphere) decays, $E[\phi](\Sigma_\tau)\leq\frac{C D}{\tau^2}$, where $C$ is a constant only depending on $n$ and $m$, and $D<\infty$ is a suitable higher order initial energy on $\Sigma_0$; moreover we improve the decay rate for the first order energy to $E[\partial_t\phi](\Sigma_\tau^R)\leq\frac{C D_\delta}{\tau^{4-2\delta}}$ for any $\delta>0$ where $\Sigma_\tau^R$ denotes the hypersurface $\Sigma_\tau$ truncated at an arbitrarily large fixed radius $R<\infty$ provided the higher order energy $D_\delta$ on $\Sigma_0$ is finite. We conclude our paper by interpolating between these two results to obtain the pointwise estimate $\lvert\phi\rvert_{\Sigma_\tau^R}\leq\frac{C D_\delta^\prime}{\tau^{\frac{3}{2}-\delta}}$.
In this work we follow the new physical-space approach to decay for the wave equation of Dafermos and Rodnianski \cite{DRNew}.
\end{abstract}

\tableofcontents

\section{Introduction}

The study of the wave equation on black hole spacetimes has generated considerable interest in recent years. This stems mainly from its role as a model problem for the nonlinear black hole stability problem \cite{DO, DRrev}, and more recent advances in the analysis of linear waves \cite{DRC}.

In this paper we study the linear wave equation on higher dimensional Schwarzschild black holes. The motivation for this problem lies --- apart from the above mentioned relation to the nonlinear stability problem (which is expected to be simpler in the higher dimensional case \cite{CBCL}; for work on the $5$-dimensional case under symmetry see also \cite{DH,H}) --- on one hand in the purely mathematical curiosity of dealing with higher dimensions and on the other hand in its interest for theories of high energy physics \cite{ER}.

In the philosophy of \cite{CK} it is understood that the resolution of the nonlinear stability problem requires an understanding of the linear equations in a sufficiently robust setting, in particular a proof of the uniform boundedness and decay of solutions to the linear wave equation based on the method of energy currents which (ideally) only uses properties of the spacetime that are stable under perturbations, and does not rely heavily on the specifics of the unperturbed metric; (for an introduction in the context of black hole spacetimes see \cite{DRC}). Correspondingly we establish in this paper on higher dimensional Schwarzschild spacetime backgrounds boundedness and decay results analogous to the current state of the art in the $3+1$-dimensional case \cite{Lid}.

The decay argument presented here departs from earlier work that either makes use of multipliers with weights in the temporal variable (notably \cite{CKlinear, AndBlue, BlueS, DRRadDecay, Lid}) which in one form or the other are due to Morawetz \cite{M}, or that relies on the exact stationarity of the spacetime (such as \cite{Ching, TLocal, DSS} based on Fourier analytic methods). Here we follow the new physical-space approach to decay of Dafermos and Rodnianski \cite{DRNew}, which only uses multipliers with weights in the radial variable. Thus our work --- especially the improvement of Section \ref{sec:iid} --- is of independent interest for the $3+1$-dimensional Schwarzschild and Minkowski case and also for a wider class of spacetimes including Kerr black hole exteriors.

\subsection{Statement of the Theorems}

We consider solutions to the wave equation
\begin{equation}
  \label{eq:wave}
  \Box_g\phi=0
\end{equation}
on higher dimensional Schwarzschild black hole spacetimes; these backgrounds are a family of $n+1$-dimensional Lorentzian manifolds $(\mathcal{M}_m^n,g)$ parametrized by the mass of the black hole $m>0$, ($n\geq 3$). They arise as spherically symmetric solutions of the vacuum Einstein equations, the governing equations of General Relativity, and are discussed as such in Section \ref{sec:geometry}; for the relevant concepts see also \cite{DRC, HE}.

\begin{figure}[bt]
\begin{center}
\input{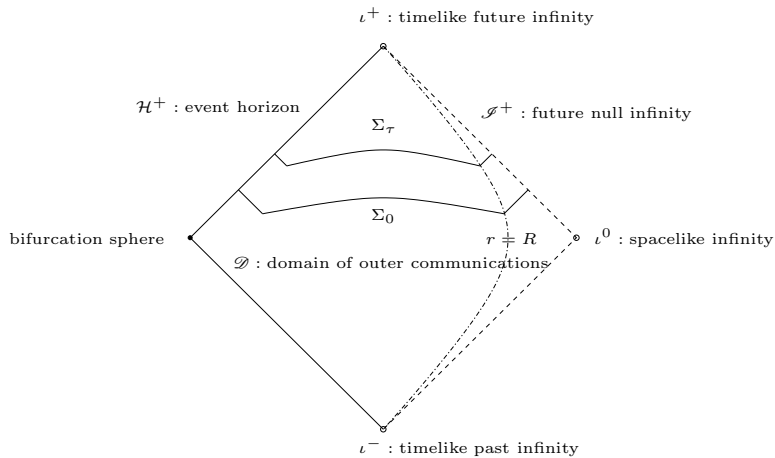}
\caption{The hypersurface $\Sigma_0$ in the domain of outer communications $\mathcal{D}$.}
\label{fig:intro}
\end{center}
\end{figure}

More precisely, we consider solutions to \eqref{eq:wave} on the domain of outer communications $\mathcal{D}$ of $\mathcal{M}$ --- which comprises the exterior up to and including the event horizons of the black hole --- with initial data prescribed on a hypersurface $\Sigma_0$ consisting of an incoming null segment crossing the event horizon to the future of the bifurcation sphere, a spacelike segment and an outgoing null segment emerging from a larger sphere of radius $R$ terminating at future null infinity; see figure \ref{fig:intro} (the exact parametrization -- which is chosen merely for technical reasons -- is given in Section \ref{sec:iled}).

In the exterior of the black hole the metric $g$ takes the classical form in $(t,r)$-coordinates \cite{TS},
\begin{equation}
g=-\bigl(1-\frac{2m}{r^{n-2}}\bigr)\ud t^2+\bigl(1-\frac{2m}{r^{n-2}}\bigr)^{-1}\ud r^2+r^2\gn\,,
\end{equation}
where $r>\rh$, $t\in(-\infty,\infty)$, and $\gn$ denotes the standard metric on the unit $n-1$-sphere; however this coordinate system breaks down on the horizon $r=\rh$ and we shall for that reason introduce in Section \ref{sec:geometry} the global geometry of $(\mathcal{M}_m^n,g)$ using a double null foliation, from which we derive an alternative double null coordinate system for the exterior of the black-hole,
\begin{equation}
g=-4\bigl(1-\frac{2m}{r^{n-2}}\bigr)\ud\us\ud\vs+r^2\gn\,,
\end{equation}
so called Eddington-Finkelstein coordinates.

In this paper both the conditions on the initial data and the statements on the decay of the solutions are formulated using the concepts of energy and the energy momentum tensor associated to \eqref{eq:wave} in particular (see Section \ref{sec:overview} and also Appendix \ref{ref:nd}):
\begin{equation}
\label{eq:intro:energymomentum}
T_{\mu\nu}[\phi]=\partial_\mu\phi\,\partial_\nu\phi-\frac{1}{2}\,g_{\mu\nu}\,\partial^\alpha\phi\partial_\alpha\phi\,.
\end{equation}
The corresponding 1-contravariant-1-covariant tensorfield fulfills the physical requirement that the linear transformation $-T:T\mathcal{M}\to T\mathcal{M}$ maps the hyperboloid of future-directed unit timelike vectors into the closure of the open future cone at each point.
Physically, \[-T\cdot u\in T_p\mathcal{M}\] is the energy-momentum density relative to an observer at $p\in\mathcal{M}$ with 4-velocity $u\in T_p\mathcal{M}$, 
and it is for this reason that we refer to \[\varepsilon=g(T\cdot u,u)=T(u,u)\geq 0\] as the energy density at $p\in\mathcal{M}$ relative to the observer with 4-velocity $u\in T_p\mathcal{M}$.
One may think of a spacelike hypersurface as a collection of locally simultaneous observers with a 4-velocity given by the normal.
The hypersurfaces relative to which we establish energy decay are simply defined by $\St{\tau}\doteq\varphi_{\tau}(\St{0}\cap\mathcal{D})$, where $\varphi_\tau$ denotes the 1-parameter group of isometries generated by $\dt$.
The energy flux through the hypersurface $\St{\tau}$ is then given by
\begin{equation}
  \label{eq:intro:energy}
  E[\phi](\Sigma_\tau)\doteq\int_\St{\tau}\Bigl(J^N[\phi],n\Bigr)
\end{equation}
where $(J^N[\phi],n)\doteq T[\phi](N,n)$, $n$ is the normal\footnote{On spacelike segments of $\St{\tau}$ the vector $n$ is indeed timelike; however, on the null segments of the hypersurfaces $\Sigma_\tau$ the ``normal'' $n$ is in fact a null vector, but the notation is kept for convenience; see Appendix \ref{sec:notation}.} to $\Sigma_\tau$ and $N$ is a timelike $\varphi_\tau$-invariant future directed vectorfield which is constructed in Section \ref{sec:redshift} for the purpose of turning $\varepsilon^N\doteq T(N,N)$ into a nondegenerate energy up to and including the horizon.
Note that the energy $E[\phi](\Sigma_\tau)$ in particular bounds a suitably defined $\mathrm{\dot{H}}^1$-norm on $\Sigma_\tau$.

The classes of solutions to \eqref{eq:wave} to which our results apply are formulated in terms of finite energy conditions on the initial data, for the purpose of which we list the following quantities:
\begin{multline}
  D^{(2)}_2(\tau_0)\doteq\int_{\tau_0+R^\ast}^\infty\!\!\!\!\!\!\ud\vs\int_\Sn\dm{\gn}\sum_{k=0}^1 r^2\sqb{\ddvs{(\rn\partial_t^k\phi)}}\Bigr\vert_{\us=\tau_0}\\+\int_\St{\tau_0}\Bigl(\sum_{k=0}^2 J^N[\partial_t^k\phi],n\Bigr)
\end{multline}
\begin{multline}
  D^{(4-\delta)}_5(\tau_0)\doteq\intU{R}{\tau_0}{\sum_{k=0}^1r^{4-\delta}\sqb{\dddvs{(\rn\partial_t^k\phi)}}\\+\sum_{k=0}^4r^2\sqb{\ddvs{(\rn\partial_t^k\phi)}}+\sum_{k=0}^3\sumn r^2\sqb{\ddvs{\rn\Omega_i\partial_t^k\phi}}}\displaybreak[0]\\
  +\int_\St{\tau_0}\Bigl(\sum_{k=0}^5J^N[\partial_t^k\phi]+\sum_{k=0}^4\sumn J^N[\Omega_i\partial_t^k\phi],n\Bigr)
\end{multline}
\begin{multline}
  D^{(4-\delta)}_7(\tau_0)\doteq\intU{R}{\tau_0}{\sum_{k=0}^2r^{4-\delta}\sqb{\dddvs{(\rn\partial_t^k\phi)}}\\+\sum_{k=0}^2\sumn r^{4-\delta}\sqb{\dddvs{(\rn\Omega_i\partial_t^k\phi)}}\displaybreak[0]\\+\sum_{k=0}^5r^2\sqb{\ddvs{(\rn\partial_t^k\phi)}}+\sum_{k=0}^5\sumn r^2\sqb{\ddvs{\rn\Omega_i\partial_t^k\phi}}\\+\sum_{k=0}^4\sum_{i,j=1}^\frac{n(n-1)}{2} r^2\sqb{\ddvs{\rn\Omega_i\Omega_j\partial_t^k\phi}}}\displaybreak[1]\\
  +\int_\St{\tau_0}\Bigl(\sum_{k=0}^6J^N[\partial_t^k\phi]+\sum_{k=0}^6\sumn J^N[\Omega_i\partial_t^k\phi]+\sum_{k=0}^5\sum_{i,j=1}^\frac{n(n-1)}{2} J^N[\Omega_i\Omega_j\partial_t^k\phi],n\Bigr)
\end{multline}
Here $\Omega_i:i=1,\ldots,\frac{n-1}{2}$ are the generators of the spherical isometries of the spacetime \M. (See also Section \ref{sec:high}.)

Among the propositions on linear waves on higher dimensional Schwarzschild black hole spacetimes proven in this paper, we wish to highlight the following conclusions\footnote{The ``redshift'' proposition, and the ``integrated local energy decay'' proposition are to be found on page \pageref{prop:localredshift} in Section \ref{sec:redshift} and page \pageref{prop:ILED} in Section \ref{sec:iled} respectively.}.
\begin{thm}[Energy decay]
\label{thm:ed}
Let $\phi$ be a solution of the wave equation $\Box_g\phi=0$ on $\mathcal{D}\subset\mathcal{M}_m^n$, where $n\geq 3$ and $m>0$, with initial data prescribed on $\St{\tau_0}\ (\tau_0>0)$.
\begin{itemize}
\item If $D\doteq D_2^{(2)}(\tau_0)<\infty$ then there exists a constant $C(n,m)$ such that \begin{equation}\label{eq:thm:ed}E[\phi](\St{\tau})\leq\frac{C\,D}{\tau^2}\quad(\tau>\tau_0)\,.\end{equation}
\item Furthermore if for some $0<\delta<\frac{1}{2}$ and $R>\sqrt[n-2]{8nm/\delta}$ also $D^\prime\doteq D_5^{(4-\delta)}(\tau_0)<\infty$ then there exists a constant $C(n,m,\delta)$ such that \begin{equation}\label{eq:thm:hed}E[\partial_t\phi](\Sigma_{\tau}^\prime)\leq\frac{C\,D^\prime}{\tau^{4-2\delta}}\quad(\tau>\tau_0)\,,\end{equation} where $\Sigma_\tau^\prime\doteq\Sigma_\tau\cap\{r\leq R\}$.
\end{itemize}
\end{thm}
While each of these energy decay statements lend themselves to prove pointwise estimates for $\phi$ and $\partial_t\phi$ respectively (see Section \ref{sec:pointwise}) we would like to emphasize that using the (refined) integrated local energy decay estimates of Section \ref{sec:iled} an interpolation argument allows to improve the pointwise bound on $\phi$ directly in the interior.\footnote{In this paper we use the term ``interior'' to refer to a region of finite radius, i.e.~the term ``interior region'' is used interchangeably with ``a region of compact $r$ (including the horizon)'', and is of course \emph{not} meant to refer to the interior of the black hole, which is not considered in this paper.}
\begin{thm}[Pointwise decay]
\label{thm:pd}
Let $\phi$ be solution of the wave equation as in Theorem \ref{thm:ed}.
If for some $0<\delta<\frac{1}{4}$, $D\doteq D_7^{(4-\delta)}(\tau_0)<\infty\quad(\tau_0>1)$ then there exists a constant $C(n,m,\delta)$ such that \begin{equation}\label{eq:thm:pd}r^\frac{n-2}{2}\lvert\phi\rvert\Bigr\vert_{\Sigma_\tau^\prime}\leq\frac{C\,D}{\tau^{\frac{3}{2}-\delta}}\qquad(\rh\leq r< R,\ \tau>\tau_0)\end{equation} where $\Sigma_\tau^\prime$ and $R$ are as in Theorem \ref{thm:ed}.
\end{thm}

\begin{rmkthm}[Decay rates and method of proof]
Theorems \ref{thm:ed} and \ref{thm:pd} extend the presently known decay results for linear waves on $3+1$-dimensional Schwarzschild black holes to higher dimensions $n>3$; for $3+1$-dimensional Schwarzschild black holes \eqref{eq:thm:ed} was first established in \cite{DRRadDecay}, and (\ref{eq:thm:hed}, \ref{eq:thm:pd}) more recently in \cite{Lid}.
However, both proofs use multipliers with weights in $t$, \cite{DRRadDecay} by using the conformal Morawetz vectorfield in the decay argument, and \cite{Lid} by using in addition the scaling vectorfield. Here we extend \eqref{eq:thm:ed} to higher dimensions $n>3$ in the spirit of \cite{DRNew} only using multipliers with weights in $r$, and provide a new proof of the improved decay results \eqref{eq:thm:hed} and \eqref{eq:thm:pd} in the $n=3$-dimensional case in particular.
\end{rmkthm}

\subsection{Overview of the Proof}
\label{sec:overview}

In this section we give an overview of the work in this paper, and present some of the ideas in the proof that lead to Theorem \ref{thm:ed}; references to previous work is made when useful, but for a more detailed account of previous work on the wave equation on Schwarzschild black hole spacetimes see \S 1.3 in \cite{DRubsrk} and references therein.

\paragraph{Energy Identities.}
Let us recall that the wave equation \eqref{eq:wave} arises as the Euler-Lagrange equation of the action
\begin{equation}
\mathscr{A}\bigl[\phi,g;\mathcal{U}\bigr]=\int_\mathcal{U}(g^{-1})^{\mu\nu}\,\partial_\mu\phi\,\partial_\nu\phi\,\dm{g}\qquad(\mathcal{U}\subset\mathcal{M})
\end{equation}
and we obtain the energy momentum tensor $T^{\mu\nu}$ from the response of the action to variations of the metric:
\begin{equation}
\dot{\mathscr{A}}\vert_\phi \doteq -\int_\mathcal{U} T^{\mu\nu}\dot{g}_{\mu\nu}\dm{g}
\end{equation}
for all variations $\dot{g}=\mathcal{L}_X g$ compactly supported in $\mathcal{U}\subset\mathcal{M}$, where $T^{\mu\nu}$ is symmetric $T^{\mu\nu}=T^{\nu\mu}$.
Thus defined the stationarity of the action yields the conservation law (more generally see \cite{ChMP})
\begin{equation}
\nabla_\nu T^{\mu\nu} = 0\,.
\end{equation}
Indeed, here we find \eqref{eq:intro:energymomentum} and by virtue of the wave equation \eqref{eq:wave}
\begin{equation}
\nabla^\mu T_{\mu\nu}=(\Box_g\phi)(\partial_\nu\phi)=0\,.
\end{equation}
The conservation of $T^{\mu\nu}$ together with the following positivity property of $T^{\mu\nu}$ are crucial for the energy estimates that are central to our approach.
\begin{prop}[positivity condition]
The energy momentum tensor \eqref{eq:intro:energymomentum} satisfies \[T(X,Y)\geq 0\] for all future-directed \emph{causal} vectors $X,Y$ at a point.
\end{prop}

Now let $X$ be a vectorfield on \M. We define the energy current $J^X[\phi]$ associated to the multiplier $X$ by
\begin{equation}
J_\mu^X[\phi] \doteq T_{\mu\nu}[\phi]X^\nu\,.
\end{equation}
Then
\begin{equation}
\begin{split}
K^X &\doteq \nabla\cdot J^X \doteq \nabla^\mu J_\mu^X \\ &= \bigl(\nabla^\mu T_{\mu\nu}\bigr) X^\nu + \frac{1}{2} T_{\mu\nu}\bigl(\nabla^\mu X^\nu+\nabla^\nu X^\mu\bigr) \\
 &= {}^{(X)}\pi^{\mu\nu}T_{\mu\nu}
\end{split}
\end{equation}
where we have used that $T_{\mu\nu}$ is conserved and symmetric.
Here
\begin{equation}
{}^{(X)}\pi(Y,Z) \doteq \frac{1}{2}(\mathcal{L}_X g)(Y,Z) = \frac{1}{2} g(\nabla_Y X,Z)+\frac{1}{2}g(Y,\nabla_Z X)
\end{equation}
is the \emph{deformation tensor} of $X$.
\begin{remark}
If $X$ is a Killing field, i.e. $X$ generates an isometry of $g$, ${}^{(X)}\pi=0$, then $K^X=0$, i.e. $J^X$ is conserved.
\end{remark}
\noindent In the following we shall refer to 
\begin{equation}
  \label{eq:ei}
  \int_\mathcal{D} K^X\dm{g}=\int_{\partial\mathcal{D}}{}^\ast J^X
\end{equation}
as the \emph{energy identity for $J^X$ (or simply $X$) on $\mathcal{D}$}, where $\mathcal{D}\subset\mathcal{M}$; (this is of course the content of Stokes' Theorem, and ${}^\ast J$ denotes the Hodge-dual of $J$, see also Appendix \ref{sec:integration}). Typically the terms are rearranged so that we have the integrals on the future boundary of $\mathcal{D}$ together with the bulk spacetime integral on the left hand side, and the corresponding integrals on the past boundary of $\mathcal{D}$ on the right hand side; often positive terms on the left hand side are dropped, and we refer to the resulting inequality as \emph{energy inequality}. Moreover we refer to $X$ in \eqref{eq:ei} as the \emph{multiplier vectorfield}.
One can say that the entire paper is concerned with the construction of vectorfields $X$, associated currents $J^X$ and their modifications, and the application of \eqref{eq:ei} and various derived energy inequalities to appropriately chosen domains $\mathcal{D}\subset\mathcal{M}$.

\smallskip
The new approach \cite{DRNew} to obtaining robust decay estimates requires us to first establish (i) uniform boundedness of energy, (ii) an integrated local energy decay estimate and (iii) good asymptotics towards null infinity.

\paragraph{Redshift effect.} The reason (i) is nontrivial as compared to Minkowski space is that the energy corresponding to the multiplier $\partial_t$ degenerates on the horizon (c.f.~\cite{DRC}); it was recognized in \cite{DRRadDecay}, and formulated more generally in \cite{DRC}, that the redshift property of Killing horizons is the key to obtaining control on the nondegenerate energy. An explicit construction a vectorfield $N$ is given in Section \ref{sec:redshift} which allows us to state the redshift property in the language of multipliers and energy currents, and a proof of the uniform boundedness of the nondegenerate energy is given (independently of other calculations in this paper) in Section \ref{sec:uniformboundedness}.

\paragraph{Integrated local energy decay.} Section \ref{sec:iled} is devoted to establishing (ii). This is achieved by the use of radial multiplier vectorfields of the form $f(\rs)\partial_\rs$ (see Section \ref{sec:radialmult}), first in the high angular frequencies regime in Section \ref{sec:high}, and then more generally in Section $\ref{sec:low}$; the difficulty here lies in overcoming the ``trapping'' obstruction, which is the insight that it is impossible to prove a local decay estimate for the integral of the energy density on spacetime regions which contain the timelike hypersurface of the photon sphere without losing derivatives (see \cite{DRC}). While the latter construction does not require a decomposition into spherical harmonics, it does use a commutation with angular momentum operators, which is then replaced by a commutation with the vectorfield $\partial_t$ only in the proof of the main integrated local energy decay estimate of Prop.~\ref{prop:ILED} in Section \ref{sec:pfILED}, where the results of Section \ref{sec:high} and \ref{sec:low} are combined. It is only for the purpose of that replacement that we recourse to the decomposition on the sphere (used to show the positive definiteness of the current associated to the multiplier for high angular frequencies), but we would like to emphasize that the decay results of Section \ref{sec:decay} albeit with a higher loss of differentiability could also be obtained just on the basis of the general current introduced in Section \ref{sec:low}.
In the context of the Schwarzschild spacetime the need for such vectorfields whose associated currents give rise to positive definite spacetime integrals was first recognised and used in \cite{BS, DRRadDecay}, and such estimates have since then been extended by many authors \cite{MMTT,AC}.

\paragraph{p-hierarchy.} In Section \ref{sec:ed} we use a multiplier of the form $r^p\partial_\vs$ that gives rise to a weighted energy inequality which we consequently exploit in a hierarchy of two steps; this approach --- which yields the corresponding quadratic decay rate in \eqref{eq:thm:ed} --- is pioneered in \cite{DRNew} for a large class of spacetimes, including the $3+1$-dimensional Schwarzschild and Kerr black hole spacetimes. In Section \ref{sec:iid} a further commutation with $\partial_\vs$ is carried out, which allows us to extend the hierarchy of commuted weighted energy inequalities to four steps, yielding the correponding decay rate for the first order energy. The argument involves dealing with an (arbitrarily small) degeneracy of the first order energy density at infinity which corresponds to the $\delta$-loss in the decay estimate \eqref{eq:thm:hed}. In both cases (iii) is ensured by the imposition of higher order finite energy conditions on the initial data.

\paragraph{Interpolation.} The pointwise decay of Theorem \ref{thm:pd} then follows from Theorem \ref{thm:ed} and the (refined) integrated local energy decay estimates of Section \ref{sec:refinements} by a simple interpolation argument in Section \ref{sec:pointwise}.

\paragraph{Final Comments.} The currents in Section \ref{sec:high} and Section \ref{sec:low} and the corresponding integrated local energy decay result already appeared in the Smith-Rayleigh-Knight essay \cite{VSE}. Independently a version of integrated local energy decay was subsequently obtained in \cite{LM}. \cite{VSE} also contained an alternative proof of \eqref{eq:thm:ed} of Theorem \ref{thm:ed} using the conformal Morawetz vectorfield.

\begin{description}
\item[Acknowledgements.]
The author would like to thank Mihalis Dafermos for suggesting this problem and his support and encouragement.
The author also thanks the UK Engineering and Physical Sciences Research Council and the Cambridge European Trust as well as the European Research Council for their financial support.
\end{description}

\section{Global causal geometry of the higher dimensional Schwarzschild solution}
\label{sec:geometry}

In this Section, we give a discussion (in the spirit of \S 3 of \cite{ChF}) of the global geometry of the $n+1$-dimensional Schwarzschild black hole spacetime \cite{TS}, the underlying manifold on which the wave equation is studied in this paper.

The $n+1$-dimensional Schwarzschild spacetime manifold $\mathcal{M}$ ($n\geq3,\ n\in\mathbb{N}$) is spherically symmetric, i.e.~$\mathrm{SO}(n)$ acts by isometry. The group orbits are $(n-1)$-spheres, and the quotient \Q$=$\M$/\mathrm{SO}(n)$ is a 2-dimensional Lorentzian manifold with boundary. The metric $g$ on \M\ assumes the form
\begin{equation}
g=\stackrel{{\scriptscriptstyle \mathcal{Q}}}{\smash[b]{g}}+\gamma_r=\stackrel{{\scriptscriptstyle \mathcal{Q}}}{\smash[b]{g}}+r^2\gn
\end{equation}
where $\stackrel{{\scriptscriptstyle \mathcal{Q}}}{\smash[b]{g}}$ is the Lorentzian metric on \Q\ to be discussed below, $\gn$ is the standard metric on $\Sn$,
and $r$ is the \emph{area radius}, (for the area of the $(n-1)$-sphere at $x\in\mathcal{Q}$ is given by $\wn r^{n-1}(x)$, where $\wn=\frac{2\pi^\frac{n}{2}}{\Gamma(\frac{n}{2})}$ is the area of the unit $(n-1)$-sphere);
or more precisely in local coordinates $x^a:a=1,2$ on \Q, and local coordinates $y^A:A=1,\dots,n-1$ on $\Sn$
\begin{equation*}g_{(x,y)}=g_{ab}(x)\ud x^a\ud x^b+r^2(x)(\gn)_{AB}\ud y^A\ud y^B\,.\end{equation*}

The $n+1$-dimensional Schwarzschild spacetime is a solution of the vacuum Einstein equations, which in other words means that its Ricci curvature vanishes identically.
This implies in particular (see derivation in \cite{VST}) that the area radius function $r$ satisfies the Hessian equations
\begin{equation}\label{eq:hessian}
\nabla_a\partial_br=\frac{(n-2)}{2r}\bigl[1-(\partial^cr)(\partial_cr)\bigr]g_{ab}\ ,
\end{equation}
as a result of which the \emph{mass function} $m$ on \Q\ defined\footnote{We choose the normalization of the mass function to be independent of the dimension $n$; this is motivated by a consideration of the \emph{mass equations} in the presence of matter, see \cite{VST}.} by
\begin{equation}\label{def:hm}
1-\frac{2m}{r^{n-2}}=g^{ab}\,\partial_ar\,\partial_br
\end{equation}
is constant, see \cite{VST}; we take this parameter $m>0$ to be positive.

\begin{figure}[ht]
\centering
\input{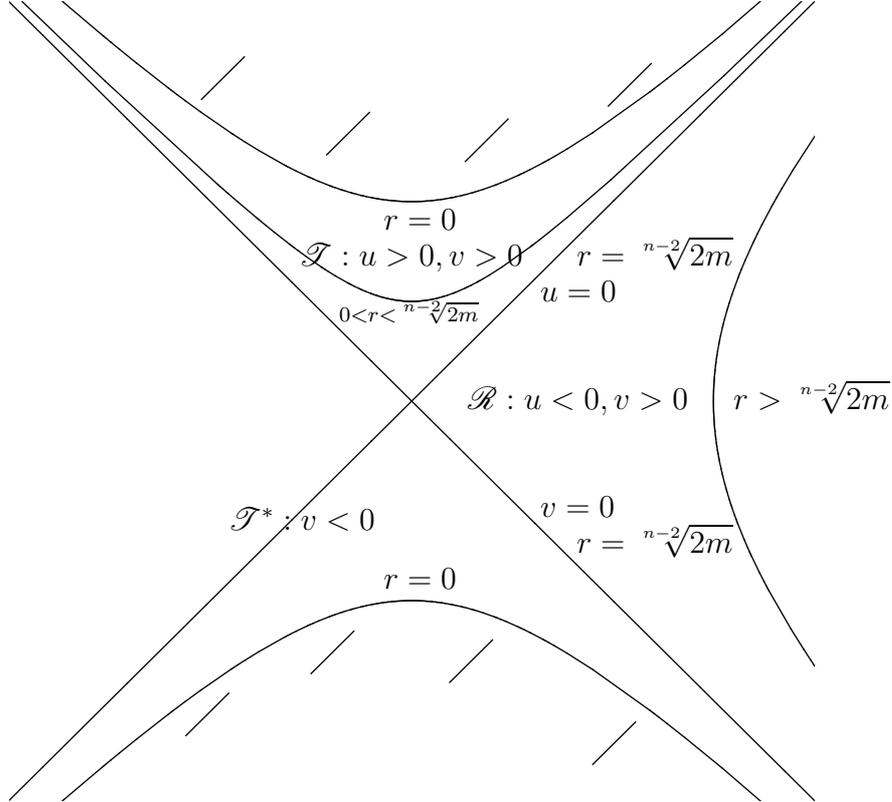}
\caption{Global causal geometry of the Schwarzschild solution.}
\label{gg}
\end{figure}

On \Q\ we choose functions $u$, $v$ whose level sets are outgoing and incoming null curves respectively, which are increasing towards the future. These functions define a null system of coordinates, in which the metric $\stackrel{{\scriptscriptstyle \mathcal{Q}}}{\smash[b]{g}}$ takes the form
\begin{equation} \label{gq}
\stackrel{{\scriptscriptstyle \mathcal{Q}}}{\smash[b]{g}}=-\Omega^2\ud u\ud v\ .
\end{equation}
The Hessian equations \eqref{eq:hessian} in null coordinates read
\begin{subequations}
\begin{align}
\frac{\partial^2r}{\partial u^2}&-\frac{2}{\Omega}\frac{\partial\Omega}{\partial u}\frac{\partial r}{\partial u}=0 \label{eq:hessian:nuu}\\
\frac{\partial^2r}{\partial u\,\partial v}&+\frac{n-2}{r}\frac{\partial r}{\partial u}\frac{\partial r}{\partial v}=-\frac{n-2}{4r}\Omega^2 \label{eq:hessian:nuv}\\
\frac{\partial^2r}{\partial v^2}&-\frac{2}{\Omega}\frac{\partial\Omega}{\partial v}\frac{\partial r}{\partial v}=0 \label{eq:hessian:nvv}\,,
\end{align}
\end{subequations}
and the defining equation for the mass function \eqref{def:hm} is
\begin{equation}\label{eq:hm:n}
1-\frac{2m}{r^{n-2}}=-\frac{4}{\Omega^2}\frac{\partial r}{\partial u}\frac{\partial r}{\partial v}\ .
\end{equation}
The system (\eqref{eq:hessian:nuv}, \eqref{eq:hm:n}) can be rewritten as the partial differential equation
\begin{equation}
  \label{eq:rsuv}
  \frac{\partial\rs}{\partial u\partial v}=0  
\end{equation}
for a new radial function $\rs(r)$ that is related to $r$ by
\begin{equation}
  \label{eq:rsr}
  \frac{\ud\rs}{\ud r}=\frac{1}{\cf}\,.
\end{equation}
A solution of (\eqref{eq:rsuv},\eqref{eq:rsr}) is given by\footnote{Here the representation in terms of null coordinates is such that $\rs=-\infty$ is contained in the $(u,v)$ plane \emph{and} the metric is non-degenerate at $r=\rh$.}
\begin{equation}\label{eq:solrs}
\rs=\frac{1}{(n-2)}\rh\log|uv|\,,
\end{equation}
or
\begin{equation}\label{eq:absuvr}
\begin{split}
|uv| &= e^{\frac{\scriptstyle{(n-2)\,\rs}}{\scriptstyle{\rh}}}\\
&= e^{\frac{\scriptstyle{(n-2)\,r}}{\scriptstyle{\rh}}}\exp\Bigl[\bigl.\int\frac{n-2}{x^{n-2}-1}\ud x\bigr\vert_{x=\frac{r}{\rh}}\Bigr]\ .
\end{split}
\end{equation}
We find more explicitly, by an elementary integration (see \cite{VST}),
\begin{equation} \label{uvr}
uv=
 \begin{cases}

  \displaystyle{e^{\frac{r}{2m}}\bigl(1-\frac{r}{2m}\bigr)}
  & \text{, } n=3\\[2ex]

  \displaystyle{e^{\frac{2r}{\sqrt{2m}}}
  \frac{\bigl(1-\frac{r}{\sqrt{2m}}\bigr)}
  {\bigl(1+\frac{r}{\sqrt{2m}}\bigr)}

  }
  & \text{, } n=4\\[4ex]

  \displaystyle{
  \begin{aligned}
   &e^{\frac{(n-2)\,r}{\rh}}\Bigl(1-\frac{r}{\rh}\Bigr)
    \begin{cases}
    \ 1 & \text{, }n\text{ odd}\\
    \bigl(1+\frac{r}{\rh}\bigr)^{-1} & \text{, }n\text{ even}
    \end{cases}
   \\
   &\quad\times\prod_{j=1}^{[\frac{n-3}{2}]}\Bigl(\frac{r^2}{(2m)^\frac{2}{n-2}}-2\cos\bigl(\frac{2\pi j}{n-2}\bigr)\frac{r}{(2m)^\frac{1}{n-2}}+1\Bigr)^{\cos(2\pi j\frac{n-3}{n-2})} \\
   &\quad\times\prod_{j=1}^{[\frac{n-3}{2}]}\exp\biggl[2\sin\bigl(2\pi j\frac{n-3}{n-2}\bigr)\arctan\Bigl(\frac{\frac{r}{\rh}-\cos(\frac{2\pi j}{n-2})}{\sin(\frac{2\pi j}{n-2})}\Bigr)\biggr]
  \end{aligned}
  }
  & \text{, } n\geq 5

 \end{cases}
\end{equation}
Note, in particular that the $u=0$ and $v=0$ lines are the constant $r=\rh$ curves, and that all other curves of constant radius are hyperbolas in the $(u,v)$ plane --- timelike for $r>\rh$, spacelike for $r<\rh$. This outlines the well-known \emph{global} causal geometry of the Schwarzschild solution (see figure \ref{gg}).

It is easy to see \cite{VST} that for \eqref{eq:solrs} the trapped region, the apparent horizon, the exterior, and the antitrapped regions respectively are given by
\begin{align*}
\mathscr{T}&\doteq\Bigl\{(u,v)\in\mathcal{Q}:\frac{\partial r}{\partial u}<0,\frac{\partial r}{\partial v}<0\Bigr\}=\Bigl\{(u,v)\in\mathcal{Q}:u>0,v>0\Bigr\}\displaybreak[0]\\
\mathscr{A}&\doteq\Bigl\{(u,v)\in\mathcal{Q}:\frac{\partial r}{\partial u}<0,\frac{\partial r}{\partial v}=0\Bigr\}=\Bigl\{(u,v)\in\mathcal{Q}:u=0,v>0\Bigr\}\displaybreak[0]\\
\mathscr{R}&\doteq\Bigl\{(u,v)\in\mathcal{Q}:\frac{\partial r}{\partial u}<0,\frac{\partial r}{\partial v}>0\Bigr\}=\Bigl\{(u,v)\in\mathcal{Q}:u<0,v>0\Bigr\}\displaybreak[0]\\
\mathscr{T}^\ast&\doteq\Bigl\{(u,v)\in\mathcal{Q}:\frac{\partial r}{\partial u}>0\Bigr\}=\Bigl\{(u,v)\in\mathcal{Q}:v<0\Bigr\}\,.
\end{align*}
Note this forms a partition of $\mathcal{Q}=\overline{\mathscr{T}\cup\mathscr{A}\cup\mathscr{R}\cup\mathscr{T}^\ast}$, and that in view of \eqref{eq:hm:n} $r<\rh$ in $\mathscr{T}$, $r=\rh$ in $\mathscr{A}$ and $r>\rh$ in $\mathscr{R}$.
We shall refer to
\begin{equation}
  \label{def:doc}
  \mathcal{D}\doteq\overline{\mathcal{R}}=\Bigl\{(u,v)\in\mathcal{Q}:u\leq0,v\geq0\Bigr\}
\end{equation}
as the \emph{domain of outer communications}.

Finally, 
\begin{align}
\Omega^2 
 &= \begin{cases}
  \displaystyle{
   4\frac{(2m)^3}{r}e^{-\frac{r}{2m}} 
  }&\text{, }n=3\\[2ex]
  \displaystyle{
   \bigl(\frac{2m}{r}\bigr)^2\bigl(\frac{r}{\sqrt{2m}}+1\bigr)^2e^{-\frac{2r}{\sqrt{2m}}} 
  }&\text{, }n=4\\[2ex]
  \displaystyle{
   \begin{aligned}
    &\bigl(\frac{2}{n-2}\bigr)^2\frac{(2m)^\frac{n}{n-2}}{r^{n-2}}\begin{cases}1&\text{, n odd}\\(\frac{r}{\rh}+1)^2&\text{, n even}\end{cases}\\
    &\quad\times\prod_{j=1}^{[\frac{n-3}{2}]}\Bigl(\frac{r^2}{(2m)^\frac{2}{n-2}}-2\cos(\frac{2\pi j}{n-2})\frac{r}{\rh}+1\Bigr)^{1-\cos(2\pi j\frac{n-3}{n-2})}\\
    &\quad\times\prod_{j=1}^{[\frac{n-3}{2}]}\exp\Bigl[-2\sin(2\pi j\frac{n-3}{n-2})\arctan\bigl(\frac{\frac{r}{\rh}-\cos(\frac{2\pi j}{n-2})}{\sin(\frac{2\pi j}{n-2})}\bigl)\Bigl]\\
    &\quad\times e^{-\frac{(n-2)\,r}{\rh}}
   \end{aligned}
  }&\text{, }n\geq 5
 \end{cases} \label{omegasqr}
\end{align}

One may now also think of $r$ as a function of $u,v$ implicitly defined by \eqref{uvr}. In $\mathscr{R}$ where $r>\rh$ (and $v-u>|u+v|$) $r$ may be complemented by
\begin{equation} \label{def:t}
t=\frac{2}{n-2}\rh\operatorname{artanh}\bigl(\frac{u+v}{v-u}\bigr)\,;
\end{equation}
note
\begin{equation}\label{eq:dt}
\ud t=\frac{1}{n-2}\rh\,(\frac{1}{v}\ud v-\frac{1}{u}\ud u)
\end{equation}
and we will denote by $\overline{\Sigma}_t$ the corresponding level sets in $\mathcal{D}$.

We find in these coordinates the classic expression for the Schwarzschild metric in the exterior region:
\begin{equation}\label{eq:sm}
g=-\bigl(1-\frac{2m}{r^{n-2}}\bigr)\ud t^2+\bigl(1-\frac{2m}{r^{n-2}}\bigr)^{-1}\ud r^2+r^2\gn\ .
\end{equation}
In Regge-Wheeler coordinates $(t,\rs)$, where $\rs$ is centered at the photon sphere $r=\rph$, 
\begin{equation}\label{rscph}\rs=\int_{\rphp}^r\frac{1}{1-\frac{2m}{{r'}^{n-2}}}\ud r'\,,\end{equation}
the metric obviously takes the conformally flat form
\begin{equation}
g=\bigl(1-\frac{2m}{r^{n-2}}\bigr)(-\ud t^2+\ud{\rs}^2)+r^2\gn\ .
\end{equation}
We shall also use the Eddington-Finkelstein coordinates
\begin{equation}
\us=\frac{1}{2}(t-\rs) \qquad\quad \vs=\frac{1}{2}(t+\rs)
\end{equation}
which are again double null coordinates:
\begin{equation}\label{eq:mefc}
g=-4\bigl(1-\frac{2m}{r^{n-2}}\bigr)\ud\us\ud\vs+r^2\gn\ .
\end{equation}
The two systems of null coordinates in $\mathscr{R}$ are related by
\begin{equation}
u=-e^{-\frac{(n-2)\,\us}{\rh}}\qquad v=e^\frac{(n-2)\,\vs}{\rh}\ .
\end{equation}

\section{The Red-Shift Effect}
\label{sec:redshift}

In this section we prove a manifestation of the \emph{local redshift effect}  in the Schwarzschild geometry of Section \ref{sec:geometry} in the framework of multiplier vectorfields.

\begin{prop}[local redshift effect]\label{prop:localredshift}
Let $\phi$ be a solution of the wave equation \eqref{eq:wave}, then there exists a $\varphi_t$-invariant future-directed smooth vectorfield $N$ on $\mathcal{D}$, two radii $\rh<r_0^{(N)}<r_1^{(N)}$, and a constant $b>0$ such that
\begin{equation}
  \label{eq:localredshift}
  K^N(\phi)\geq b\,(J^N(\phi),N)\qquad(\rh\leq r<r_0^{(N)})
\end{equation}
and $N=T\ (r\geq r_1^{(N)})$.
\end{prop}

The vectorfield $N$ will be constructed explicitly with the following vectorfields.

\paragraph{$T$-vectorfield.}

Here $\varphi_t$ is the 1-parameter group of diffeomorphisms generated by the vectorfield
\begin{equation}
T=\frac{1}{2}\frac{n-2}{\rh}\bigl(v\frac{\partial}{\partial v}-u\frac{\partial}{\partial u}\bigr)\,;
\end{equation}
note that in $\mathscr{R}$ where $r>\rh$ (recall \eqref{eq:dt}) \[T=\frac{\partial}{\partial t}\,.\]
$T$ is a Killing vectorfield, \begin{equation}\deformt{T}=0\,.\label{Tkilling}\end{equation}
For,
\begin{gather}
\deformt{T}_{uu}=0\qquad\deformt{T}_{vv}=0\notag\\
\deformt{T}_{uv}=\frac{1}{4}\frac{n-2}{\rh}g_{uv}\bigl(v\frac{\partial r}{\partial v}-u\frac{\partial r}{\partial u}\bigr)\frac{\partial\log\Omega^2}{\partial r}=0\notag\\
\deformt{T}_{aA}=0\\
\deformt{T}_{AB}=\frac{1}{2}\frac{n-2}{\rh}\bigl(v\frac{\partial r}{\partial v}-u\frac{\partial r}{\partial u}\bigr)\,r\,(\gn)_{AB}=0\,.\notag
\end{gather}
$T$ is timelike in the exterior, spacelike in the interior of the black hole and \emph{null on the horizon},
\begin{equation}
g(T,T) = \frac{1}{4}\frac{(n-2)^2}{(2m)^\frac{2}{n-2}}\,uv\,\Omega^2 = -\bigl(1-\frac{2m}{r^{n-2}}\bigr)  \begin{cases} <0 & r>\rh \\ =0 & r=\rh \\ >0 & r<\rh \end{cases}
\end{equation}
In particular,
\begin{gather}
T\onh=\frac{1}{2}\frac{n-2}{\rh}\,v\frac{\partial}{\partial v}\\
T\vert_{\mathcal{H}^+\cap\mathcal{H}^-}=0\,.\notag
\end{gather}

\paragraph{$Y$-vectorfield.}

Let us also define a vectorfield $Y$ on \Hp conjugate to $T$:
\begin{equation}
Y\onh=-\frac{2}{\frac{\partial r}{\partial u}}\frac{\partial}{\partial u}
\end{equation}
Indeed, \begin{equation}g(T,Y)\onh=-2\end{equation} because \[\Omega^2\onh=-4\frac{\rh}{n-2}\frac{1}{v}\frac{\partial r}{\partial u}\,.\]
Furthermore, as a consequence of \eqref{eq:hessian:nuv}
\begin{equation*}
\Onh{\frac{\partial^2 r}{\partial u\,\partial v}}=-\Onh{\frac{n-2}{4r}\Omega^2} = \Onh{\frac{1}{v}\frac{\partial r}{\partial u}}
\end{equation*}
we have
\begin{equation} \label{bTY}
\begin{split}
[T,Y]\onh &= [T,Y]^u \Onh{\frac{\partial}{\partial u}}+[T,Y]^v\Onh{\frac{\partial}{\partial v}} \\
 &= \frac{n-2}{\rh}\frac{1}{\frac{\partial r}{\partial u}}\Bigl[v\frac{1}{\frac{\partial r}{\partial u}}\frac{\partial^2 r}{\partial u\,\partial v}-1\Bigr]\Onh{\frac{\partial}{\partial u}}\\
 &= 0\,.
\end{split}
\end{equation}

We can now prove that the surface gravity of the event horizon is \emph{positive}; this is essential for the existence of the redshift effect, (see more generally \cite{DRC}, and also \cite{Ar} for work where this is not the case).

\begin{lemma}[surface gravity] \label{lemma:surfacegravity}
On \Hp \begin{equation} \nabla_T T=\sg T \end{equation}
with \begin{equation}\label{eq:sg} \sg=\frac{1}{2}\frac{n-2}{\rh}>0\,.\end{equation}
$\sg$ is called the \emph{surface gravity}.
\end{lemma}

\begin{note} $T=\sg(v\frac{\partial}{\partial v}-u\frac{\partial}{\partial u})$ \end{note}
\begin{proof}
Since $\deformt{T}=0$, we have \[g(\nabla_X T,Y)=-g(\nabla_Y T,X)\] for all vectorfields $X,Y$.
Therefore,
\begin{gather*}
g(\nabla_T T,T)=-g(\nabla_T T,T)=0\\
g(\nabla_T T, E_A) = -g(\nabla_{E_A}T,T) = -\frac{1}{2}E_A\cdot g(T,T) = 0 \quad :A=1,\ldots,n-1,
\end{gather*}
because $g(T,T)=0$ on \Hp, and similarly \[g(\nabla_T T,Y) = -\frac{1}{2}Y\cdot g(T,T)\,.\]
Now, \[g(T,T)=-\frac{n-2}{\rh}\,u\frac{\partial r}{\partial u}\]
so \[Y\cdot g(T,T)\onh = 2 \frac{n-2}{\rh}\,.\]
We obtain
\begin{equation*}
\begin{split}
\nabla_T T &= -\frac{1}{2}g(\nabla_T T,T)Y-\frac{1}{2}g(\nabla_T T,Y)T+\sum_{j=1}^{n-1}g(\nabla_T T,E_A)E_A \\
 &= \frac{1}{2}\frac{n-2}{\rh}\,T\,.\qedhere
\end{split}
\end{equation*}
\end{proof}

\noindent Alternatively, $\sg$ is characterized by \begin{equation} \nabla_T Y = - \sg Y \end{equation} on \Hp.
Clearly \[g(\nabla_T Y,Y)=\frac{1}{2}T\cdot g(Y,Y)=0\] since $Y$ is null along \Hp, and
\[g(\nabla_T Y,T)\stackrel{\eqref{bTY}}{=} g(\nabla_Y T,T) \stackrel{\eqref{Tkilling}}{=} -g(\nabla_T T,Y) = 2\sg\,;\]
also \[g(\nabla_T Y,E_A) \stackrel{\eqref{bTY}}{=} g(\nabla_Y T, E_A) \stackrel{\eqref{Tkilling}}{=} -g(\nabla_{E_A} T,Y) =0 \quad :A=1,\ldots,n-1,\] because $\nabla_{E_A} T = 0$.
Note, for later use, \begin{equation} \nabla_{E_A} Y = -\frac{2}{\rh}\,E_A \end{equation} on \Hp.

We defined $Y$ on \Hp conjugate to $T$, $g(T,Y)\onh=-2$. Next we extend $Y$ to a neighborhood of the horizon by
\begin{equation*}
\nabla_Y Y=-\sigma(Y+T) \qquad (\sigma>\frac{16}{n-2}(2m)^\frac{3}{n-2})
\end{equation*}
and then we extend $Y$ to $\mathscr{R}$ by Lie-transport along the integral curves of $T$: \[[T,Y]=0\,.\]

\begin{prop}[redshift] \label{prop:redshift}
For the future-directed timelike vectorfield
\begin{equation} \label{def:N} N=T+Y \end{equation}
there is a $b>0$ such that on \Hp 
\begin{equation} \label{eq:lreH} K^N\geq b\,(J^N,N)\,.\end{equation}
\end{prop}
\begin{proof}
Let us calculate
\begin{equation*}
\begin{split}
K^Y =& \deformt{Y}^{\mu\nu}T_{\mu\nu} \\
 =& \frac{1}{4}\Bigl\{\deformt{Y}(T,T)\,T(Y,Y)+2\deformt{Y}(T,Y)\,T(Y,T)+\deformt{Y}(Y,Y)\,T(T,T)\Bigr\} \\
&-\sum_{A=1}^{n-1}\Bigl\{\deformt{Y}(E_A,Y)\,T(E_A,T)+\deformt{Y}(E_A,T)\,T(E_A,Y)\Bigr\} \\
&+\sum_{A,B=1}^{n-1}\deformt{Y}(E_A,E_B)\,T(E_A,E_B)
\end{split}
\end{equation*}
Now, on one hand, on \Hp,
\begin{gather*}
\deformt{Y}(T,T)=g(\nabla_T Y,T)=2\sg\\
\deformt{Y}(T,Y)=\frac{1}{2}g(\nabla_T Y,Y)+\frac{1}{2}g(T,\nabla_Y Y)=\sigma\\
\deformt{Y}(Y,Y)=g(\nabla_Y Y,Y)=2\sigma\\
\deformt{Y}(E_A,Y)=\frac{1}{2}g(\nabla_{E_A} Y,Y)+\frac{1}{2}g(E_A,\nabla_Y Y)=0\\
\deformt{Y}(E_A,T)=\frac{1}{2}g(\nabla_{E_A} Y,T)+\frac{1}{2}g(E_A,\nabla_T Y)=0\\
\deformt{Y}(E_A,E_B)=\frac{1}{2}g(\nabla_{E_A} Y,E_B)+\frac{1}{2}g(E_A,\nabla_{E_B} Y)=-\frac{2}{\rh}\delta_{AB}\,.
\end{gather*}
Thus \[K^Y=\frac{1}{2}\sg\,T(Y,Y)+\frac{1}{2}\sigma\,T(Y+T,T)-\frac{2}{\rh}\sum_{A=1}^{n-1}T(E_A,E_A)\,.\]
On the other hand, on \Hp,
\begin{gather*}
T(Y,Y)=\Bigl(\frac{2}{\frac{\partial r}{\partial u}}\frac{\partial\phi}{\partial u}\Bigr)^2\\
T(Y,T)=\normsph{\nablab\phi}^2\\
T(T,T)=\bigl(\sg v \frac{\partial\phi}{\partial v}\bigr)^2
\end{gather*}
and, on \Hp,
\begin{multline*}T(E_A,E_B)=(E_A\cdot\phi)(E_B\cdot\phi)-\frac{1}{2}(2m)^\frac{2}{n-2}\delta_{AB}\normsph{\nablab\phi}\\-\frac{1}{2}(n-2)\rhp \frac{v}{\frac{\partial r}{\partial u}}\delta_{AB}(\frac{\partial\phi}{\partial u})(\frac{\partial\phi}{\partial v})\,.\end{multline*}
Using Cauchy's inequality, on \Hp,
\begin{equation*}
\begin{split}
-\frac{2}{\rh}&\sum_{A=1}^{n-1}T(E_A,E_A)=\\
=&(n-3)\rhp\normsph{\nablab\phi}^2+(n-2)(n-1)\frac{v}{\frac{\partial r}{\partial u}}(\frac{\partial\phi}{\partial u})(\frac{\partial\phi}{\partial v})\\
\geq&(n-3)\rhp T(Y,T)-\frac{1}{4}\sg T(Y,Y)\\&\quad-\frac{1}{\sg}\frac{2(n-1)}{(n-2)}(2m)^\frac{2}{n-2}\,T(T,T)\\
\geq&-\frac{1}{4}\sg T(Y,Y)-\frac{n-1}{\sg^2}\rhp T(T,T)\,.
\end{split}
\end{equation*}
Since we have chosen $\sigma > 2\frac{n-1}{\sg^2}\rhp$, $K^Y$ has a sign,
\[K^Y\geq\frac{1}{4}\sg\,T(Y,Y)+\sigma^\prime\,T(Y+T,T)\]
for $0<\sigma^\prime<\frac{\sigma}{2}-\frac{n-1}{\sg^2}\rhp$, or
\[K^Y\geq b\,T(Y+T,Y+T)\] for $0<b<\min\{\frac{\sg}{4},\frac{\sigma^\prime}{2}\}$.
This yields the result \[K^N=K^Y\geq b\,T(N,N)=b\,(J^N,N)\,.\qedhere\]
\end{proof}

Finally, we find an explicit expression for $Y$.
Consider the vectorfield
\begin{equation*}
\hat{Y}=-\frac{2}{\frac{\partial r}{\partial u}}\frac{\partial}{\partial u}
\end{equation*}
on $\mathscr{R}\cup\mathscr{A}$ formally defined by the expression for $Y$ on \Hp.
In $\mathscr{R}$ \[\hat{Y}=\frac{2}{1-\frac{2m}{r^{n-2}}}\dus\,.\]
$\hat{Y}$ generates geodesics, this being a consequence of the Hessian equations \eqref{eq:hessian:nuu},
\[\nabla_{\hat{Y}}\hat{Y}=\bigl(\frac{2}{\frac{\partial r}{\partial u}}\bigr)^2\Bigl[-\frac{1}{\frac{\partial r}{\partial u}}\frac{\partial^2r}{\partial u^2}+\frac{2}{\Omega}\frac{\partial\Omega}{\partial u}\Bigr]\frac{\partial}{\partial u}=0\,,\]
and is Lie-transported by $T$:
\begin{equation*}
[T,\hat{Y}]=\frac{2}{(\frac{\partial r}{\partial u})^2}\bigl([T,\frac{\partial}{\partial u}]\cdot r\bigr)\frac{\partial}{\partial u}-\frac{2}{\frac{\partial r}{\partial u}}[T,\frac{\partial}{\partial u}]=-\sg\hat{Y}+\sg\hat{Y}=0
\end{equation*}
because $[T,\frac{\partial}{\partial u}]=\sg\frac{\partial}{\partial u}$.
$Y$ as constructed above coincides with
\begin{equation} \label{Yco}
Y=\alpha(r)\hat{Y}+\beta(r)T
\end{equation}
where
\begin{gather*}
\alpha(r)=1+\frac{\sigma}{4\sg}\bigl(1-\frac{2m}{r^{n-2}}\bigr)\\
\beta(r)=\frac{\sigma}{4\sg}\bigl(1-\frac{2m}{r^{n-2}}\bigr)\,.
\end{gather*}
Indeed, on \Hp,
\[Y\onh=\hat{Y}\onh=-\frac{2}{\frac{\partial r}{\partial u}}\Onh{\frac{\partial}{\partial u}}\]
and
\begin{equation*}
\begin{split}
\Onh{\nabla_Y Y} &= \Onh{\nabla_{\hat{Y}}Y}=(\hat{Y}\cdot\alpha)\hat{Y}\onh+\Onh{\nabla_{\hat{Y}}\hat{Y}}+(\hat{Y}\cdot\beta)T\onh\\
&= -\sigma\,(Y+T)\onh
\end{split}
\end{equation*}
since
\begin{gather*}
\hat{Y}\cdot\alpha\onh=\frac{\sigma}{4\sg}(n-2)\frac{2m}{r^{n-1}}\hat{Y}\cdot r\onh=-\sigma\\
\hat{Y}\cdot\beta\onh=-\sigma
\end{gather*}
and $Y$ remains Lie-transported by $T$:
\begin{equation*}
[T,Y]=(T\cdot\alpha)\hat{Y}+(T\cdot\beta)T+\alpha\,[T,\hat{Y}]+\beta\,[T,T]=0
\end{equation*}
since \[T\cdot\alpha=0=T\cdot\beta\,.\]
Thus the vectorfield $Y$ is given explicitly by
\begin{equation}\label{Y}
Y=\begin{cases}
-\frac{2}{\frac{\partial r}{\partial u}}\frac{\partial}{\partial u} & \text{on \Hp} \\
\Bigl[1+\frac{\sigma}{4\sg}\bigl(1-\frac{2m}{r^{n-2}}\bigr)\Bigr]\frac{2}{1-\frac{2m}{r^{n-2}}}\dus+\frac{\sigma}{4\sg}\bigl(1-\frac{2m}{r^{n-2}}\bigr)\frac{\partial}{\partial t} & \text{in } \mathscr{R}
\end{cases}
\end{equation}

Clearly, by continuity, we can choose two values $\rh<r_0^{(N)}<r_1^{(N)}<\infty$ and set
\begin{equation*}
N=\begin{cases} T+Y & \rh\leq r\leq r_0^{(N)} \\ T & r\geq r_1^{(N)} \end{cases}
\end{equation*}
with a smooth $\varphi_t$-invariant transition of the timelike vectorfield $N$ in $r_0^{(N)}\leq r\leq r_1^{(N)}$, such that \eqref{eq:lreH} extends to the neighborhood $\rh<r<r_0^{(N)}$ of the event horizon.

\begin{remark} For a geometric interpretation of Prop.~\ref{prop:redshift} see \cite{VST}, and also \cite{DRC}.
\end{remark}

\section{Integrated Local Energy Decay}
\label{sec:iled}

In this section we prove several \emph{integrated local energy decay} statements, i.e.~estimates on the energy density of solutions of \eqref{eq:wave} integrated on (bounded) \emph{space-time} regions; this in an essential ingredient for the decay mechanism employed in Section \ref{sec:decay}.

\begin{figure}[bt]
  \begin{center}
    \input{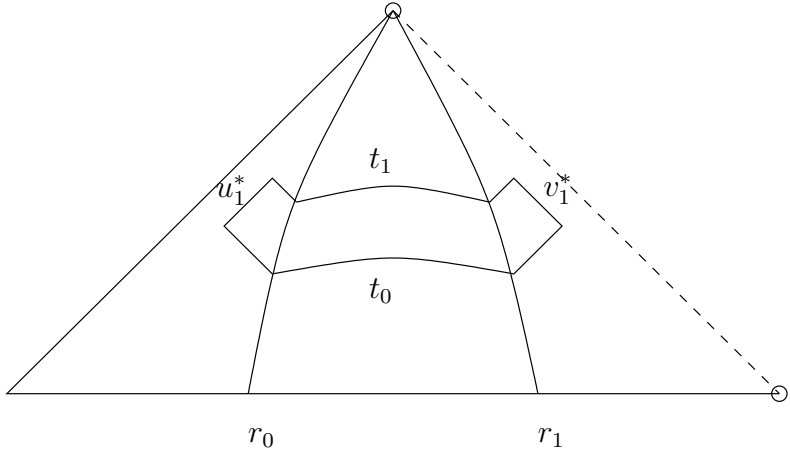}
    \caption{The region $\mathcal{R}_{r_0,r_1}(t_0,t_1,u_1^\ast,v_1^\ast)$.}
    \label{fig:trapchar}
  \end{center}
\end{figure}

Let $\mathcal{R}_{r_0,r_1}(t_0, t_1, u_1^\ast, v_1^\ast)$ be the region composed
of a trapezoid and characteristic rectangles as follows, (see figure
\ref{fig:trapchar}):
\begin{align}
  \label{def:Rfinite}
  \mathcal{R}_{r_0,r_1}(t_0,t_1,u_1^\ast,v_1^\ast) \doteq &\phantom{\cup}\Bigl\{(t,r):t_0\leq t\leq t_1, r_0\leq r\leq r_1\Bigr\}\\
  &\cup\Bigl\{(t,r):r\leq r_0,\frac{1}{2}(t-\rs)\leq u_1^\ast,t_0+r_0^\ast\leq t+\rs\leq t_1+r_0^\ast\Bigr\}\notag\\
  &\cup\Bigl\{(t,r):r\geq r_1,\frac{1}{2}(t+\rs)\leq
  v_1^\ast,t_0-r_1^\ast\leq t-\rs\leq t_1-r_1^\ast\Bigr\}\notag
\end{align}
We denote by
\begin{equation}\label{def:Rinfty}
  \mathcal{R}^\infty_{r_0,r_1}(t_0)\doteq\bigcup_{t_1\geq t_0}\ \bigcup_{u_1^\ast\geq\frac{1}{2}(t_1-r_0^\ast)}\ \bigcup_{v_1^\ast\geq\frac{1}{2}(t_1+r_1^\ast)}\mathcal{R}(t_0,t_1,u_1^\ast,v_1^\ast)
\end{equation}
and its past boundary by
\begin{equation}
  \label{def:Sigmatau}
  \Sigma_{\tau_0}\doteq\partial^-\mathcal{R}^\infty_{r_0,r_1}(t_0)\qquad \tau_0=\frac{1}{2}(t_0-r_1^\ast)\,.
\end{equation}

We shall first state the central estimate.

\begin{prop}[integrated local energy decay estimate]\label{prop:ILED}
  Let $\phi$ be a solution of the wave equation $\Box_g\phi=0$. Then
  there exist $\rhp<r_0<r_1$ and a constant $C(n,m)$ depending on the dimension $n$ and the
  mass $m$, such that
  \begin{multline}\label{eq:ILED}
    \int_{\mathcal{R}_{r_0,r_1}^\infty(t_0)}\Biggl\{\frac{1}{r^n}\bigl(\frac{\partial\phi}{\partial\rs}\bigr)^2+\frac{1}{r^{n+1}}\Bigl(\frac{\partial\phi}{\partial t}\Bigr)^2+\frac{1}{r^3}\bigl(\cf\bigr)\normsph{\nablab\phi}^2\Biggr\}\dm{g}\\
    \leq C(n,m)\int_\St{\tau_0}\Bigl(J^T(\phi)+J^T(T\cdot\phi),n\Bigr)
  \end{multline}
 for any $t_0\geq 0$, where $\tau_0=\frac{1}{2}(t_0-r_1^\ast)$.
\end{prop}

The degeneracy at infinity can in fact be improved:

\begin{prop}[improved integrated local energy decay estimate]\label{prop:IILED}
  Let $\phi$ be a solution of the wave equation $\Box_g\phi=0$, then
  there exists a constant $C(n,m,\delta)$ for each $0<\delta<1$ such
  that
  \begin{multline}
    \int_{\mathcal{R}_{r_0,r_1}^\infty(t_0)}\biggl\{\frac{1}{r^{1+\delta}}\bigl(\ddrs{\phi}\bigr)^2+\frac{1}{r^{1+\delta}}\bigl(\ddt{\phi}\bigr)^2+\frac{1}{r}\bigl(\cf\bigr)\normsph{\nablab\phi}^2\biggl\}\dm{g}\leq\\
    \leq
    C(n,m,\delta)\int_{\St{\tau_0}}\Bigl(J^T(\phi)+J^T(T\cdot\phi),n\Bigr)
  \end{multline}
    for any $t_0\geq 0$, where $r_0<r_1$ are as above, and $\tau_0=\frac{1}{2}(t_0-r_1^\ast)$.
  \end{prop}

As a consequence of the redshift effect of Section \ref{sec:redshift}, and the uniform boundedness of the nondegenerate energy (which is proven independently in Section \ref{sec:uniformboundedness}), we can infer in a more geometric formulation:

\begin{cor}[nondegenerate integrated local energy decay]\label{cor:nILED}
  Let $\phi$ be a solution of \eqref{eq:wave}, then for any $R>\rh$ there exists a constant $C(n,m,R)$ such that
  \begin{equation}
    \label{eq:nILED}
    \int_{\tau^\prime}^\tau\ud\taub\int_\Stp{\taub}\ned{N}{\phi}\leq
    C(n,m,R)\int_\St{\tau^\prime}\Bigl(J^N(\phi)+J^T(T\cdot\phi),n\Bigr)\,,
  \end{equation}
  for all $\tau^\prime<\tau$, where $\Stp{\tau}\doteq\St{\tau}\cap\{r\leq R\}$.
\end{cor}

\begin{proof}
Let \[\mathcal{R}^\prime(\taup,\tau)\doteq\mathrm{J}^-(\Stp{\tau})\cap\mathrm{J}^+(\St{\taup})\,.\]
In $\mathcal{R}^\prime(\taup,\tau)\cap\{r<r_0^{(N)}\}$ we have by Prop.~\ref{prop:localredshift} \[\ned{N}{\phi}\leq\frac{1}{b}K^N(\phi)\,,\]
and in $\mathcal{R}^\prime(\taup,\tau)\cap\{r\geq r_1^{(N)}\}$ trivially $(J^N(\phi),n)\leq(J^T(\phi),n)$.
Therefore using the energy identity for $N$ on $\mathcal{R}^\prime(\taup,\tau)$ the estimate \eqref{eq:nILED} follows from Prop.~\ref{prop:uniformboundedness} and Prop.~\ref{prop:ILED}.
\end{proof}

In the above, no control is obtained on a spacetime integral of $\phi^2$ itself; however, all that is needed for the decay argument of Section \ref{sec:decay} is an estimate for the integal of $\phi^2$ on timelike boundaries.
\begin{prop}[zeroth order terms on timelike boundaries]\label{prop:zeroth}
  Let $\phi$ be solution of the wave equation \eqref{eq:wave}, and $R>\sqrt[n-2]{8nm}$. 
  Then there is a constant $C(n,m,R)$ such that
  \begin{multline}
    \int_{2\tau^\prime+R^\ast}^{2\tau+R^\ast}\ud t\int_\Sn\dm{\gn}\,\phi^2\vert_{r=R}\leq\\
    \leq C(n,m,R)\int_{2\tau^\prime+R^\ast}^{2\tau+R^\ast}\ud t\int_\Sn\dm{\gn}\,\Bigl\{\sqb{\ddrs{\phi}}+\sqv{\nablab\phi}\Bigr\}\bigr\vert_{r=R}\\
    +C(n,m,R)\int_\St{\tau^\prime}\ned{T}{\phi}
  \end{multline}
for all $\tau^\prime<\tau$.
\end{prop}

The central result of Prop.~\ref{prop:ILED} combines results for two different regimes, that of high angular frequencies and that of low angular frequencies.
First we will use radial multiplier vectorfields to construct positive definite currents to deal with the former regime, and then a more general current using a commutation with angular momentum operators for the latter.

\begin{remark} The specific parametrization \eqref{def:Sigmatau} has technical advantages, but $\St{\tau}$ can in principle be replaced by a foliation of strictly \emph{spacelike} hypersurfaces terminating at future null infinity and crossing the event horizon to the future of the bifurcation sphere.
\end{remark}

\subsection{Radial multiplier vectorfields}
\label{sec:radialmult}
A \emph{radial multiplier} is a vectorfield of the form
\begin{equation}
  X=f(\rs)\frac{\partial}{\partial\rs}\,.
\end{equation}
We would like the associated current to be positive, however we find in general, as it is shown below:
\begin{equation}\label{eq:KX}
\begin{split}
K^X &= \frac{f^\prime}{1-\frac{2m}{r^{n-2}}}\bigl(\frac{\partial\phi}{\partial\rs}\bigr)^2+\frac{f}{r}\Bigl(1-\frac{nm}{r^{n-2}}\Bigr)\normsph{\nablab\phi}^2\\
 &\quad-\frac{1}{2}\biggl[f^\prime+(n-1)\frac{f}{r}\Bigl(1-\frac{2m}{r^{n-2}}\Bigr)\biggr]\partial^\alpha\phi\,\partial_\alpha\phi
\end{split}
\end{equation}
\begin{note}
For the definiteness of the expression \eqref{eq:KX} the \emph{photon sphere} at $r=\rph$ plays a distinguished role.
\end{note}

\paragraph{Calculation of the deformation tensor $\deformt{X}$.}
It is convenient to work in Eddington-Finkelstein coordinates
\begin{equation}
  X=\frac{1}{2}f(\rs)\frac{\partial}{\partial\vs}-\frac{1}{2}f(\rs)\frac{\partial}{\partial\us}\,.
\end{equation}
For the connection coefficients of \eqref{eq:mefc} one obtains
\begin{align}
\nabla_\dus\dus&=-(n-2)\,\frac{2m}{r^{n-1}}\,\dus\notag\\
\nabla_\dvs\dvs&=\phantom{+}(n-2)\,\frac{2m}{r^{n-1}}\,\dvs\notag\\
\nabla_{E_A}E_B&=\nablab_{E_A}E_B+\frac{r}{2}(\gn)_{AB}\dus-\frac{r}{2}(\gn)_{AB}\dvs\displaybreak[0]\\
\nabla_\dus E_B&=-\frac{1}{r}\bigl(1-\frac{2m}{r^{n-2}}\bigr)E_B\notag\\
\nabla_\dvs E_B&=\phantom{+}\frac{1}{r}\bigl(1-\frac{2m}{r^{n-2}}\bigr)E_B\,.\notag
\end{align}
Therefore
\begin{gather}
\deformt{X}_{\us\us}=g(\nabla_\dus X,\dus)=\bigl(1-\frac{2m}{r^{n-2}}\bigr)f^\prime\notag\\
\deformt{X}_{\vs\vs}=g(\nabla_\dvs X,\dvs)=\bigl(1-\frac{2m}{r^{n-2}}\bigr)f^\prime\notag\\
\begin{split}
\deformt{X}_{\us\vs}&=\frac{1}{2}g(\nabla_\dus X,\dvs)+\frac{1}{2}g(\dus,\nabla_\dvs X)\\
 &=-\bigl(1-\frac{2m}{r^{n-2}}\bigr)\bigl(f^\prime+(n-2)\frac{2m}{r^{n-1}}f\bigr)
\end{split}\\
\deformt{X}_{aA}=0\notag\displaybreak[0]\\
\begin{split}
\deformt{X}_{AB}&=\frac{1}{2}g(\nabla_{E_A} X,E_B)+\frac{1}{2}g(E_A,\nabla_{E_B} X)\\
 &=f\,r\,\bigl(1-\frac{2m}{r^{n-2}}\bigr)(\gn)_{AB}\notag
\end{split}
\end{gather}

The formula for $K^X$ above \eqref{eq:KX} is now obtained by writing out  (see also Appendix \ref{ref:nd})\[K^X=\deformt{X}^{\alpha\beta}\,T_{\alpha\beta}\] and rearranging the terms so as to complete $(\frac{\partial\phi}{\partial\us})^2+(\frac{\partial\phi}{\partial\vs})^2$ to $(\frac{\partial\phi}{\partial\rs})^2$.
This rearrangement is also related to the following modification of currents; for observe that
\begin{equation}
  \Box(\phi^2)=2(\partial^\alpha\phi)(\partial_\alpha\phi)
\end{equation}
if $\Box\phi=0$.

\paragraph{First modified current.}
\label{par:firstmodifiedcurrent}
Denoting by
\begin{equation}
  J_\mu^{X,0}=T_{\mu\nu}X^\nu
\end{equation}
define the first modified current by
\begin{multline} \label{eq:firstmodifiedcurrent}
J_\mu^{X,1}=J_\mu^{X,0}+\frac{1}{4}\Bigl(f^\prime+(n-1)\frac{f}{r}\bigl(1-\frac{2m}{r^{n-2}}\bigr)\Bigr)\partial_\mu(\phi^2)\\
-\frac{1}{4}\partial_\mu\Bigl(f^\prime+(n-1)\frac{f}{r}\bigl(1-\frac{2m}{r^{n-2}}\bigr)\Bigr)\phi^2\,.
\end{multline}
Consequently the divergences are
\begin{align}
K^{X,0}&=\nabla^\mu J_\mu^{X,0}=K^X\\
K^{X,1}&=\nabla^\mu J_\mu^{X,1}=K^X+\frac{1}{4}\Bigl(f^\prime+(n-1)\frac{f}{r}\bigl(1-\frac{2m}{r^{n-2}}\bigr)\Bigr)\Box(\phi^2)\notag\\
 &\qquad-\frac{1}{4}\Box\Bigl(f^\prime+(n-1)\frac{f}{r}\bigl(1-\frac{2m}{r^{n-2}}\bigr)\Bigr)\phi^2\displaybreak[0]\notag\\
&=\frac{f^\prime}{1-\frac{2m}{r^{n-2}}}\bigl(\frac{\partial\phi}{\partial\rs}\bigr)^2+\frac{f}{r}\bigl(1-\frac{nm}{r^{n-2}}\bigr)\normsph{\nablab\phi}^2\notag\\
&\qquad-\frac{1}{4}\Box\Bigl(f^\prime+(n-1)\frac{f}{r}\bigl(1-\frac{2m}{r^{n-2}}\bigr)\Bigr)\phi^2
\end{align}
Since for any function $w$
\begin{align}
\Box(w)&=(g^{-1})^{\mu\nu}\nabla_\mu\partial_\nu w\notag\\
&=-\frac{1}{1-\frac{2m}{r^{n-2}}}\partial_\us\partial_\vs w-\frac{n-1}{2r}\bigl(\partial_\us w-\partial_\vs w\bigr)+\laplacesph w\,,
\end{align}
a straight-forward calculation for
\begin{equation}
w=f^\prime+(n-1)\frac{f}{r}\bigl(1-\frac{2m}{r^{n-2}}\bigr)
\end{equation}
shows
\begin{multline}
\Box\Bigl(f^\prime+(n-1)\frac{f}{r}\bigl(1-\frac{2m}{r^{n-2}}\bigr)\Bigr)=\\
=\frac{1}{1-\frac{2m}{r^{n-2}}}\tprime{f}+2(n-1)\frac{\dprime{f}}{r}+(n-1)\Bigl[(n-3)+(n-1)\frac{2m}{r^{n-2}}\Bigr]\frac{\sprime{f}}{r^2}\\
+(n-1)\biggl[\Bigl((n-1)(n-2)-(n-3)\Bigr)\bigl(\frac{2m}{r^{n-2}}\bigr)^2-n\frac{2m}{r^{n-2}}-(n-3)\biggr]\frac{f}{r^3}\,.
\end{multline}
Thus we finally obtain
\begin{multline}\label{eq:KX1}
K^{X,1}=\frac{\sprime{f}}{1-\frac{2m}{r^{n-2}}}\bigl(\frac{\partial\phi}{\partial\rs}\bigr)^2+\frac{f}{r}\bigl(1-\frac{nm}{r^{n-2}}\bigr)\normsph{\nablab\phi}^2\\
-\frac{1}{4}\frac{\tprime{f}}{1-\frac{2m}{r^{n-2}}}\phi^2-\frac{n-1}{2}\frac{\dprime{f}}{r}\phi^2-\frac{n-1}{4}\Bigl[(n-3)+(n-1)\frac{2m}{r^{n-2}}\Bigr]\frac{\sprime{f}}{r^2}\phi^2\\
-\frac{n-1}{4}\Bigl[(n-1)^2\bigl(\frac{2m}{r^{n-2}}\bigr)^2-n\frac{2m}{r^{n-2}}-(n-3)\Bigr]\frac{f}{r^3}\phi^2\,.
\end{multline}

\paragraph{Applications of the first modified current.}
The proofs of Prop.~\ref{prop:IILED} and Prop.~\ref{prop:zeroth} are applications of this formula, as it appears in the energy identity for $J^{X,1}$ on $\dD{R}{\tau_1}{\tau_2}$, see Appendix \ref{sec:integration}.
\label{pg:pfzero}
\begin{proof}[Proof of Prop.~\ref{prop:zeroth}]
Choose $f=1$ identically, then
\begin{multline}\label{KX1f1}
K^{X,1}=\frac{1}{r}\bigl(1-\frac{nm}{r^{n-2}}\bigr)\normsph{\nablab\phi}^2\\
+\frac{n-1}{4}\Bigl[(n-3)+n\frac{2m}{r^{n-2}}-(n-1)^2\bigl(\frac{2m}{r^{n-2}}\bigr)^2\Bigr]\frac{1}{r^3}\phi^2\,.
\end{multline}
Since precisely
\begin{multline}
g\bigl(J^{X,1},\drs\bigr)=\frac{1}{4}\bigl(\ddvs{\phi}\bigr)^2+\frac{1}{4}\bigl(\ddus{\phi}\bigr)^2-\frac{1}{2}\bigl(\cf\bigr)\normsph{\nablab\phi}^2\\
+\frac{n-1}{2r}\bigl(\cf\bigr)\phi\ddrs{\phi}+\frac{n-1}{4r^2}\Bigl[1-(n-1)\frac{2m}{r^{n-2}}\Bigr]\bigl(\cf\bigr)\phi^2
\end{multline}
we deduce from the energy identity for $J^{\drs,1}$ in ${}^R\mathcal{D}_{\taup}^{\tau}$ that
\begin{multline}\label{boundaryintegralineq}
\int_{R^\ast+2\taup}^{R^\ast+2\tau}\ud t\int_\Sn\dm{\gn} r^{n-1}\times\\\times\biggl\{\frac{1}{4}\bigl(\ddvs{\phi}\bigr)^2+\frac{1}{4}\bigl(\ddus{\phi}\bigr)^2+\frac{n-1}{4R^2}\Bigl[\frac{1}{2}-(n-1)\frac{2m}{R^{n-2}}\Bigr]\bigl(1-\frac{2m}{R^{n-2}}\bigr)\phi^2\biggr\}\vert_{r=R}\\
+\int_{{}^R\mathcal{D}_{\taup}^{\tau}}\frac{n-1}{4r}\Bigl[(n-3)+n\frac{2m}{r^{n-2}}-(n-1)^2\bigl(\frac{2m}{r^{n-2}}\bigr)^2\Bigr]\frac{1}{r^2}\phi^2\dm{g}\leq\displaybreak[1]\\
\leq\int_{R^\ast+2\taup}^{R^\ast+2\tau}\ud t\int_\Sn\dm{\gn}\,r^{n-1}\times\\\times\biggl\{\frac{1}{2}\bigl(\cf\bigr)\normsph{\nablab\phi}^2+\frac{n-1}{2}\bigl(\cf\bigr)\bigl(\ddrs{\phi}\bigr)^2\biggr\}\vert_{r=R}\\
+C(n,m)\int_\St{\taup}\ned{T}{\phi}\,,
\end{multline}
where we have used Prop.~\ref{prop:boundary:JX1} for the boundary terms on $\partial{}^R\mathcal{D}_{\taup}^{\tau}\setminus\{r=R\}$;
note that \[(n-3)+n\frac{2m}{r^{n-2}}-(n-1)^2\bigl(\frac{2m}{r^{n-2}}\bigr)^2>0\qquad(R>\sqrt[n-2]{8nm})\,.\qedhere\]
\end{proof}

\noindent\emph{Proof of Prop.~\ref{prop:IILED}.}
On one hand we need $\sprime{f}=\mathcal{O}(\frac{1}{r^{1+\delta}})$ in view of \eqref{eq:KX1}
while on the other we already know from the proof of
Prop.~\ref{prop:zeroth} that $f=1$ generates a positive bulk term for
$r$ large enough. We choose
\begin{equation}
  f=1-\bigl(\frac{R}{r}\bigr)^\delta  
\end{equation}
(where $R>\rh$ is chosen suitably in the last step of the proof) and indeed find
\begin{multline}
  K^{X,1}=\delta\frac{R^\delta}{r^{1+\delta}}\bigl(\ddrs{\phi}\bigr)^2+\frac{f}{r}\bigl(1-\frac{nm}{r^{n-2}}\bigr)\normsph{\nablab\phi}^2\displaybreak[0]\\
  +\Biggl\{\frac{n-1}{4}(n-3)\Bigl[1-\bigl(\frac{R}{r}\bigr)^\delta(1+\delta)\Bigr]+\frac{1}{4}\bigl(\frac{R}{r}\bigr)^\delta\Bigl[2(n-1)-(2+\delta)\Bigr]\delta(1+\delta)\\
  +\biggl[\frac{n-1}{4}n\Bigl[1-\bigl(\frac{R}{r}\bigr)^\delta\Bigr]-\frac{\delta}{4}\bigl(\frac{R}{r}\bigr)^\delta\Bigl[n(n+\delta)-2(1+\delta)^2\Bigr]\biggr]\frac{2m}{r^{n-2}}\\
  -\biggl[\frac{(n-1)^3}{4}\Bigl[1-\bigl(\frac{R}{r}\bigr)^\delta\Bigr]-\frac{\delta}{4}\bigl(\frac{R}{r}\bigr)^\delta\Bigl[\bigl(n-(1+\delta)\bigr)(n-1)-\delta^2\Bigr]\biggr]\bigl(\frac{2m}{r^{n-2}}\bigr)^2\Biggr\}\frac{1}{r^3}\phi^2\\
  \geq 0
\end{multline}
for $r\geq R_1>R$, $R_1=R_1(n,m)>\rh$ chosen large enough.
This gives control on $\ddrs{\phi}$ and the angular derivatives:
\begin{equation*}
  \int_{{}^{R_1}\mathcal{D}_{\tau_1}^{\tau_2}}\biggl\{\delta\frac{R^\delta}{r^{1+\delta}}\bigl(\ddrs{\phi}\bigr)^2+\frac{f(R_1)}{r}\bigl(1-\frac{nm}{r^{n-2}}\bigr)\normsph{\nablab\phi}^2\biggr\}\leq\int_{{}^{R_1}\mathcal{D}_{\tau_1}^{\tau_2}}K^{X,1}
\end{equation*}
Here and in the following
$\tau_2>\tau_1>\frac{1}{2}\bigl(t_0-R^\ast)$.  For $\ddt{\phi}$ we use
the auxiliary current (see also Appendix \ref{sec:boundary})
\[J_\mu^\text{aux}=\frac{1}{2}\bigl(\cf\bigr)\delta\frac{R^\delta}{r^{1+\delta}}\partial_\mu(\phi^2)\]
to find easily
\begin{multline*}
  \intD{R_1}\delta\frac{R^\delta}{r^{1+\delta}}\bigl(\ddt{\phi}\bigr)^2\leq \intD{R_1}\biggl\{\delta(n+\delta)\frac{R^\delta}{r^{1+\delta}}\bigl(\ddrs{\phi}\bigr)^2\\
  +\delta\frac{R^\delta}{r^{1+\delta}}\bigl(\cf\bigr)\normsph{\nablab{\phi}}^2+\delta(n+\delta)\frac{R^\delta}{r^{3+\delta}}\phi^2+K^\text{aux}\biggr\}
\end{multline*}
Note that for $r\geq R_1$ in particular \[\frac{1}{4}\Bigl[2(n-1)-(2+\delta)\Bigr]\delta(1+\delta)\frac{R^\delta}{r^{3+\delta}}\phi^2\leq K^{X,1}\] hence
\begin{multline*}
  \intD{R_1}\delta\frac{R^\delta}{r^{1+\delta}}\Bigl\{\bigl(\ddt{\phi}\bigr)^2+\bigl(\ddrs{\phi}\bigr)^2\Bigr\}\leq
  C(n,m,\delta)\intD{R_1}\Bigl\{K^{X,1}+K^\text{aux}\Bigr\}\leq\displaybreak[0]\\ \leq
  C(n,m,\delta)\intD{R}\Bigl\{K^{X,1}+K^\text{aux}\Bigr\}\\+C(n,m,\delta)\int_{{}^R\mathcal{D}_{\tau_1}^{\tau_2}\cap\{R<r<R_1\}}\Bigl\{\frac{1}{r^{1+\delta}}\bigl(\ddt{\phi}\bigr)^2+\frac{1}{r^3}\phi^2\Bigr\}
\end{multline*}
By Prop.~\ref{prop:boundary:JX1} (also \eqref{eq:appendix:boundary})
\begin{multline*}
  \intpD{R}{}^\ast J^{X,1}\leq C(n,m,\delta)\int_\St{\tau_1}\ned{T}{\phi}\displaybreak[0]\\
  +C(n,m,\delta)\int_{R^\ast+2\tau_1}^{R^\ast+2\tau_2}\ud
  t\int_\Sn\dm{\gn}r^{n-1}\times\\\times\biggl\{\frac{1}{2}\bigl(\ddvs{\phi}\bigr)^2+\frac{1}{2}\bigl(\ddus{\phi}\bigr)^2+\frac{1}{2}\bigl(\cf\bigr)\normsph{\nablab\phi}^2+\frac{1}{r^2}\phi^2\biggr\}\vert_{r=R}
\end{multline*}
and by Prop.~\ref{prop:boundary:aux}
\begin{multline*}
  \intpD{R}{}^\ast J^\text{aux}\leq C(n,m,\delta)\int_\St{\tau_1}\ned{T}{\phi}\displaybreak[0]\\
  +\int_{R^\ast+2\tau_1}^{R^\ast+2\tau_2}\ud
  t\int_\Sn\dm{\gn}r^{n-1}\biggl\{\frac{\delta}{2}\frac{1}{2}\Bigl[\bigl(\ddus{\phi}\bigr)^2+\bigl(\ddvs{\phi}\bigr)^2\Bigr]+\frac{\delta}{2}\frac{R^{2\delta}}{r^{2+2\delta}}\phi^2\biggr\}_{r=R}\,.
\end{multline*}
Therefore by the energy identity for $J^{X,1}$ and $J^\text{aux}$ on ${}^R\mathcal{D}_{\tau_1}^{\tau_2}$:
\begin{multline*}
  \intD{R}\Bigl\{K^{X,1}+K^\text{aux}\Bigr\}\leq C(n,m,\delta)\int_{\Sigma_{\tau_1}}\Bigl(J^T(\phi),n\Bigr)\\
  +C(n,m,\delta)\int_{R^\ast+2\tau_1}^{R^\ast+2\tau_2}\ud  t\int_\Sn\dm{\gn}r^{n-1}\times\\\times\biggl\{\frac{1}{2}\bigl(\ddvs{\phi}\bigr)^2+\frac{1}{2}\bigl(\ddus{\phi}\bigr)^2+\frac{1}{2}\bigl(\cf\bigr)\normsph{\nablab\phi}^2+\frac{1}{r^2}\phi^2\biggr\}\vert_{r=R}
\end{multline*}
Our earlier \eqref{boundaryintegralineq} derived from the current $J^{\drs,1}$ now allows us to control the $\ddvs{\phi}$, $\ddus{\phi}$ derivatives and $\phi^2$ on the $r=R$ boundary together with the $\phi^2$ term in the region $R\leq r\leq R_1$ in one step:
\begin{multline*}
  \intD{R_1}\frac{R^\delta}{r^{1+\delta}}\biggl\{\bigl(\ddt{\phi}\bigr)^2+\bigl(\ddrs{\phi}\bigr)^2\biggr\}\leq C(n,m,\delta)\int_{\Sigma_{\tau_1}}\Bigl(J^T(\phi),n\Bigr)\displaybreak[0]\\
  +C(n,m,\delta)\int_{R^\ast+2\tau_1}^{R^\ast+2\tau_2}\ud t\int_\Sn\dm{\gn}r^{n-1}\times\\\times\biggl\{\frac{1}{2}\bigl(\cf\bigr)\normsph{\nablab\phi}^2+\frac{n-1}{2}\bigl(\ddrs{\phi}\bigr)^2\biggr\}\vert_{r=R}\displaybreak[0]\\
  +C(n,m,\delta)\int_{{}^R\mathcal{D}_{\tau_1}^{\tau_2}\cap\{R<r<R_1\}}\frac{1}{r^{1+\delta}}\bigl(\ddt{\phi}\bigr)^2
\end{multline*}
With $t_0$ fixed, we can now choose $R$ by Prop.~\ref{prop:ILED} such that
\begin{equation*}
\intD{R_1}\frac{1}{r^{1+\delta}}\biggl\{\bigl(\ddt{\phi}\bigr)^2+\bigl(\ddrs{\phi}\bigr)^2\biggr\}\leq C(n,m,\delta)\int_{\Sigma_{\tau_1}}\Bigl(J^T(\phi)+J^T(T\cdot\phi),n\Bigr)\,.
\end{equation*}
\qed

While it is possible to find simple functions $f\geq 0$ to ensure the positivity of $K^{X,1}$ asymptotically, this is not possible in the entire domain of outer communication; this is a consequence of \emph{trapping}, which more concretely appears as the indefiniteness of sign in \eqref{eq:KX1} at the photon sphere $r=\rph$.

In the following our strategy will be to prove non-negativity of $K^{X,1}$ not pointwise but by using Poincar\'e inequalities after integration over the spheres (the group orbits of $\mathrm{SO}(n)$). This is achieved in two alternative constructions: in Section \ref{sec:high} with a decomposition into spherical harmonics, and in Section \ref{sec:low} by a commutation with angular momentum operators.

\subsection{High angular frequencies}
\label{sec:high}
Here the idea is to control with the second term in \eqref{eq:KX1} after a decomposition of $\phi$ into spherical harmonics all other terms of order $\phi^2$. For this dominant term to be positive we evidently need \[f(\rs)\begin{cases}<0&r<\rph\\=0&r=\rph\\>0&r>\rph\end{cases}\,.\] Since $f$ should also be bounded one may guess that \[f(\rs)=\arctan\bigl(\frac{(n-1)\,\rs}{\rph}\bigr)\] is a good choice; while it can ensure positivity at the photon sphere, it fails to do so away from the photon sphere in the intermediate regions near the horizon and in the asymptotics. After having briefly recalled the decomposition into spherical harmonics, we will therefore give a more refined construction of $f$, nonetheless guided by the overall characteristics of this function, which will in particular allow us to track the dependence of the lowest spherical harmonic number (for which we can establish non-negativity) on the dimension $n$.

\paragraph{Fourier expansion on the sphere $\Sn$.}
We know all eigenvalues of \[-\laplacessph+\bigl(\frac{n-2}{2}\bigr)^2\] on $\Sn$ are given by \[\bigl(l+\frac{n-2}{2}\bigr)^2\qquad(l\geq 0)\,.\] Let $E_l$, $l\geq 0$, be the corresponding eigenspace in $\mathrm{L}^2(\Sn)$. Recall \[\dim_\mathbb{C}E_l=\bigl(l+\frac{n-2}{2}\bigr)\frac{2}{l}\binom{n-2+l-1}{l-1}\] and furthermore \[\mathrm{L}^2(\Sn)=\bigoplus_{l\geq 0}E_l\,.\]
Denote by $\pi_l$ the orthogonal projection of $\mathrm{L}^2(\Sn)$ onto $E_l$, then for $\phi\in\mathrm{L}^2(\Sn)$
\begin{equation}
  \phi=\sum_{l\geq 0}\pi_l\phi\,.
\end{equation}
This is the Fourier expansion on the sphere $\Sn$.
We find 
\begin{equation}
\label{eigenvaluessphere}
\laplacessph\pi_l\phi=-l(l+n-2)\pi_l\phi\,.
\end{equation}
Now,
\begin{equation*}
\begin{split}
l(l+n-2)\int_{\Sn}(\pi_l\phi)^2\dm{\gn}&=-\int_{\Sn}(\pi_l\phi)(\laplacessph\pi_l\phi)\dm{\gn}\\
&=\int_{\Sn}\vert\nablassph\pi_l\phi\vert^2\dm{\gn}
\end{split}
\end{equation*}
and assuming that \[\pi_l\phi=0\qquad(0\leq l<L)\] for some $L>0$,
\begin{equation*}
\begin{split}
L(L+n-2)\frac{1}{r^2}\int_{S_r}\phi^2\dm{\gamma_r}&\leq\frac{1}{r^2}\sum_{l\geq L}l(l+n-2)\int_{\Sn}(\pi_l\phi)^2 r^{n-1}\dm{\gn}\\
&=\frac{1}{r^2}\sum_{l\geq L}\int_{\Sn}\vert\nablassph\pi_l\phi\vert^2\dm{\gn}r^{n-1}\\
&=\frac{1}{r^2}\sum_{l\geq L}\int_{\Sn}\vert\pi_l\nablassph\phi\vert^2\dm{\gn}r^{n-1}\\
&=\int_{\Sn}\frac{1}{r^2}\vert\nablassph\phi\vert^2\,r^{n-1}\dm{\gn}\\
&=\int_{S_r}\normsph{\nablab\phi}^2\dm{\gamma_r}\,,
\end{split}
\end{equation*}
where for the commutation of $\pi_l$ with $\nablassph$ we have used that the projection is of the form
\begin{equation}
\label{projectionform}
(\pi_l\phi)(r\xi)=\int_{\Sn}\pi_l(\langle\xi,\xi^\prime\rangle)\phi(r\xi^\prime)\dm{\gn}(\xi^\prime)\,.
\end{equation}
We have proven the following Poincar\'e-type inequality:
\begin{lemma}[Poincar\'e inequality]\label{lemma:poincare}
Let $\phi\in \mathrm{H}^1(S_r),\,S_r=(\Sn,r^2\gn)$, have vanishing projection to $E_l$, $0\leq l<L$, for some $L\in\mathbb{N}$, i.e. \[\pi_l\phi=0\qquad(0\leq l<L)\,,\] then
\begin{equation*}
\int_{S_r}\bigl\vert\nablab\phi\bigr\vert^2\dm{\gamma_r}\geq L(L+n-2)\frac{1}{r^2}\int_{S_r}\phi^2\dm{\gamma_r}\,.
\end{equation*}
\end{lemma}

\paragraph{Construction of the multiplier function for high angular frequencies.}
The idea is to prescribe the 3\raisebox{1ex}{\scriptsize rd} derivative of $f$ and to find its 2\raisebox{1ex}{\scriptsize nd} and 1\raisebox{1ex}{\scriptsize st} derivatives by integration with boundary values and parameters that ensure that $f$ remains bounded.
Let \begin{equation}\label{def:alpha}\alpha=\frac{n-1}{\rphp} \end{equation}
and $\gamma\geq 2,\,\gamma\in\mathbb{N}$.
Consider
\begin{equation}
f^{\text{III}}_{\gamma,\alpha}(\rs)=\begin{cases}
-1 & |\rs|\leq\frac{1}{\gamma\alpha}\\
\phantom{+}1 & \frac{1}{\gamma\alpha}<|\rs|\leq b_{\gamma,\alpha}\\[1ex]
\displaystyle{\bigl(\frac{b_{\gamma,\alpha}}{\rs}\bigr)^6} & |\rs|\geq b_{\gamma,\alpha}
\end{cases}
\end{equation}
where \begin{equation}b_{\gamma,\alpha}=\frac{5}{6}\frac{2}{\gamma\alpha}\,.\end{equation} Note that $b_{\gamma,\alpha}$ is chosen so that
\begin{equation}
  \int_0^\infty f^{\text{III}}_{\gamma,\alpha}(\rs)\ud\rs=0\,.
\end{equation}
Now define 
\begin{equation}
  f_{\gamma,\alpha}^{\text{II}}(\rs)=\int_0^{\rs}f_{\gamma,\alpha}^{\text{III}}(t)\ud t\,.
\end{equation}
Obviously $f_{\gamma,\alpha}^{\text{II}}(-\rs)=-f_{\gamma,\alpha}^{\text{II}}(\rs)$ and in explicit form
\begin{equation}
f_{\gamma,\alpha}^{\text{II}}(\rs)=\begin{cases}
-\rs & |\rs|\leq\frac{1}{\gamma\alpha}\\
\rs-\frac{2}{\gamma\alpha} & \frac{1}{\gamma\alpha}<\rs\leq b_{\gamma,\alpha} \\
\rs+\frac{2}{\gamma\alpha} & -b_{\gamma,\alpha}\leq\rs<-\frac{1}{\gamma\alpha}\\[1ex]
\displaystyle{-\frac{b_{\gamma,\alpha}^6}{5\rs^5}} & |\rs|\geq b_{\gamma,\alpha}
\end{cases}\,.
\end{equation}
The functions $f^{\text{II}}_{\gamma,\alpha}$ and $f^{\text{III}}_{\gamma,\alpha}$ are sketched in figure \ref{fIIfIII}.
\begin{figure}[bp]
\centering
\input{fIIIfII.pstex_t}
\caption{Sketch of the functions $f^{\text{II}}_{\gamma,\alpha}$ and $f^{\text{III}}_{\gamma,\alpha}$, and the adjusted functions (dot-dashed) for $\rs\leq 0$.} 
\label{fIIfIII}
\end{figure}
Next define \begin{equation}f_{\gamma,\alpha}^{\text{I}}=\int_{-\infty}^{\rs}f_{\gamma,\alpha}^{\text{II}}(t)\ud t\,.\end{equation}
Here we find
\begin{equation}
f_{\gamma,\alpha}^{\text{I}}(\rs)=\begin{cases}
\displaystyle{\frac{b_{\gamma,\alpha}^6}{20\rs^4}} & \rs\leq -b_{\gamma,\alpha}\\[2ex]
\displaystyle{\frac{b_{\gamma,\alpha}^2}{20}+\frac{1}{2}(\rs^2-b_{\gamma,\alpha}^2)+\frac{2}{\gamma\alpha}(\rs+b_{\gamma,\alpha})} & -b_{\gamma,\alpha}\leq\rs\leq -\frac{1}{\gamma\alpha}\\[2ex]
\displaystyle{\frac{13}{12}\frac{1}{(\gamma\alpha)^2}-\frac{\rs^2}{2}} & -\frac{1}{\gamma\alpha}\leq\rs\leq 0
\end{cases}
\end{equation}
and $f_{\gamma,\alpha}^{\text{I}}(\rs)=f_{\gamma,\alpha}^{\text{I}}(-\rs)$, as sketched in figure \ref{fIf}.
\begin{figure}[tbp]
\centering
\input{fIf0.pstex_t}
\caption{Sketch of the functions $f^{\text{I}}_{\gamma,\alpha}$ and $f^{\text{0}}_{\gamma,\alpha}$, and the adjusted functions (dot-dashed) for $\rs\leq 0$.}
\label{fIf}
\end{figure}
Finally define 
\begin{equation}
  f_{\gamma,\alpha}^{\text{0}}(\rs)=\int_0^{\rs}f_{\gamma,\alpha}^{\text{I}}(t)\ud t\,.
\end{equation}
Here again $f_{\gamma,\alpha}^{\text{0}}(-\rs)=-f_{\gamma,\alpha}^{\text{0}}(\rs)$ and in particular \begin{equation}f_{\gamma,\alpha}^\text{0}(\frac{1}{\gamma\alpha})=\int_0^\frac{1}{\gamma\alpha}\bigl(\frac{13}{12}\frac{1}{(\gamma\alpha)^2}-\frac{t^2}{2}\bigr)\ud t=\frac{11}{12}\frac{1}{(\gamma\alpha)^3}\,.
\end{equation}
Moreover the calculus yields
\begin{gather}
f(b_{\gamma,\alpha})>\frac{1}{(\gamma\alpha)^3}\notag\\
\lim_{\rs\to\infty}f_{\gamma,\alpha}^\text{0}(\rs)<\frac{3}{2}\frac{1}{(\gamma\alpha)^3}\,.
\end{gather}
The function $f_{\gamma,\alpha}^\text{0}$ is sketched in figure \ref{fIf}.
While this function would suffice in the region $\rs\geq-\frac{1}{\gamma\alpha}$ it does not fall-off fast enough as $\rs\to-\infty$.
\begin{lemma}
With $\rs$ defined by \eqref{rscph} we have for all $n\geq 3$ \[\lim_{\rs\to-\infty}\bigl(1-\frac{2m}{r^{n-2}}\bigr)(-\rs)=0\,.\]
In fact, \[\bigl(1-\frac{2m}{r^{n-2}}\bigr)\leq\frac{\rhp}{(-\rs)}\] for all $\rs<0$.
\end{lemma}
\begin{proof}See Appendix \ref{ref:rrs}.\end{proof}
It is easy to convince oneself that one can make an adjustment to $f^\text{III}$ on $\rs\leq 0$ that introduces faster decay while keeping the area under the graph of $f^\text{III}$ \emph{and} $f^\text{II}$ fixed \cite{VST}. 
In other words, there are constants \begin{equation}b_{\gamma,\alpha}\leq b\leq\frac{4}{\gamma\alpha}\qquad \frac{1}{4}\leq c\leq 1\end{equation} such that if we redefine $f^\text{III}_{\gamma,\alpha}$ for $\rs\leq 0$ as
\begin{equation}
f^\text{III}_{\gamma,\alpha}(\rs)=\begin{cases}
-1 & -\frac{1}{\gamma\alpha}\leq\rs\leq 0\\
\phantom{+}c & -b\leq\rs\leq\frac{1}{\gamma\alpha}\\
\displaystyle{\bigl(1-\frac{2m}{r^{n-2}}\bigr)^6\bigl(\frac{b}{\rhp}\bigr)^6} & \rs\leq -b
\end{cases}
\end{equation}
then
\begin{gather*}
\int_0^{-\infty}f_{\gamma,\alpha}^\text{III}(\rs)\ud\rs=0\\
\int_{-\infty}^0\int_0^{\rs}f_{\gamma,\alpha}^\text{III}(t)\ud t\ud\rs=\int_{-\infty}^0\int_0^{-\rs}(-f_{\gamma,\alpha}^\text{III}(t))\ud t\ud\rs\,.
\end{gather*}

\noindent The adjusted functions in comparison the the old are also sketched in figures \ref{fIIfIII} and \ref{fIf}.
Note in particular that for $\rs\leq 0$
\begin{gather}
f^\text{II}(\rs)\leq\frac{1}{\gamma\alpha}\\
f^\text{I}(\rs)\leq f^\text{I}(-\rs)\leq\frac{13}{12}\frac{1}{(\gamma\alpha)^2}
\end{gather}
and for $\rs\leq -\frac{1}{\gamma\alpha}$
\begin{equation}
\frac{11}{12}\frac{1}{(\gamma\alpha)^3}\leq\vert f^\text{0}(\rs)\vert\leq f^\text{0}(-\rs)<\frac{3}{2}\frac{1}{(\gamma\alpha)^3}\,.
\end{equation}

\begin{remark}
In order to deal with smooth functions one could use (e.g. at the level of second derivatives) a convolution with a Gaussian on the scale given by $\gamma\alpha$ (or finer). I.e.~one could define
\begin{equation*}
\dprime{f}_{\gamma,\alpha}(\rs)=\frac{\gamma\alpha}{\sqrt{\pi}}\int_{-\infty}^{\infty} e^{-(\gamma\alpha)^2 (\rs-t)^2} f_{\gamma,\alpha}^\text{II}(t)\ud t
\end{equation*}
and find $\tprime{f}_{\gamma,\alpha}=\frac{\ud}{\ud\rs}\dprime{f}_{\gamma,\alpha}$ by differentiation, and $\sprime{f}_{\gamma,\alpha}$ and $f_{\gamma,\alpha}$ by integration with the boundary values $\sprime{f}_{\gamma,\alpha}(-\infty)=0$, $f_{\gamma,\alpha}(0)=0$ as above. However, I choose not to do so (as it does not give further insight) and work directly with the step-functions, i.e. define \[\dprime{f}_{\gamma,\alpha}=f_{\gamma,\alpha}^\text{III}\,.\]
\end{remark}

We are now in the position to prove a non-negativity property of the terms occuring in \eqref{eq:KX1} which we will denote by ${}^0K^{X,1}$,
\begin{equation}
  K^{X,1}=\frac{\sprime{f}}{1-\frac{2m}{r^{n-2}}}\bigl(\frac{\partial\phi}{\partial\rs}\bigr)^2+{}^0K^{X,1}\,.
\end{equation}

\begin{prop}[Positivity of the current $J^{X_{\gamma,\alpha},1}$]
For $n\geq 3$, \[X_{\gamma,\alpha}=f_{\gamma,\alpha}\frac{\partial}{\partial\rs}\qquad(\text{where we choose }\gamma=12)\,,\] and $\phi\in\mathrm{H}^1(S)$ satisfy
\begin{equation*}
\int_S {}^0K^{X_{\gamma,\alpha},1} \dm{\gamma}\geq 0
\end{equation*}
provided \[\pi_l\phi=0\qquad(0\leq l<L)\] for a fixed $L\geq(6\gamma n)^2$.
\label{posproposition}
\end{prop}
\noindent\emph{Proof.}
By Lemma \ref{lemma:poincare}
\begin{multline}
\int_S {}^0K^{X_{\gamma,\alpha},1}\dm{\gamma}\geq\int_S\biggl\{L(L+n-2)\frac{f_{\gamma,\alpha}}{r^3}\bigl(1-\frac{nm}{r^{n-2}}\bigr)\\
 -\frac{1}{4}\frac{\tprime{f}_{\gamma,\alpha}}{1-\frac{2m}{r^{n-2}}}-\frac{n-1}{2}\frac{\dprime{f}_{\gamma,\alpha}}{r}-\frac{n-1}{4}\Bigl[(n-3)+(n-1)\frac{2m}{r^{n-2}}\Bigr]\frac{\sprime{f}_{\gamma,\alpha}}{r^2}\\
 -\frac{n-1}{4}\Bigl[(n-1)^2\bigl(\frac{2m}{r^{n-2}}\bigr)^2-n\frac{2m}{r^{n-2}}-(n-3)\Bigr]\frac{f_{\gamma,\alpha}}{r^3}\biggr\}\phi^2\dm{\gamma}\,.
\end{multline}
We divide into the five regions \[-\infty<-\frac{4}{\gamma\alpha}<-\frac{1}{\gamma\alpha}<\frac{1}{\gamma\alpha}<b_{\gamma,\alpha}<\infty\,.\]
\subsubsection*{Step 1. \textnormal{(near the photon sphere, $\vert\rs\vert<\frac{1}{\gamma\alpha}$)}}
\begin{lemma}
In the region $|\rs|<\frac{1}{\gamma\alpha}$ the corresponding value of $r$ lies in the interval \[\sqrt[n-2]{\delta n m}<r<\frac{n}{\alpha}\] where $\delta=\max\{\frac{1}{3},\frac{4}{3}\frac{2}{n}\}$.
\end{lemma}
\begin{proof}Omitted. See \cite{VST}.\end{proof}
\noindent Recalling the graphs of $f_{\gamma,\alpha}$ and its derivatives we then find in the region $|\rs|<\frac{1}{\gamma\alpha}$:
\begin{multline*}
\int_S {}^0K^{X_{\gamma,\alpha},1}\dm{\gamma}\geq\\
\quad\geq\int_S\biggl\{\frac{1}{4}-\frac{1}{2}\frac{\alpha}{\delta^\frac{1}{n-2}}\frac{1}{\gamma\alpha}-\frac{1}{4}\frac{\alpha^2}{\delta^\frac{2}{n-2}}\frac{1}{n-1}\Bigl[(n-3)+\frac{1}{\delta}(n-1)\frac{2}{n}\Bigr]\frac{13}{12}\frac{1}{(\gamma\alpha)^2}\\-\frac{1}{4}\frac{\alpha^3}{\delta^\frac{3}{n-2}}\frac{2}{\delta n} \frac{3}{2}\frac{1}{(\gamma\alpha)^3}\biggr\}\phi^2\dm{\gamma}
\end{multline*}
\begin{equation*}
\begin{split}
\phantom{\int_S {}^0K^{X_{\gamma,\alpha},1}\dm{\gamma}}
&\quad\geq\int_S\biggl\{\frac{1}{4}-\frac{1}{2}\frac{3}{\gamma}-\frac{3}{4}\frac{13}{12}\bigl(\frac{3}{\gamma}\bigr)^2-\frac{3}{4}\bigl(\frac{3}{\gamma}\bigr)^3\biggr\}\phi^2\dm{\gamma}\\
&\quad\geq\int_S\frac{1}{4}\frac{1}{8}\,\phi^2\,\dm{\gamma}
\end{split}
\end{equation*}
because $\gamma=12$.

\subsubsection*{Step 2. \textnormal{(in the intermediate region, $\frac{1}{\gamma\alpha}\leq\rs\leq\frac{5}{6}\frac{2}{\gamma\alpha}$})}
\begin{lemma} In the region $\frac{1}{\gamma\alpha}\leq\rs\leq\frac{5}{6}\frac{2}{\gamma\alpha}$ we for the corresponding value of $r$,
\[\Bigl(1+\frac{1}{3\gamma(n-1)}\Bigr)\rphp\leq r\leq\frac{n}{\alpha}\,.\]
\end{lemma}
\begin{proof}Omitted. See \cite{VST}.\end{proof}
\noindent Collecting the first and the last term, we find in this region,
\begin{multline*}
\int_S {}^0K^{X_{\gamma,\alpha},1}\dm{\gamma}\geq\\
\int_S\Biggl\{\frac{\alpha^3}{n^3}\frac{11}{12}\frac{1}{(\gamma\alpha)^3}\Bigl[\bigl(1-\frac{nm}{r^{n-2}}\bigr)L(L+n-2)-\frac{n-1}{4}(n-1)^2\bigl(\frac{2}{n}\bigr)^2+\frac{1}{4}(n-1)(n-3)\Bigr]\\
\shoveright{-\frac{1}{4}\frac{1}{1-\frac{2}{n}}+\frac{\alpha}{3}\frac{1}{3}\frac{1}{\gamma\alpha}-\frac{1}{4}\alpha^2\frac{1}{n-1}\Bigl[(n-3)+(n-1)\frac{2}{n}\Bigr]\frac{1}{(\gamma\alpha)^2}\Biggr\}\phi^2\dm{\gamma}}\displaybreak[0]\\
\geq\int_S\biggl\{\frac{11}{12}\frac{1}{(n\gamma)^3}\Bigl[\frac{1}{6\gamma(n-1)}L(L+n-2)-(n-1)+\frac{1}{4}(n-1)(n-3)\Bigr]\\\shoveright{-\frac{3}{4}-\frac{1}{4}\frac{1}{\gamma^2}\biggr\}\phi^2\dm{\gamma}}\\
\geq\int_S\biggl\{\frac{11}{12}\frac{1}{6}\Bigl(\frac{(6\gamma n)^2}{\gamma^2 n^2}\Bigr)^2-1\biggr\}\phi^2\dm{\gamma}\geq\int_S \phi^2\dm{\gamma}
\end{multline*}
because $L\geq(6\gamma n)^2$, where we have used that for $\frac{1}{\gamma\alpha}\leq\rs\leq\frac{5}{6}\frac{2}{\gamma\alpha}$, \[1-\frac{nm}{r^{n-2}}\geq \frac{1}{6\gamma(n-1)}\,.\]

\subsubsection*{Step 3. \textnormal{(in the asymptotics, $\rs\geq b_{\gamma,\alpha}$)}}
Given the general fact Prop.~\ref{prop:limitrsr} we here only need the weaker statement
\begin{lemma}
For $\rs\geq\frac{5}{6}\frac{2}{\gamma\alpha}$, \[\frac{r}{\rs}\leq 2\gamma n\,.\]
\end{lemma}
\begin{proof}Omitted. See \cite{VST}.\end{proof}
\noindent Here
\begin{multline*}
\int_S {}^0K^{X_{\gamma,\alpha},1}\dm{\gamma}\geq\\
\int_S\biggl\{\frac{1}{(\gamma\alpha)^3}\Bigl[\frac{1}{6\gamma(n-1)}L(L+n-2)-\frac{3}{2}(n-1)\Bigr]\frac{1}{r^3}-\frac{1}{4}\frac{1}{1-\frac{2}{n}}\bigl(\frac{5}{6}\frac{2}{\gamma\alpha\rs}\bigr)^6\\\shoveright{-\frac{1}{4}\alpha^2\frac{1}{n-1}\Bigl[(n-3)+(n-1)\frac{2}{n}\Bigr]\frac{1}{20}\frac{\bigl(\frac{5}{6}\frac{2}{\gamma\alpha}\bigr)^6}{\rs^4}\biggr\}\phi^2\dm{\gamma}}\displaybreak[0]\\
\geq\int_S\Bigl\lbrack\frac{L^2}{6\gamma^4 n}+\frac{L}{6\gamma^4 n}(n-2)-\frac{3}{2}\frac{1}{\gamma^3}(n-1)-\frac{3}{4}\bigl(\frac{r}{\rs}\bigr)^3\bigl(\frac{5}{6}\frac{2}{\gamma}\bigr)^6\frac{1}{(\alpha\rs)^3}\\\shoveright{-\frac{1}{4}\frac{1}{20}\bigl(\frac{r}{\rs}\bigr)^3\bigl(\frac{5}{6}\frac{2}{\gamma}\bigr)^6\frac{1}{\alpha\rs}\Bigr\rbrack\frac{1}{(\alpha r)^3}\phi^2\dm{\gamma}}\\
\geq\int_S\Bigl\lbrack(6n)^3-(4n)^3\Bigr\rbrack\frac{1}{(\alpha r)^3}\phi^2\dm{\gamma}\geq\int_S\bigl(\frac{n}{\alpha r})^3\phi^2\dm{\gamma}\\
\geq\int_S\biggl(\frac{\rphp}{r}\biggr)^3\phi^2\dm{\gamma}
\end{multline*}
where in the third bound we have again used $L\geq(6\gamma n)^2$ and the Lemma.

\subsubsection*{Step 4. \textnormal{(in the intermediate region, $-\frac{4}{\gamma\alpha}\leq\rs\leq-\frac{1}{\gamma\alpha}$)}}
Recall $\gamma=12$.
\begin{lemma}
For $k\leq\gamma$, $k\in\mathbb{N}$, \[\Bigl(1-\frac{2m}{r^{n-2}}\Bigr)^{-1}\vert_{\rs=-\frac{k}{\gamma\alpha}}\leq 17\]
and consequently \[-\Bigl(1-\frac{nm}{r^{n-2}}\Bigr)\vert_{\rs=-\frac{1}{\gamma\alpha}}\geq\frac{1}{20}\frac{1}{2\gamma}\,.\]
\end{lemma}
\begin{proof}Omitted. See \cite{VST}.\end{proof}
\noindent In the region $-\frac{4}{\gamma\alpha}\leq\rs\leq-\frac{1}{\gamma\alpha}$ we directly apply the Lemma to see that,
\begin{multline*}
\int_S {}^0K^{X_{\gamma,\alpha},1}\dm{\gamma}\geq\\
\int_S\biggl\{L(L+n-2)\frac{1}{(nm)^\frac{3}{n-2}}\frac{11}{12}\frac{1}{(\gamma\alpha)^3}\frac{1}{20}\frac{1}{2\gamma}-\frac{1}{4}17-\frac{n-1}{2}\frac{1}{\rhp}\frac{1}{\gamma\alpha}\\-\frac{n-1}{2}\Bigl[(n-3)+(n-1)\Bigr]\frac{1}{(2m)^\frac{2}{n-2}}\frac{13}{12}\frac{1}{(\gamma\alpha)^2}-\frac{n-1}{4}\Bigl[n+(n-3)\Bigr]\frac{1}{(2m)^\frac{3}{n-2}}2\frac{1}{(\gamma\alpha)^3}\biggr\}\phi^2\dm{\gamma}\displaybreak[0]\\
\shoveleft{\geq\int_S\biggl\{\frac{1}{(3\gamma)^4}\frac{1}{(n-1)^3}L(L+n-2)}\\\shoveright{-\frac{17}{4}-\frac{3}{2}\frac{1}{2\gamma}-\frac{13}{12}\frac{1}{\gamma^2}\bigl(\frac{n}{2}\bigr)^\frac{2}{n-2}-\frac{1}{n-1}\frac{1}{\gamma^3}\bigl(\frac{n}{2}\bigr)^\frac{3}{n-2}\biggr\}\phi^2\dm{\gamma}}\\
\geq\int_S\biggl\{2^4 n-\frac{23}{4}\biggr\}\phi^2\dm{\gamma}\geq\int_S\phi^2\dm{\gamma}
\end{multline*}
because $L\geq(6\gamma n)^2$.

\subsubsection*{Step 5. \textnormal{(near the horizon, $\rs\leq-b$)}}
Finally we see for $\rs\leq-b$, recalling the adjustment to faster fall-off,
\begin{multline*}
\int_S {}^0K^{X_{\gamma,\alpha},1}\dm{\gamma}\geq\\
\int_S\biggl\{L(L+n-2)\frac{1}{(nm)^\frac{3}{n-2}}\frac{11}{12}\frac{1}{(\gamma\alpha)^3}\frac{1}{20}\frac{1}{2\gamma}
-\frac{1}{4}\bigl(1-\frac{2}{n}\bigr)^5-\frac{n-1}{2}\frac{1}{\rhp}\frac{1}{\gamma\alpha}\\\shoveright{-\frac{(n-1)^2}{4}\frac{1}{\rhp}\frac{1}{(\gamma\alpha)^2}-\frac{n-1}{4}\Bigl[n+(n-3)\Bigr]\frac{1}{(2m)^\frac{3}{n-2}}2\frac{1}{(\gamma\alpha)^3}\biggr\}\phi^2\dm{\gamma}}\displaybreak[0]\\
\shoveleft{\geq\int_S\biggl\{\frac{1}{(3\gamma)^4}\frac{1}{(n-1)^3}L(L+n-2)}\\\shoveright{-\frac{1}{4}-\frac{1}{2\gamma}\bigl(\frac{n}{2}\bigr)^\frac{1}{n-2}-\frac{1}{(2\gamma)^2}\bigl(\frac{n}{2}\bigr)^\frac{2}{n-2}-\frac{4}{n-1}\frac{1}{(2\gamma)^3}\bigl(\frac{n}{2}\bigr)^\frac{3}{n-2}\biggr\}\phi^2\dm{\gamma}}\\
\geq\int_S\biggl\{2^4 n-\frac{5}{4}\biggr\}\phi^2\dm{\gamma}\geq\int_S\phi^2\dm{\gamma}
\end{multline*}
where we have used that here
\begin{equation*}
\frac{\tprime{f}}{1-\frac{2m}{r^{n-2}}}=\Bigl(1-\frac{2m}{r^{n-2}}\Bigr)^5\bigl(\frac{b}{\rhp}\bigr)^6\leq\bigl(1-\frac{2}{n}\bigr)^5\leq 1\,.
\end{equation*}
\qed

In fact, we have shown more, because all lower bounds in Step 1-5 are minorized by $\frac{1}{4}\frac{1}{8}\frac{(2m)^\frac{3}{n-2}}{r^3}$.
\begin{cor}\label{cor:xestimate:L}
Let $\phi\in\mathrm{H}^2$ be a solution of the wave equation, \[\Box_g\phi=0\,,\] satisfying \[\pi_l\phi=0\qquad(0\leq l<L)\] on the standard sphere $S=(\Sn,r^2\gn)$ for a fixed $L\geq(6\gamma n)^2$. Then
\begin{multline*}
\int_S\biggl\{\frac{1}{4}\frac{1}{8}\frac{(2m)^\frac{3}{n-2}}{r^3}\phi^2\\+\frac{1}{(20\gamma^2)^3}\frac{1}{(n-2)^2(n-1)^6}\Bigl(1-\frac{2m}{r^{n-2}}\Bigr)^5\frac{(2m)^\frac{6}{n-2}}{r^4}\bigl(\frac{\partial\phi}{\partial\rs}\bigr)^2\biggr\}\dm{\gamma}\\
\leq\int_S K^{X_{\gamma,\alpha},1}\dm{\gamma}
\end{multline*}
\end{cor}
\begin{proof}
It remains to be shown that
\begin{equation}\label{tbs:fprime}
\frac{1}{20}\frac{1}{(4\cdot5(n-2))^2}\Bigl(1-\frac{2m}{r^{n-2}}\Bigr)^6\frac{b_{\gamma,\alpha}^6}{r^4}\leq\sprime{f}_{\gamma,\alpha}\,.\tag{$\ast$}
\end{equation}
First \[\int_{-\infty}^{\rs}\Bigl(1-\frac{2m}{r^{n-2}}\Bigr)^6\ud\rs=\int_{\rhp}^r\Bigr(1-\frac{2m}{r^{n-2}}\Bigr)^5\ud r\] because ${\ud\rs}/{\ud r}=\bigl(1-\frac{2m}{r^{n-2}}\bigr)^{-1}$.
Now choose $\rh<r_0<r$ so close to $r$ as to satisfy \[\frac{r-r_0}{r_0}=\frac{1}{2}\frac{1}{5(n-2)}\Bigl(1-\frac{2m}{r^{n-2}}\Bigr)\] then by the mean value theorem
\begin{equation*}
\begin{split}
\int_{\rhp}^r\bigl(1-\frac{2m}{r^{n-2}}\bigr)^5\ud r&\geq \bigl(1-\frac{2m}{r_0^{n-2}}\bigr)^5(r-r_0)\\
& \geq\bigl(1-\frac{2m}{r^{n-2}}\bigr)^5\Bigl[1-5(n-2)\frac{1}{1-\frac{2m}{r^{n-2}}}\frac{r-r_0}{r_0}\Bigr](r-r_0)\\
& \geq\frac{1}{4}\frac{1}{5(n-2)}\bigl(1-\frac{2m}{r^{n-2}}\bigr)^6\rhp\,.
\end{split}
\end{equation*}
We conclude for $\rs\leq-b$,
\begin{equation*}
\begin{split}
\sprime{f}_{\gamma,\alpha}(\rs)&=\int_{-\infty}^{\rs}\int_{-\infty}^{s^\ast}\bigl(1-\frac{2m}{r^{n-2}}\bigr)\vert_{\rs=s^\ast}\bigl(\frac{b}{\rhp}\bigr)^6\ud s^\ast\ud\rs\\
&\geq\frac{1}{4}\frac{1}{5(n-2)}\int_{-\infty}^{\rs}\bigl(1-\frac{2m}{r^{n-2}}\bigr)^6\ud\rs\,\frac{b^6}{(2m)^\frac{5}{n-2}}\\
&\geq\bigl(\frac{1}{4}\frac{1}{5(n-2)}\bigr)^2\bigl(1-\frac{2m}{r^{n-2}}\bigr)^6\frac{b^6}{(2m)^\frac{4}{n-2}}\\
&\geq\frac{1}{(4\cdot 5(n-2))^2}\bigl(1-\frac{2m}{r^{n-2}}\bigr)^6\,\frac{b_{\gamma,\alpha}^6}{r^4}\,.
\end{split}
\end{equation*}
Second for $\rs\geq 0$
\begin{equation*}
\frac{1}{(4\cdot 5(n-2))^2}\frac{1}{r^4}=\frac{1}{(4\cdot 5(n-2))^2}\bigl(\frac{\rs}{r}\bigr)^4\frac{1}{\rs^4}\\
\leq\frac{1}{\rs^4}
\end{equation*}
Since, thirdly, \[\frac{b_{\gamma,\alpha}}{r}\leq 1\,,\] we have established \eqref{tbs:fprime} for the regions $\rs\leq-b$, $\rs\geq b_{\gamma,\alpha}$, $-b\leq\rs\leq b_{\gamma,\alpha}$, respectively.
\end{proof}

\begin{remark} This estimate of the zeroth order term $\phi^2$ suffices to obtain an estimate for all derivatives using a commutation with the vectorfield $T$; see Proof of Prop.~\ref{prop:ILED} in Section \ref{sec:pfILED}.
\end{remark}

\subsection{Low angular frequencies and commutation}
\label{sec:low}

While the current constructed in Section \ref{sec:high} required a decomposition into spherical harmonics we will now altogether avoid a recourse to the Fourier expansion on the sphere. The key to the positivity property was Poincar\'e's inequality which states in more generality:

\begin{lemma}[Poincar\'e inequality]\label{lemma:gpoincare}
Let $(S,\gamma)$ be a compact Riemannian manifold, and $\phi\in\mathrm{H}^1(S)$ a function on $S$ with mean value
\begin{equation*}\bar{\phi}=\frac{1}{\int_S\dm{\gamma}}\int_S\phi\dm{\gamma}\,.\end{equation*}
Then
\begin{equation*}
\int_S(\phi-\bar{\phi})^2\dm{\gamma}\leq\frac{1}{\lambda_1(S)}\int_S\vert\nablab\phi\vert^2\dm{\gamma}
\end{equation*}
where $\lambda_1(S)$ is the first eigenvalue of the negative Laplacian, $-\triangle\!\!\!\!/\,=-\nablab^a\nablab_a$, on $S$. ($\nablab$ denotes covariant differentiation on S.)
\end{lemma}

\noindent Now let $(S,\gamma)=(\Sn,\gn)$ then we read off from \eqref{eigenvaluessphere} here \begin{equation}\lambda_1(\Sn)=n-1\,.\end{equation}
Choose a basis of the Lie algebra of $\mathrm{SO}(n)$, \begin{equation}\label{generatorrotations}\Omega_i:i=1,\ldots,\frac{n(n-1)}{2}\,,\end{equation} and apply Lemma \ref{lemma:gpoincare} to the functions $\Omega_i\phi$ of vanishing mean: \begin{equation}\int_\Sn\Omega_i\phi\,\dm{\gn}=0\,.\end{equation}
Then we obtain
\begin{equation}
\int_\Sn\vert\nablab\Omega_i\phi\vert^2\dm{\gn}\geq(n-1)\int_\Sn(\Omega_i\phi)^2\dm{\gn}
\end{equation}
or on $(S,\gamma)=(S_r,\gamma_r)=(\Sn,r^2\gn)$:
\begin{equation}\label{poincareSn}
\int_{S_r}\vert\nablab\Omega_i\phi\vert^2\dm{\gamma_r}\geq\frac{n-1}{r^2}\int_{S_r}(\Omega_i\phi)^2\dm{\gamma_r}\,.
\end{equation}
Also note
\begin{equation}\label{sumgenerators}
\sum_{i=1}^\frac{n(n-1)}{2}(\Omega_i\phi)^2=r^2\normsph{\nablab\phi}^2\,.
\end{equation}

\paragraph{Second modified current.} Recall we are considering vectorfields of the form \[X=f(\rs)\frac{\partial}{\partial\rs}\,.\]
Define
\begin{equation}
J_\mu^{X,2}=J_\mu^{X,1}+\frac{f^\prime}{f(1-\frac{2m}{r^{n-2}})}\beta\,X_\mu\,\phi^2
\end{equation}
where $\beta=\beta(\rs)$ is a function to be chosen below.
Then
\begin{equation*}
K^{X,2}=K^{X,1}+\nabla^\mu\Bigl(\frac{f^\prime}{f(1-\frac{2m}{r^{n-2}})}\beta\,X_\mu\,\phi^2\Bigr)
\end{equation*}\\[-4.0ex]
\begin{multline}
=\frac{f^\prime}{1-\frac{2m}{r^{n-2}}}\bigl(\frac{\partial\phi}{\partial\rs}+\beta\phi\bigr)^2
+\frac{f}{r}\bigl(1-\frac{nm}{r^{n-2}}\bigr)\normsph{\nablab\phi}^2\displaybreak[1]\\
-\frac{1}{4}\frac{\tprime{f}}{1-\frac{2m}{r^{n-2}}}\phi^2+\frac{\dprime{f}}{1-\frac{2m}{r^{n-2}}}\Bigl[\beta-\frac{n-1}{2r}\bigl(1-\frac{2m}{r^{n-2}}\bigr)\Bigr]\phi^2\displaybreak[0]\\
-\frac{\sprime{f}}{1-\frac{2m}{r^{n-2}}}\biggl[\beta^2-\sprime{\beta}-\frac{n-1}{r}\beta\bigl(1-\frac{2m}{r^{n-2}}\bigr)+\frac{n-1}{4r^2}\Bigl((n-3)+(n-1)\frac{2m}{r^{n-2}}\Bigr)\bigl(1-\frac{2m}{r^{n-2}}\bigr)\biggr]\phi^2\\
-\frac{n-1}{4}\Bigl[(n-1)^2\bigl(\frac{2m}{r^{n-2}}\bigr)-n\frac{2m}{r^{n-2}}-(n-3)\Bigr]\frac{f}{r^3}\phi^2
\end{multline}
Now choose
\begin{equation}
\beta=\frac{n-1}{2r}\bigl(1-\frac{2m}{r^{n-2}}\bigr)+\delta
\end{equation}
then
\begin{equation}
\beta^2-\sprime{\beta}-\frac{n-1}{r}\beta\bigl(1-\frac{2m}{r^{n-2}}\bigr)+\frac{n-1}{4r^2}\Bigl((n-3)+(n-1)\frac{2m}{r^{n-2}}\Bigr)\bigl(1-\frac{2m}{r^{n-2}}\bigr)=-\sprime{\delta}+\delta^2
\end{equation}
and
\begin{multline}
K^{X,2}=\frac{f^\prime}{1-\frac{2m}{r^{n-2}}}\bigl(\frac{\partial\phi}{\partial\rs}+\beta\phi\bigr)^2
+\frac{f}{r}\bigl(1-\frac{nm}{r^{n-2}}\bigr)\normsph{\nablab\phi}^2\displaybreak[0]\\
-\frac{1}{1-\frac{2m}{r^{n-2}}}\biggl\{\frac{1}{4}\tprime{f}-\delta\dprime{f}+\bigl(\delta^2-\sprime{\delta}\bigr)\sprime{f}\biggr\}\phi^2\\
-\frac{n-1}{4}\Bigl[(n-1)^2\bigl(\frac{2m}{r^{n-2}}\bigr)^2-n\frac{2m}{r^{n-2}}-(n-3)\Bigr]\frac{f}{r^3}\phi^2
\end{multline}
\begin{note}
Suppose outside a compact interval $[-\alpha,\alpha]\subset\mathbb{R}$ $\sprime{f}$ is of the form $\sprime{f}(\rs)=\frac{1}{\rs^2}\ (|\rs|>\alpha)$.
Then we could choose $\delta=-\frac{1}{\rs}\ (|\rs|>\alpha)$ so that $\delta\dprime{f}=\frac{2}{\rs^4}\geq 0$ and $-\sprime{\delta}+\delta^2=0\,.$
\end{note}

\paragraph{Definition of the current $J^{(\alpha)}$.} Let $\alpha>0$ and introduce a shifted coordinate 
\begin{equation}
  x=\rs-\alpha-\sqrt{\alpha}\,.
\end{equation}
The modification we choose is 
\begin{equation}
\delta=-\frac{x}{\alpha^2+x^2}
\end{equation}
so that 
\begin{equation}
-\sprime{\delta}+\delta^2=\frac{\alpha^2}{(\alpha^2+x^2)^2}\,.
\end{equation}
Let 
\begin{equation}\label{eq:faC}
f^a=-\frac{C}{\alpha^2 r^{n-1}}\qquad(C>0)
\end{equation}
and 
\begin{equation}
\sprime{(f^b)}=\frac{1}{\alpha^2+x^2}\qquad (f^b)(\rs)=\int_0^\rs\frac{1}{\alpha^2+x(t^\ast)^2}\ud t^\ast\,.
\end{equation}
Note that then
\begin{equation}\label{identityfa}
\sprime{(f^a)}+(n-1)\frac{f^a}{r}\bigl(1-\frac{2m}{r^{n-2}}\bigr)=0
\end{equation}
and
\begin{equation}\label{identityfb}
\frac{1}{4}\tprime{(f^b)}-\delta\dprime{(f^b)}+(\delta^2-\sprime{\delta})\sprime{(f^b)}=-\frac{1}{2}\frac{x^2-\alpha^2}{(x^2+\alpha^2)^3}\,.
\end{equation}
Our current is built from the multiplier vectorfields 
\begin{equation}
X^a=f^a\frac{\partial}{\partial\rs}\qquad X^b=f^b\frac{\partial}{\partial\rs}
\end{equation}
by setting
\begin{equation}\label{currentalpha}
J_\mu^{(\alpha)}(\phi)\doteq J_\mu^{X^a,0}(\phi)+\sum_{i=1}^\frac{n(n-1)}{2}J_\mu^{X^b,2}(\Omega_i\phi)
\end{equation}
and will be shown to have the property that its divergence 
\begin{equation}
K^{(\alpha)}\doteq\nabla^\mu J_\mu^{(\alpha)}
\end{equation}
is nonnegative upon integration over the spheres.

\begin{prop}[Positivity of the current $J^{(\alpha)}$]\label{positivitypropalpha}
For $n\geq 3$, and $\phi\in\mathrm{H}^1(S)$ \[\int_S K^{(\alpha)}\dm{\gamma}\geq 0\] provided $\alpha$ is chosen sufficently large, and $C(n,m,\alpha)$ set to be \eqref{Cstar} below.
\end{prop}
\noindent\emph{Proof.}
In view of \eqref{identityfa} and \eqref{identityfb}
\begin{multline}\label{Kalphalowerbound}
K^{(\alpha)}\geq\frac{\sprime{(f^a)}}{1-\frac{2m}{r^{n-2}}}\bigl(\frac{\partial\phi}{\partial\rs}\bigr)^2+\frac{f^a}{r}\bigl(1-\frac{nm}{r^{n-2}}\bigr)\normsph{\nablab\phi}^2\displaybreak[0]\\
+\sum_{i=1}^\frac{n(n-1)}{2}\frac{f^b}{r}\Bigl(1-\frac{nm}{r^{n-2}}\Bigr)\normsph{\nablab\Omega_i\phi}^2+\sum_{i=1}^\frac{n(n-1)}{2}\,F\,(\Omega_i\phi)^2\\
+\sum_{i=1}^\frac{n(n-1)}{2}\frac{n-1}{4r^3}\Bigl[(n-3)+n\frac{2m}{r^{n-2}}-(n-1)^2\bigl(\frac{2m}{r^{n-2}}\bigr)^2\Bigr]\,f^b\,(\Omega_i\phi)^2
\end{multline}
where
\begin{equation}
F\doteq\frac{1}{2}\frac{1}{1-\frac{2m}{r^{n-2}}}\frac{x^2-\alpha^2}{(x^2+\alpha^2)^3}\,.
\end{equation}
So by Poincar\'e's inequality \eqref{poincareSn} and \eqref{sumgenerators}
\begin{multline} \label{intKalphalowerbound}
\int_S K^{(\alpha)} \dm{\gamma}\geq\int_S\biggl\{\frac{C(n-1)}{\alpha^2 r^n}\bigl(\frac{\partial\phi}{\partial\rs}\bigr)^2+\\
+\Bigl[(n-1)\frac{f^b}{r}\bigl(1-\frac{nm}{r^{n-2}}\bigr)+F\,r^2+\frac{1}{r}H\Bigr]\normsph{\nablab\phi}^2\biggr\}\dm{\gamma}
\end{multline}
where
\begin{equation}
H\doteq\frac{n-1}{4}\Bigl[(n-3)+n\frac{2m}{r^{n-2}}-(n-1)^2\bigl(\frac{2m}{r^{n-2}}\bigr)^2\Bigr]f^b-\frac{C}{\alpha^2 r^{n-1}}\bigl(1-\frac{nm}{r^{n-2}}\bigr)\,.
\end{equation}
\subsubsection*{Step 1. $H\geq 0$}
It is equivalent to show that \[\breve{H}\doteq r^{n-1}\,H\:\frac{r^{n-2}}{2m}\] is nonnegative. We consider $\breve{H}$ to be a function of \[\rho\doteq\frac{r^{n-2}}{2m}\] so
\begin{equation*}
\breve{H}=\frac{n-1}{4}(2mr)\Bigl[(n-3)\rho^2+n\rho-(n-1)^2\Bigr]f^b-\frac{C}{\alpha^2}\bigl(\rho-\frac{n}{2}\bigr)
\end{equation*}
Note that \[r=\sqrt[n-2]{nm}\iff\rho=\frac{n}{2}\iff\rs=0\] and \[\breve{H}(\frac{n}{2})=0\,.\]
Moreover we choose the constant $C$ such that
\[\frac{\ud\breve{H}}{\ud\rho}\vert_{\rho=\frac{n}{2}}=0\,.\]
\begin{multline*}
\frac{\ud\breve{H}}{\ud\rho}=\frac{n-1}{4}(2mr)\Bigl[\frac{(n-3)(2n-3)}{n-2}\rho+\frac{n-1}{n-2}n-\frac{(n-1)^2}{n-2}\frac{1}{\rho}\Bigr]f^b\\
+\frac{n-1}{4(n-2)}\frac{2mr^2}{\rho-1}\Bigl[(n-3)\rho^2+n\rho-(n-1)^2\Bigr]\sprime{(f^b)}-\frac{C}{\alpha^2}
\end{multline*}
where we have used \[\frac{\ud r}{\ud\rho}=\frac{r}{(n-2)\rho}\qquad\quad\frac{\ud\rs}{\ud\rho}=\frac{1}{\rho-1}\frac{r}{n-2}\,.\]
Hence we choose
\begin{equation}\label{Cstar}
C=\frac{(n-1)^2}{4(n-2)}\frac{(\frac{n}{2})^2-(n-1)}{\frac{n}{2}-1}\,2m\,(nm)^\frac{2}{n-2}\,\frac{\alpha^2}{\alpha^2+(\alpha+\sqrt{\alpha})^2}\,.\tag{$\ast$}
\end{equation}
Note that then also \[\frac{\ud H}{\ud r}\vert_{r=\sqrt[n-2]{nm}}=0\,.\]
Now returning to the expression for $\breve{H}$ let us denote by $1\leq\rho_0\leq\frac{n}{2}$ the value of $\rho$ for which \[(n-3)\rho_0+n-(n-1)^2\frac{1}{\rho_0}=0\,,\] i.e.
\begin{equation*}
\rho_0=\frac{2(n-1)^2}{n+\sqrt{n^2+4(n-1)^2(n-3)}}\,.
\end{equation*}
We divide into the four regions \[1<\rho_0<\frac{n}{2}<\rho^\ast<\infty\] where $\rho^\ast$ is to be chosen large enough below.
\paragraph{Step 1a. \textnormal{(near the horizon, $1\leq\rho\leq\rho_0$)}}
Clearly $\breve{H}\geq 0$ termwise, because $f^b\leq 0$.
\paragraph{Step 1b. \textnormal{(near the photon sphere, $\rho_0\leq\rho\leq\frac{n}{2}$)}}
We show $H=H(r)$ is convex on $r_0\leq r\leq\sqrt[n-2]{nm}$ where \[r_0=\sqrt[n-2]{\frac{4(n-1)^2m}{n+\sqrt{n^2+4(n-1)^2(n-3)}}}\,.\]
Differentiating twice yields
\begin{multline*}
\frac{\ud^2 H}{\ud r^2}=\frac{n-1}{4}\frac{1}{\bigl(1-\frac{2m}{r^{n-2}}\bigr)^2}\dprime{(f^b)}\Bigl[(n-3)+n\frac{2m}{r^{n-2}}-(n-1)^2\bigl(\frac{2m}{r^{n-2}})^2\Bigr]\\
+\frac{n-1}{2}\frac{1}{1-\frac{2m}{r^{n-2}}}\sprime{(f^b)}(n-2)\Bigl[2(n-1)^2\frac{2m}{r^{n-2}}-n\Bigr]\frac{2m}{r^{n-1}}\displaybreak[0]\\
-\frac{n-1}{4}\frac{1}{\bigl(1-\frac{2m}{r^{n-2}}\bigr)^2}\sprime{(f^b)}(n-2)\Bigl[(n-3)+n\frac{2m}{r^{n-2}}-(n-1)^2\bigl(\frac{2m}{r^{n-2}}\bigr)^2\Bigr]\frac{2m}{r^{n-1}}\displaybreak[1]\\
+\frac{n-1}{4}(f^b)\Bigl[(n-2)(n-1)n\frac{2m}{r^n}-2(2n-3)(n-2)(n-1)^2\bigl(\frac{2m}{r^{n-1}}\bigr)^2\Bigr]\\
-\frac{(n-1) n C}{\alpha^{n-1}r^{n+1}}\bigl(1-\frac{nm}{r^{n-2}}\bigr)+3\frac{(n-1)(n-2)C}{\alpha^{n-1}r^n}\frac{nm}{r^{n-1}}
\end{multline*}
Since $\dprime{(f^b)}\geq 0$ we further have in this region the bound
\begin{multline*}
\frac{\ud^2 H}{\ud r^2}\geq\frac{n-1}{2}\frac{1}{1-\frac{2m}{r^{n-2}}}\times\\\times\biggl[2(n-1)^2\frac{2m}{r^{n-2}}-n-\frac{1}{2}\frac{1}{1-\frac{2m}{r_0^{n-2}}}\Bigl((n-3)+n\frac{2m}{r^{n-2}}-(n-1)^2\bigl(\frac{2m}{r^{n-2}}\bigr)^2\Bigl)\biggr]\times\\\times\frac{2m}{r^{n-1}}(n-2)\sprime{(f^b)}\displaybreak[0]\\
+\frac{n-1}{4}\frac{2m}{r^{n-2}}\Bigl[1-\frac{2(2n-3)(n-1)}{n}\bigl(\frac{2m}{r^{n-2}}\bigr)\Bigr]\frac{(f^b)}{r^2}
\end{multline*}
Since for $n\geq 3$
\begin{multline*}
2(n-1)^2\frac{2}{n}-n\\-\frac{1}{2}\frac{2(n-1)^2}{2(n-1)^2-n-\sqrt{n^2+4(n-1)^2(n-3)}}\Bigl((n-3)+2-\bigl(2\frac{n-1}{n}\bigr)^2\Bigr)\\\geq 1
\end{multline*}
and \[1-\frac{2(2n-3)(n-1)}{n}\frac{2}{n}\leq -1\]
we finally obtain in this region
\begin{equation*}
\frac{\ud^2 H}{\ud r^2}\geq\frac{(n-1)(n-2)}{2r}\frac{1}{\rho-1}\sprime{(f^b)}>0\,.
\end{equation*}
\paragraph{Step 1c. \textnormal{(in the intermediate region, $\frac{n}{2}\leq\rho\leq\rho^\ast$)}}
We show $\breve{H}=\breve{H}(\rho)$ is convex on $\frac{n}{2}\leq\rho\leq\rho^\ast$ for $\rs(\rho=\rho^\ast)\leq\alpha$.
\begin{multline*}
\frac{\ud^2\breve{H}}{\ud\rho^2}=\frac{(n-1)^2}{4(n-2)^2}\frac{2mr}{\rho^2}\Bigl[(n-3)(2n-3)\rho^2+n\rho+(n-3)(n-1)\Bigr](f^b)\displaybreak[0]\\
+\frac{(n-1)^2}{4(n-2)^2}\frac{2mr^2}{(\rho-1)^2}\times\\\times\biggl[3(n-3)\rho^2-3(n-5)\rho+(n-1)(n-5)-n\frac{2n-1}{n-1}+3(n-1)\frac{1}{\rho}\biggr]\sprime{(f^b)}\displaybreak[0]\\
+\frac{n-1}{4(n-2)^2}\frac{2mr^3}{(\rho-1)^2}\Bigl[(n-3)\rho^2+n\rho-(n-1)^2\Bigr]\dprime{(f^b)}
\end{multline*}
Since for $\rho\geq\frac{n}{2}$, and $n\geq 3$, \[3(n-3)\rho(\rho-1)+6\rho+(n-1)(n-5)-n\frac{2n-1}{n-1}+3(n-1)\frac{1}{\rho}\geq 1\]
and \[(n-3)\rho^2+n\rho-(n-1)^2\geq 0\] we have
\begin{equation*}
\frac{\ud^2\breve{H}}{\ud\rho^2}\geq\frac{(n-1)^2}{4(n-2)^2}\frac{2mr^2}{(\rho-1)^2}\sprime{(f^b)}>0
\end{equation*}
because $(f^b)\geq 0$ for $\rs\geq 0$, and $\dprime{(f^b)}\geq 0$ for $x\leq 0$.
\paragraph{Step 1d. \textnormal{(in the asymptotics, $\rho\geq\rho^\ast$)}} We show directly $H(r)>0$ for $\rs\geq R^\ast\doteq\rs(\rho=\rho^\ast)$ and $\rho^\ast$ chosen large enough.
Let $\rs\geq R^\ast$, $R^\ast\leq\alpha$ then
\begin{equation}
f^b\geq\int_0^{R^\ast}\sprime{(f^b)}\ud\rs=\frac{1}{\alpha}\int_{-\bigl(1+\frac{1}{\sqrt{\alpha}}\bigr)}^\frac{R^\ast-\alpha-\sqrt{\alpha}}{\alpha}\frac{1}{1+{t^\ast}^2}\ud t^\ast\geq\frac{R^\ast}{5\alpha^2}
\end{equation}
provided $\alpha\geq 1$, and of course \[f^b\leq\frac{1}{\alpha}\arctan t^\ast\vert_{-(1+\frac{1}{\sqrt{\alpha}})}^0\leq\frac{\pi}{2\alpha}\,.\]
Thus
\begin{align*}
H = &\phantom{+}\frac{(n-1)(n-3)}{4}+\Bigl[\frac{(n-1)n}{4}f^b-\frac{C}{\alpha^2 2m}\frac{1}{r}\Bigr]\frac{2m}{r^{n-2}}\\
  &-\Bigl[\frac{(n-1)^3}{4}f^b-\frac{C n}{\alpha^2 4m}\frac{1}{r}\Bigr]\bigl(\frac{2m}{r^{n-2}}\bigr)^2\displaybreak[0]\\
 \geq&\phantom{+}\frac{1}{\alpha^2}\Bigl[\frac{(n-1)n}{4}\frac{R^\ast}{5}-\frac{C}{2m}\frac{1}{r}\Bigr]\frac{2m}{r^{n-2}}\\
  &-\frac{(n-1)^3}{4}\frac{\pi}{2\alpha}\bigl(\frac{2m}{r^{n-2}}\bigr)^2\\
 >&\ 0
\end{align*}
for $R^\ast$ (and consequently $\alpha$) chosen large enough.

\subsubsection*{Step 2.\ \eqref{fvsF}}
Since $\bigl(1-\frac{nm}{r^{n-2}}\bigr)f^b\geq 0$ and $F\geq 0$ for $|x|\geq\alpha$ we need to show
\begin{equation}\label{fvsF}
(n-1)(f^b)\bigl(1-\frac{nm}{r^{n-2}}\bigr)+F\,r^3\geq 0
\end{equation}
for \[-\alpha\leq x\leq\alpha \iff \sqrt{\alpha}\leq\rs\leq\sqrt{\alpha}+2\alpha\,.\]
In this whole region, in view of Prop.~\ref{prop:limitrsr},
\begin{gather*}
\lim_{\alpha\to\infty}\frac{\rs}{r}=1\\
\lim_{\alpha\to\infty}\bigl(1-\frac{2m}{r^{n-2}}\bigr)=\lim_{\alpha\to\infty}\bigl(1-\frac{nm}{r^{n-2}}\bigr)=1\,.
\end{gather*}
\framebox[1.2\width]{$n\geq 4$:} Since
\begin{equation*}
f^b(\rs) \geq \int_{\sqrt{\alpha}}^{\rs} \frac{1}{\alpha^2+x^2}\ud\rs \geq\frac{x+\alpha}{2\alpha^2}
\end{equation*}
it suffices to show
\begin{equation*}
(n-1)\frac{x+\alpha}{2\alpha^2}+\frac{1}{2}\frac{x^2-\alpha^2}{(x^2+\alpha^2)^3}r^3 \geq 0
\end{equation*}
which is implied by
\begin{equation*}
\frac{\alpha-x}{n-1}\frac{(x+\alpha+\sqrt{\alpha})^3}{(x^2+\alpha^2)^2}\leq 1\,.
\end{equation*}
For $-\alpha\leq x\leq 0$ \[(x+\alpha+\sqrt{\alpha})^3\leq\alpha^3(1+\frac{1}{\sqrt{\alpha}})^3\leq\frac{4}{3}\alpha^3\] for $\alpha$ large enough, thus
\begin{equation*}
\frac{\alpha-x}{n-1}\frac{(x+\alpha+\sqrt{\alpha})^3}{(x^2+\alpha^2)^2}\leq\frac{1}{n-1}\frac{2\alpha}{\alpha^4}\frac{4}{3}\alpha^3\leq\frac{8}{9}\,.
\end{equation*}
For $0\leq x\leq\alpha$ we have to show \[\frac{\alpha}{n-1}\frac{(x+\alpha+\sqrt{\alpha})^3}{(x^2+\alpha^2)^2}\leq 1\,.\]
Since \[(x+\alpha+\sqrt{\alpha})^3\leq 2^\frac{3}{2}(1+\frac{1}{\sqrt{\alpha}})^3(x^2+\alpha^2)^\frac{3}{2}\] we have for $\alpha$ large enough
\begin{equation*}
\frac{\alpha}{n-1}\frac{(x+\alpha+\sqrt{\alpha})^3}{(x^2+\alpha^2)^2}\leq
\frac{\alpha}{n-1}\frac{2^\frac{3}{2}(1+\frac{1}{\sqrt{\alpha}})^3}{(x^2+\alpha^2)^\frac{1}{2}}\leq
\frac{2^\frac{3}{2}}{3}(1+\frac{1}{\sqrt{\alpha}})^3<1\,.
\end{equation*}
\framebox[1.2\width]{$n=3$:} See \cite{DRNote}.
\qed
\\
\par Given the \emph{strict} inequalities proven in Step 2 of the proof of Prop.~\ref{positivitypropalpha} for $\alpha$ chosen large enough we can keep a fraction of the manifestly nonnegative $\vert\nablab\Omega_i\phi\vert^2$ term in \eqref{Kalphalowerbound}. Furthermore we have obtained control on the $\vert\nablab\phi\vert^2$ term from \eqref{intKalphalowerbound}.
\begin{cor}\label{cor:xestimate:alpha}
Let $\phi\in\mathrm{H}^2(S)$ be a solution of the wave equation \eqref{eq:wave}. Then there exists a constant $C(n,m)$ and a current $K$ such that
\begin{multline}\label{eq:xestimate:alpha}
\int_S\biggl\{
\frac{1}{r^n}\bigl(\frac{\partial\phi}{\partial\rs}\bigr)^2
+\frac{1}{r^{n+1}}\bigl(\frac{\partial\phi}{\partial t}\bigr)^2
+r\bigl(1-\frac{nm}{r^{n-2}}\bigr)^2\normsph{\nablab^2\phi}^2
\\+\frac{r^2}{(1-\frac{2m}{r^{n-2}})(1+\rs^2)^2}\normsph{\nablab\phi}^2
\biggr\}\dm{\gamma}\\
\leq C(n,m)\int_S K\dm{\gamma}\,.
\end{multline}

\end{cor}
\begin{proof}
Set $K=K^{(\alpha)}+K^{\text{aux}}$ and choose $\alpha$ large enough.\\
Here we retrieve the time derivatives with the auxiliary current \[K^\text{aux}=\nabla^\mu J_\mu^\text{aux};\quad J^\text{aux}=J^{X^\text{aux},0};\quad X^\text{aux}=f^\text{aux}\frac{\partial}{\partial\rs}\,,\]
where $f^\text{aux}=-\frac{1}{r^n}$ satisfies \[\sprime{(f^\text{aux})}+(n-1)\frac{f^\text{aux}}{r}\bigl(1-\frac{2m}{r^{n-2}}\bigr)=\frac{1}{r^{n+1}}\bigl(1-\frac{2m}{r^{n-2}}\bigr)\,;\] 
for in view of \eqref{eq:KX}
\begin{equation*}
\frac{1}{r^{n+1}}\bigl(\frac{\partial\phi}{\partial t}\bigr)^2\leq 2\,K^\text{aux}+3\,\frac{1}{r^{n+1}}\normsph{\nablab\phi}^2\,.\qedhere
\end{equation*}
\end{proof}

\subsection{Boundary Terms}
In this section we first prove Prop.~\ref{prop:ILED} and then a refinement thereof for finite regions, which requires to estimate the boundary terms of the currents introduced in Section \ref{sec:high} and \ref{sec:low}.
\subsubsection{Proof of Prop.~\ref{prop:ILED}}
\label{sec:pfILED}

We can now combine our ealier results Cor.~\ref{cor:xestimate:L} and Cor.~\ref{cor:xestimate:alpha} to prove the \emph{integrated local energy decay estimate} \eqref{eq:ILED};
note that there is no restriction on the spherical harmonic number, and that no commutation with angular momentum operators is required.

\smallskip\noindent\emph{Proof of Prop.~\ref{prop:ILED}.}
Write
\begin{equation}
  \phi=\pi_{<L}\phi+\pi_{\geq L}\phi
\end{equation}
with
\begin{equation}
 \pll=\sum_{l=0}^{L-1}\pi_l\phi\qquad\phl=\sum_{l=L}^\infty\pi_l\phi
\end{equation}
where $L=(6\gamma n)^2$ is fixed (recall here $\gamma=12$ from Section \ref{sec:high}).

\subsubsection*{Step 1. \textnormal{(High spherical harmonics)}}
By Cor.~\ref{cor:xestimate:L}
\begin{equation}
\int_{\mathcal{R}(t_0,t_1,u_1^\ast,v_1^\ast)}\frac{1}{4}\frac{1}{8}\frac{(2m)^\frac{3}{n-2}}{r^3}\bigl(\phl\phi)^2\leq\int_{\mathcal{R}(t_0,t_1,u_1^\ast,v_1^\ast)}K^{X_{\gamma,\alpha},1}(\phl\phi)\,.
\end{equation}

It remains to estimate the boundary terms of the current $J^{X_{\gamma,\alpha},1}$, and to use this estimate to recover all derivatives using a commutation with the Killing vectorfield $T$.
\paragraph{Step 1a. \textnormal{(Boundary terms)}}
We may assume $\vert r_{0,1}^\ast\vert\geq\frac{4}{\gamma\alpha}$, $r_{0,1}$ entering the definition \eqref{def:Sigmatau}. Recalling the properties of $f_{\gamma,\alpha}$ away from the photon sphere we find
\begin{multline*}
\vert(J^{X_{\gamma,\alpha},1}(\phl\phi),\dvs)\vert\leq\vert(J^{X_{\gamma,\alpha}}(\phl\phi),\dvs)\vert\\+\frac{1}{2}\Bigl\vert\sprime{f_{\gamma,\alpha}}+(n-1)\frac{f_{\gamma,\alpha}}{r}\bigl(\cf\bigr)\Bigr\vert\bigl(\phl\phi\bigr)\bigl(\ddvs{\phl}\bigr)\\+\frac{1}{4}\Bigl\vert\Bigl(\sprime{f_{\gamma,\alpha}}+(n-1)\frac{f_{\gamma,\alpha}}{r}\bigl(\cf\bigr)\Bigr)^\prime\Bigr\vert\bigl(\phl\phi\bigr)^2\\
\leq\frac{n+1}{(\gamma\alpha)^3}\Bigl(J^T(\phl\phi),\dvs\Bigr)+\frac{1}{(\gamma\alpha)^6}\frac{1}{|\rs|^4}\bigl(\ddvs{\phl\phi}\bigr)^2\\
+\frac{1}{(\gamma\alpha)^6}\frac{1}{\vert\rs\vert^4}\Bigl[1+\frac{1}{|\rs|}\Bigr]\bigl(\phl\phi\bigr)^2\\+\frac{n-1}{(\gamma\alpha)^3}\Bigl[n+\frac{4}{(\gamma\alpha)^6}\frac{r}{|\rs|^4}\Bigr]\frac{1}{2r ^2}\bigl(\cf\bigr)\bigl(\phl\phi\bigr)^2
\end{multline*}
and by Lemma \ref{lemma:poincare}
\begin{equation*}
  \int_{S_r}\frac{1}{2}\frac{1}{r^2}\bigl(\cf\bigr)\sqb{\pi_{\geq L}\phi}\leq\frac{1}{(6\gamma n)^4}\int_{S_r}\Bigl(J^T,\dvs\Bigr)\,;
\end{equation*}
similarly for \[\biggl\vert\Bigl(J^{X_{\gamma,\alpha},1}(\phl\phi),\dus\Bigr)\biggr\vert\,.\]
Since also by Lemma \ref{ineq:vhardy} and Lemma \ref{lemma:poincare},
\begin{multline*}
\int_{\frac{1}{2}(t_0-r_0^\ast)}^{u_1^\ast}\!\!\!\!\ud\us\int_\Sn\dm{\gn}\,r^{n-1}\frac{1}{\rs^4}(\phl\phi)^2\vert_{\vs=\frac{1}{2}(t_0+r_0^\ast)}\leq\displaybreak[0]\\
\leq\frac{8}{|r_0^\ast|^4}\frac{(1+|r_0^\ast|^2)^2}{|r_0^\ast|^2}\int_{\frac{1}{2}(t_0-r_0^\ast)}^\infty\int_\Sn\dm{\gn}\,r^{n-1}\bigl(\ddus{\phl\phi}\bigr)^2\ud\us\\
+2\pi\frac{1+|r_0^\ast|^2}{|r_0^\ast|^4}\Bigl[1+\frac{(nm)^\frac{2}{n-2}}{(6\gamma n)^4}\Bigr]\int_{\frac{1}{2}(t_0-r_0^\ast)}^{\frac{1}{2}(t_0-r_0^\ast)+1}\int_\Sn\dm{\gn}\,r^{n-1}\Bigl\{\sqv{\nablab\phl\phi}+\bigl(\ddus{\phl\phi}\bigr)^2\Bigr\}\ud\us
\end{multline*}
there is a constant $C(n,m)$ (recall $\gamma=12$, $\alpha=(n-1)/\rphp$) such that
\begin{multline*}
\int_{\frac{1}{2}(t_0-r_0^\ast)}^{u_1^\ast}\ud\us\int_{\Sn}\dm{\gn}\,r^{n-1}\Bigl\vert\bigl(J^{X_{\gamma,\alpha},1}(\phl\phi),\dus\bigr)\bigr\vert\leq\\
\leq C(n,m)\int_{\frac{1}{2}(t_0-r_0^\ast)}^\infty\ud\us\int_{\Sn}\dm{\gn}\,r^{n-1}\Bigl(J^T(\phl\phi),\dus\Bigr)\Bigr\vert_{\vs=\frac{1}{2}(t_0+r_0^\ast)}\,.
\end{multline*}
To establish
\begin{multline*}
\int_{\Sn}\Bigl\vert\bigl(J^{X_{\gamma,\alpha},1}(\phl\phi),\dt\bigr)\Bigr\vert\,r^{n-1}\dm{\gn}\leq\\
\leq C(n,m)\int_{\Sn}\bigl(J^T(\phl\phi),\dt\bigr)\,r^{n-1}\dm{\gn}
\end{multline*}
note that
\begin{multline*}
\Bigl\vert\bigl(J^{X_{\gamma,\alpha},1}(\phl\phi),\dt\bigr)\Bigr\vert\leq\Bigl\vert\bigl(J^{X_{\gamma,\alpha}}(\phl\phi),\dt\bigr)\Bigr\vert\\
+\frac{1}{2}\Bigl\vert\sprime{f_{\gamma,\alpha}}+(n-1)\frac{f_{\gamma,\alpha}}{r}\bigl(\cf\bigr)\Bigr\vert(\phl\phi)\bigl(\ddt{\phl\phi}\bigr)\leq\\
\leq\vert f_{\gamma,\alpha}\vert\,\vert T(\phl\phi)(\drs,\dt)\vert\\
+\frac{1}{2}\Bigl[\frac{1}{2}r^2\vert\sprime{f_{\gamma,\alpha}}\vert+\frac{3}{2}\frac{n-1}{2}\frac{1}{(\gamma\alpha)^3}\bigl(\cf\bigr)^2\Bigr]\frac{1}{r^2}(\phl\phi)^2\\
+\frac{1}{2}\frac{1}{(\gamma\alpha)^2}\Bigl[1+\frac{3}{2}\frac{n-1}{2}\frac{1}{\gamma\alpha}\Bigr]\bigl(\ddt{\phl\phi}\bigr)^2
\end{multline*}
and by Lemma \ref{lemma:poincare}
\begin{equation*}
\int_{S_r}\frac{1}{2}\frac{1}{r^2}(\phl\phi)^2\dm{\gamma}\leq\frac{1}{(6\gamma n)^4}\bigl(1-\frac{2m}{r_0^{n-2}}\bigr)^{-1}\int_{S_r}\bigl(J^T(\phl\phi),\dt\bigr)\dm{\gamma}
\end{equation*}
which suffices in view of the properties of $f_{\gamma,\alpha}$ in particular that there is a constant $r^2\vert\sprime{f_{\gamma,\alpha}}\vert\leq C(n,m)$.
For the boundary term
\begin{multline*}
\int_{\Sn}\Bigl\vert\bigl(J^{X_{\gamma,\alpha},1}(\phl\phi),\dvs\bigr)\Bigr\vert_{\us=u_1^\ast}\,r^{n-1}\dm{\gn}\leq\\
\leq\int_{\Sn}\biggl\{\frac{1}{(\gamma\alpha)^3}\Bigl[n+1+\frac{1}{2}\frac{1}{(\gamma\alpha)^2}+\frac{n-1}{(6\gamma n)^4}\bigl(n+\frac{4\rphp}{(\gamma\alpha)^2}\bigr)\Bigr]\bigl(J^T(\phl\phi),\dvs\bigr)\\
+\frac{1}{(\gamma\alpha)^6}\frac{1}{|\rs|^4}\bigl(1+\frac{\gamma\alpha}{4}\bigr)(\phl\phi)^2\biggr\}\,r^{n-1}\dm{\gn}
\end{multline*}
we find (using the boundedness of $\phi$ on the horizon, see Section \ref{sec:uniformboundedness}) in the limit $u_1^\ast\to\infty$ a constant $C(n,m)$ such that
\begin{multline*}
\int_{\Sn}\Bigl\vert\bigl(J^{X_{\gamma\alpha},1}(\phl\phi),\dvs\bigr)\Bigr\vert\,r^{n-1}\dm{\gn}\vert_{\us=\infty}\leq\\
\leq C(n,m)\int_{\Sn}\bigl(J^T(\phl\phi),\dvs\bigr)\,r^{n-1}\dm{\gn}\vert_{\us=\infty}\,.
\end{multline*}
We conclude that there is a constant $C(n,m)$ such that
\begin{equation}
  \int_{\mathcal{R}_{r_0,r_1}^\infty(t_0)}\frac{(2m)^\frac{3}{n-2}}{r^3}\bigl(\phl\phi)^2\leq C(n,m)\int_{\St{\tau_0}}\bigl(J^T(\phl\phi),n)\ ;
\end{equation}
where $\tau_0=\frac{1}{2}(t_0-r_1^\ast)$ because
\begin{equation}
  \Box_g(\phl\phi)=0\qquad K^T(\phl\phi)=0\,.
\end{equation}
\paragraph{Step 1b. \textnormal{(Commutation with $T$)}}
Since
\begin{equation}
  \Box_g\bigl(T\cdot\phl\phi\bigr)=0
\end{equation}
we also have
\begin{equation}
\int_{\mathcal{R}_{r_0,r_1}^\infty(t_0)}\frac{(2m)^\frac{3}{n-2}}{r^3}\bigl(\ddt{\phl\phi}\bigr)^2\leq C(n,m)\int_{\St{\tau_0}}\bigl(J^T(T\cdot\phl\phi\bigr),n\bigr)\,.
\end{equation}
This is enough to control the remaining derivatives, too;
for the auxiliary current \eqref{eq:auxcurrent} yields
\begin{equation}
  K^\text{aux}=\phi\,(\partial^\mu h)(\partial_\mu\phi)+h\,\partial^\alpha\phi\,\partial_\alpha\phi
\end{equation}
which upon choosing
\begin{equation}
  h=\bigl(\cf\bigr)\frac{(2m)^\frac{3}{n-2}}{r^3}
\end{equation}
presents us with
\begin{multline}
K^\text{aux}=\phi\frac{\partial h}{\partial r}\frac{\partial\phi}{\partial\rs}-\frac{(2m)^\frac{3}{n-2}}{r^3}\bigl(\ddt{\phi}\bigr)^2+\frac{(2m)^\frac{3}{n-2}}{r^3}\bigl(\ddrs{\phi}\bigr)^2\\
+\frac{(2m)^\frac{3}{n-2}}{r^3}\bigl(\cf\bigr)\normsph{\nablab\phi}^2\,.
\end{multline}
Using Cauchy's inequality for the first term we can bound
\begin{multline}
K^\text{aux}\geq\frac{1}{2}\frac{(2m)^\frac{3}{n-2}}{r^3}\bigl(\ddrs{\phi}\bigr)^2+\frac{(2m)^\frac{3}{n-2}}{r^3}\bigl(\cf\bigr)\normsph{\nablab\phi}^2\\
-\frac{(2m)^\frac{3}{n-2}}{r^3}\bigl(\ddt{\phi}\bigr)^2-2\frac{n^2}{(2m)^\frac{2}{n-2}}\frac{(2m)^\frac{3}{n-2}}{r^3}\phi^2\,.
\end{multline}
Therefore
\begin{multline}
\int_{\mathcal{R}_{r_0,r_1}^\infty(t_0)}\biggl\{\frac{1}{2}\frac{(2m)^\frac{3}{n-2}}{r^3}\bigl(\ddrs{\phl\phi}\bigr)^2+\frac{(2m)^\frac{3}{n-2}}{r^3}\bigl(\cf\bigr)\normsph{\nablab\phl\phi}^2\biggr\}\leq\\
\leq\int_{\mathcal{R}_{r_0,r_1}^\infty(t_0)}\biggl\{K^\text{aux}(\phl\phi)+\frac{(2m)^\frac{3}{n-2}}{r^3}\bigl(\ddt{\phl\phi}\bigr)^2+2\frac{n^2}{(2m)^\frac{2}{n-2}}\frac{(2m)^\frac{3}{n-2}}{r^3}(\phl\phi)^2\biggr\}\,.
\end{multline}
The boundary terms are controlled using Prop.~\ref{prop:boundary:aux}:
\begin{equation}
\label{eq:auxboundary}
\int_{\mathcal{R}_{r_0,r_1}^\infty(t_0)}K^\text{aux}(\phl\phi)\leq C(n,m)\int_\St{\tau_0}\bigl(J^T(\phl\phi),n\bigr)\,.
\end{equation}
Hence
\begin{multline}
\int_{\mathcal{R}_{r_0,r_1}^\infty(t_0)}\frac{(2m)^\frac{3}{n-2}}{r^3}\biggl\{\Bigl(\ddrs{\phl\phi}\Bigr)^2+\bigl(\cf\bigr)\normsph{\nablab\phl\phi}^2\biggr\}\leq\\
\leq C(n,m)\int_\St{\tau_0}\Bigl(J^T(\phl\phi)+J^T(T\cdot\phl\phi),n\Bigr)\,.
\end{multline}

\subsubsection*{Step 2. \textnormal{(Low spherical harmonics)}}
Now recall the $J^{(\alpha)}$ current \eqref{currentalpha}; we will show in a first step that
\begin{equation}
\int_{\mathcal{R}_{r_0,r_1}^\infty(t_0)}K^{(\alpha)}(\phi)\leq C(n,m)\int_\St{\tau_0}\Bigl(J^T(\phi)+\sum_{i=1}^\frac{n(n-1)}{2}J^T(\Omega_i\phi),n\Bigr)\,.
\end{equation}
Then in particular by Cor.~\ref{cor:xestimate:alpha}
\begin{multline}
\int_{\mathcal{R}_{r_0,r_1}^\infty(t_0)}\biggl\{\frac{1}{r^n}\Bigl(\ddrs{\pll\phi}\Bigr)^2+\frac{1}{r^{n+1}}\Bigl(\ddt{\pll\phi}\Bigr)^2\\+\frac{r^2}{(\cf)(1+|\rs|^2)^2}\normsph{\nablab\pll\phi}^2\biggr\}\leq\\
\leq C(n,m)\int_\St{\tau_0}\Bigl(J^T(\pll\phi)+\sumn J^T(\Omega_i\cdot\pll\phi),n\Bigr)\,.
\end{multline}
But in a second step we will show that in fact there exists a constant $C(n)$ such that
\begin{equation}
\int_\St{\tau_0}\sumn\Bigl(J^T(\Omega_i\cdot\pll\phi),n\Bigr)\leq C(n)\int_\St{\tau_0}\Bigl(J^T(\pll\phi),n\Bigr)\,.
\end{equation}

\paragraph{Step 2a. \textnormal{(Boundary Terms)}}
The energy identity for $J^{(\alpha)}$ on the domain \eqref{def:Rfinite} implies more explicitly:
\begin{multline*}
  \int_{\mathcal{R}(t_0,t_1,u_1^\ast,v_1^\ast)}K^{(\alpha)}\leq\int_{\frac{1}{2}(t_0+r_0^\ast)}^{\frac{1}{2}(t_1+r_0^\ast)}\int_\Sn\vert(J^{(\alpha)},\dvs)\vert\,r^{n-1}\vert_{\us=u_1^\ast}\ud\vs\dm{\gn}\displaybreak[0]\\
+\int_{\frac{1}{2}(t_1-r_0^\ast)}^{u_1^\ast}\int_\Sn\vert(J^{(\alpha)},\dus)\vert\,r^{n-1}\vert_{\vs=\frac{1}{2}(t_1+r_0^\ast)}\ud\us\dm{\gn}\displaybreak[0]\\
+\int_{r_0^\ast}^{r_1^\ast}\int_\Sn\vert(J^{(\alpha)},T)\vert\,r^{n-1}\vert_{t=t_1}\ud\rs\dm{\gn}\displaybreak[0]\\
+\int_{\frac{1}{2}(t_1+r_1^\ast)}^{v_1^\ast}\int_\Sn\vert(J^{(\alpha)},\dvs)\vert\,r^{n-1}\vert_{\us=\frac{1}{2}(t_1-r_1^\ast)}\ud\vs\dm{\gn}\displaybreak[0]\\
+\int_{\frac{1}{2}(t_0-r_1^\ast)}^{\frac{1}{2}(t_1-r_1^\ast)}\int_\Sn\vert(J^{(\alpha)},\dus)\vert\,r^{n-1}\vert_{\vs=v_1^\ast}\ud\vs\dm{\gn}\displaybreak[0]\\
+\int_{\frac{1}{2}(t_0-r_0^\ast)}^{u_1^\ast}\int_\Sn\vert(J^{(\alpha)},\dus)\vert\,r^{n-1}\vert_{\vs=\frac{1}{2}(t_0+r_0^\ast)}\ud\us\dm{\gn}\displaybreak[0]\\
+\int_{r_0^\ast}^{r_1^\ast}\int_\Sn\vert(J^{(\alpha)},T)\vert\,r^{n-1}\vert_{t=t_0}\ud\rs\dm{\gn}\displaybreak[0]\\
+\int_{\frac{1}{2}(t_0+r_1^\ast)}^{v_1^\ast}\int_\Sn\vert(J^{(\alpha)},\dvs)\vert\,r^{n-1}\vert_{\us=\frac{1}{2}(t_0-r_1^\ast)}\ud\vs\dm{\gn}
\end{multline*}
For the boundary integrals on the $t$-const hypersurfaces, we will use $(ii)$ of the following Lemma.
\begin{lemma}[Boundary terms of $J^{(\alpha)}$ current on $t$-const hypersurfaces]
\label{lemma:boundary:alphacurrent:tconst}
On each $\bar{\Sigma}_t$
\begin{itemize}
\item[(i)] there exists a constant $C(n,m,\alpha)$ such that \[\int_\mathbb{R}\vert(J^{(\alpha)},T)\vert\,r^{n-1}\ud\rs\leq C(n,m,\alpha)\int_\mathbb{R}(J^T(\phi)+\sum_{i=1}^\frac{n(n-1)}{2}J^T(\Omega_i\phi),T)\,r^{n-1}\ud\rs\]
\item[(ii)] for $r\geq r_0$ a constant $C(n,m,\alpha,r_0)$ such that\[\vert(J^{(\alpha)},T)\vert\leq C(n,m,\alpha,r_0)\Bigl(J^T(\phi)+\sum_{i=1}^\frac{n(n-1)}{2}J^T(\Omega_i\phi),T\Bigr)\,.\]
\end{itemize}
\end{lemma}
\begin{proof}
Using the definition \eqref{currentalpha},
\begin{multline*}
(J^{(\alpha)},T)=f^a\bigl(\ddt{\phi}\bigr)\bigl(\ddrs{\phi}\bigr)+\sumn f^b\bigl(\ddt{\Omega_i\phi}\bigr)\bigl(\ddrs{\Omega_i\phi}\bigr)\\
+\frac{1}{4}\sumn\biggl(\sprime{(f^b)}+(n-1)\frac{f^b}{r}\bigl(\cf\bigr)\biggr)2\bigl(\Omega_i\phi\bigr)\bigl(\partial_t\Omega_i\phi\bigr)
\end{multline*}
because \[\partial_t\Bigl(\sprime{(f^b)}+(n-1)\frac{f^b}{r}\bigl(\cf\bigr)\Bigr)=0\]
and $g(T,\drs)=0$. By Cauchy's inequality
\begin{multline*}
  |(J^{(\alpha)},T)|\leq\frac{C}{\alpha^2 r^{n-1}}\Bigl[\frac{1}{2}\bigl(\ddt{\phi}\bigr)^2+\frac{1}{2}\bigl(\ddrs{\phi}\bigr)^2\Bigr]\\
+\sumn\frac{\pi}{\alpha}\biggl[\frac{1}{2}\bigl(\ddt{\Omega_i\phi}\bigr)^2+\frac{1}{2}\bigl(\ddrs{\Omega_i\phi}\bigr)^2\biggr]\displaybreak[0]\\
+\frac{1}{4}\sumn\Bigl(\frac{r}{\alpha^2+x^2}+(n-1)\frac{\pi}{\alpha}\bigl(\cf\bigr)\Bigr)\biggl[\frac{1}{r^2}\bigl(\Omega_i\phi\bigr)^2+\bigl(\ddt{\Omega_i\phi}\bigr)^2\biggr]
\end{multline*}
which proves (ii) in view of \[\Bigl(J^T(\phi),T\Bigr)=\frac{1}{2}\bigl(\ddt{\phi}\bigr)^2+\frac{1}{2}\bigl(\ddrs{\phi}\bigr)^2+\frac{1}{2}\bigl(\cf\bigr)|\nablab\phi|^2\,;\]
here we have also used
\begin{equation*}
  f^b=\int_0^{\rs}\frac{1}{\alpha^2+(t^\ast-\alpha-\sqrt{\alpha})^2}\ud t^\ast=\frac{1}{\alpha}\arctan x\vert_\frac{-\alpha-\sqrt{\alpha}}{\alpha}^\frac{\rs-\alpha-\sqrt{\alpha}}{\alpha}\leq\frac{\pi}{\alpha}\quad(\rs\geq 0)\,.
\end{equation*}
To establish (i) it is enough to infer
\begin{equation*}
\begin{split}
\int_{-\infty}^\infty\frac{r}{\alpha^2+x^2}|\nablab\phi|^2\,r^{n-1}\ud\rs &= \sumn\int_{-\infty}^\infty\frac{r^{n-2}}{\alpha^2+x^2}\bigl(\Omega_i\phi\bigr)^2\ud\rs\\
&\leq C\sumn\int_{-\infty}^\infty\bigl(\ddrs{\Omega_i\phi}\bigr)^2\,r^{n-1}\ud\rs\\
&\leq C\sumn\int_{-\infty}^\infty\Bigl(J^T(\Omega_i\phi),T\Bigr)\,r^{n-1}\ud\rs\,;
\end{split}
\end{equation*}
this is a standard Hardy inequality, cf. Proof of Prop.10.2 in \cite{DRRadDecay}.
\end{proof}
The following Lemma will be applied to the boundary terms of the $J^{(\alpha)}$-current on the null hypersurfaces in the region $r\leq r_0$.
\begin{lemma}[Boundary terms of the $J^{(\alpha)}$ current on null hypersurfaces]
\label{lemma:boundary:currentalpha:null}
\ 
\begin{itemize}
\item[(i)] On any segment of the outgoing null hypersurface $\us=u_1^\ast\geq 0$
\begin{multline*}\vert(J^{(\alpha)},\dvs)\vert\leq C(n,m,\alpha)\Bigl(J^T(\phi)+\sumn J^T(\Omega_i\phi),\dvs\Bigr)\\+\epsilon(u_1^\ast)\,\bigl(J^N(\phi),\dvs\bigr)\end{multline*}
where $C(n,m,\alpha)$ is a constant, and $\epsilon(u_1^\ast)\to 0$ as $u_1^\ast\to\infty$.
\item[(ii)] Let $v_0^\ast\geq 1$, and $u_0^\ast(\vs)$ such that $r(u_0^\ast(\vs),\vs)=r_0$ (in particular $u_0^\ast(v_0^\ast)\geq 1$). Then on the ingoing null hypersurface $\vs=v_0^\ast$:
\begin{multline*}
\int_{u_0^\ast}^\infty\vert(J^{(\alpha)},\dus)\vert\,r^{n-1}\ud\us\leq\\
\leq C(n,m,\alpha)\int_{u_0^\ast}^\infty\Bigl(J^T(\phi)+\sumn J^T(\Omega_i\phi),\dus\Bigr)\,r^{n-1}\ud\us
\end{multline*}
\end{itemize}
\end{lemma}
\begin{proof}
Using the definition \eqref{currentalpha} we find
\begin{multline*}
(J^{(\alpha)},\dus)=f^a\,T(\phi)(\drs,\dus)+\sumn\biggl\{f^b\,T(\Omega_i\phi)(\drs,\dus)\displaybreak[0]\\
+\frac{1}{4}\Bigl(\sprime{(f^b)}+(n-1)\frac{f^b}{r}\bigl(\cf\bigr)\Bigr)\,2(\Omega_i\phi)\bigl(\ddus{\Omega_i\phi}\bigr)\displaybreak[0]\\
+\frac{1}{4}\frac{1}{2}\Bigl(\sprime{(f^b)}+(n-1)\frac{f^b}{r}\bigl(\cf\bigr)\Bigr)^\prime\,(\Omega_i\phi)^2
-\sprime{(f^b)}\,\beta\,(\Omega_i\phi)^2\biggr\}
\end{multline*}
and therefore
\begin{multline*}
  \vert(J^{(\alpha)},\dus)\vert\leq\frac{C}{\alpha^2 (2m)^\frac{n-1}{n-2}}\biggl[\frac{1}{2}\bigl(\ddus{\phi}\bigr)^2+\frac{1}{2}\bigl(\cf\bigr)\vert\nablab\phi\vert^2\biggr]\displaybreak[0]\\
+\sumn\frac{\pi}{\alpha}\biggl[\frac{1}{2}\bigl(\ddus{\Omega_i\phi}\bigr)^2+\frac{1}{2}\bigl(\cf\bigr)\vert\nablab\Omega_i\phi\vert^2\biggr]\displaybreak[0]\\
+\sumn\frac{1}{2}\Bigl(\frac{r}{\alpha^2+x^2}+(n-1)\frac{\pi}{\alpha}\bigl(\cf\bigr)\Bigr)\frac{1}{2}\bigl(\ddus{\Omega_i\phi}\bigr)^2\displaybreak[0]\\
+\biggl(\frac{n-1}{2}\frac{\pi}{\alpha}+\frac{n-1}{4}\frac{r}{\alpha^2+x^2}+\frac{n-1}{4}\frac{\pi}{\alpha}\bigl(\cf\bigr)\\+\frac{(n-1)(n-2)}{4}\frac{\pi}{\alpha}+(n-1)\frac{r}{\alpha^2+x^2}\biggr)\,\frac{1}{2}\bigl(\cf\bigr)\vert\nablab\phi\vert^2\displaybreak[0]\\
+\biggl(\frac{1}{4}\frac{r}{\alpha^2+x^2}+\frac{5}{4}\frac{|x|r^2}{(\alpha^2+x^2)^2}\biggr)\,\vert\nablab\phi\vert^2\,.
\end{multline*}
Similarly for $\vert(J^{(\alpha)},\dvs)\vert$.
Clearly, (ii) now follows from
\begin{multline*}
(J^T,\dvs)=\frac{1}{2}\bigl(\ddvs{\phi}\bigr)^2+\frac{1}{2}\bigl(\cf\bigr)\vert\nablab\phi\vert^2\displaybreak[0]\\
(J^N,\dvs)=\Bigl[1+\frac{\sigma}{4\kappa}\bigl(\cf\bigr)\Bigr]\,T(\frac{2}{\cf}\dus+\dt,\dvs)\\\geq 2\vert\nablab\phi\vert^2+\frac{1}{2}\bigl(\ddvs{\phi}\bigr)^2\,.
\end{multline*}
In case (i) we only have
\begin{gather*}
  (J^T,\dus)=\frac{1}{2}\bigl(\ddus{\phi}\bigr)^2+\frac{1}{2}\bigl(\cf\bigr)\vert\nablab\phi\vert^2\\
(J^N,\dus)\geq\frac{1}{2}\bigl(\cf\bigr)\vert\nablab\phi\vert^2\,;
\end{gather*}
but using the Hardy inequality of Lemma \ref{ineq:vhardy}:
\begin{multline*}
  \int_{u_0^\ast}^\infty\frac{r}{\alpha^2+x^2}\vert\nablab\phi\vert^2\,r^{n-1}\ud\us\leq\\
\leq\sumn\int_{u_0^\ast}^\infty\frac{1+\us^2}{\alpha^2+(\us+\alpha+\sqrt{\alpha}-\vs)^2}r_0^{n-2}\frac{1}{1+\us^2}(\Omega_i\phi)^2\ud\us\displaybreak[0]\\
\leq 8 C(n,m,\alpha)\frac{1+{u_0^\ast}^2}{{u_0^\ast}^2}\sumn\int_{u_0^\ast}^\infty\bigl(\ddus{\Omega_i\phi}\bigr)^2\,r^{n-1}\ud\us\\
+2\pi\,C(n,m,\alpha)\sumn\int_{u_0^\ast}^{u_0^\ast+1}\Bigl\{(\Omega_i\phi)^2+\bigl(\ddus{\Omega_i\phi}\bigr)^2\Bigr\}\ud\us\displaybreak[0]\\
\leq C(n,m,\alpha)\int_{u_0^\ast}^\infty\Bigl(J^T(\phi)+\sumn J^T(\Omega_i\phi),\dus\Bigr)\,r^{n-1}\ud\us\,.
\end{multline*}
Obviously the same bound holds for
\[\int_{u_0^\ast}^\infty\frac{|x|r^2}{(\alpha^2+x^2)^2}\vert\nablab\phi\vert^2\,r^{n-1}\ud\us\,.\qedhere\]
\end{proof}

\paragraph{Step 2b. \textnormal{(Commutation with $\Omega_i$)}}
Since
\begin{gather}
[\Omega_i,\dt]=0\\
\sumn\Bigl(\dt\Omega_i\cdot\pll\phi\Bigr)^2=r^2\normsph{\nablab\ddt{\pll\phi}}^2
\end{gather}
and since also
\begin{gather}
[\Omega_i,\frac{\partial}{\partial r}]=0\\
\sumn\Bigl(\drs\Omega_i\cdot\pll\phi\Bigr)^2=r^2\Bigl\vert\nablab\ddrs{\pll\phi}\Bigr\vert^2\,.
\end{gather}
Moreover
\begin{equation}
  [\pi_l,\dt]=0\,,
\end{equation}
so that
\begin{multline*}
\int_{S_r}\sumn\Bigl(\dt\Omega_i\cdot\pll\phi\Bigr)^2\dm{\gamma}=
\\=\int_{S_r}r^2\Bigl\vert\nablab\ddt{\pll\phi}\Bigr\vert^2\dm{\gamma}\leq L(L+n+2)\int_{S_r}\Bigl(\ddt{\pll\phi}\Bigr)^2\dm{\gamma}\,.
\end{multline*}
Since also (cf \eqref{projectionform})
\begin{equation*}
\int_{S_r}\normsph{\nablab\Omega_i\cdot\pll\phi}^2\dm{\gamma}\leq\frac{L(L+n-2)}{r^2}\int_{S_r}\bigl(\Omega_i\cdot\pll\phi\bigr)^2\dm{\gamma}
\end{equation*}
we have
\begin{equation*}
\int_{S_r}\sumn\normsph{\nablab\Omega_i\cdot\pll\phi}^2\dm{\gamma}\leq L(L+n-2)\int_{S_r}\normsph{\nablab\pll\phi}^2\dm{\gamma}\,.
\end{equation*}
Therefore indeed,
\begin{multline*}
\int_{S_r}\sumn\Bigl(J^T(\Omega_i\cdot\pll\phi),\dt\Bigr)=\frac{1}{2}\sumn\int_{S_r}\biggl\{\Bigl(\ddt{\Omega_i\cdot\pll\phi}\Bigr)^2\\+\Bigl(\ddrs{\Omega_i\cdot\pll\phi}\Bigr)^2+\bigl(\cf\bigr)\normsph{\nablab\Omega_i\cdot\pll\phi}^2\biggr\}\dm{\gamma}\leq\\
\leq\frac{1}{2}(6\gamma n)^2\bigl((6\gamma n)^2+n-2\bigr)\int_{S_r}\Bigl(J^T(\pll\phi),\dt\Bigr)\dm{\gamma}
\end{multline*}
because $L=(6\gamma n)^2$ is fixed; similarly, of course, for $(J^T,\dus)$ and $(J^T,\dvs)$.

We conclude the statement of the proposition with the treatment of the two regimes in Step 1 and Step 2 above from
\begin{multline*}
\int_{\mathcal{R}_{r_0,r_1}^\infty(t_0)}\Bigl\{\frac{1}{r^n}\bigl(\ddrs{\phi}\bigr)^2+\frac{1}{r^{n+1}}\bigl(\ddt{\phi}\bigr)^2+\frac{1}{r^3}\bigl(\cf\bigr)\normsph{\nablab\phi}^2\Bigr\}\leq\displaybreak[0]\\
\leq 2\int_{\mathcal{R}_{r_0,r_1}^\infty(t_0)}\Bigl\{\frac{1}{r^n}\bigl(\ddrs{\pll\phi}\bigr)^2+\frac{1}{r^{n+1}}\bigl(\ddt{\pll\phi}\bigr)^2+\frac{1}{r^2}\normsph{\nablab\pll\phi}^2\Bigr\}\\
+2 \int_{\mathcal{R}_{r_0,r_1}^\infty(t_0)}\Bigl\{\frac{1}{r^3}\bigl(\ddrs{\phl\phi}\bigr)^2+\frac{1}{r^{3}}\bigl(\ddt{\phl\phi}\bigr)^2+\frac{1}{r^3}\bigl(\cf\bigr)\normsph{\nablab\phl\phi}^2\Bigr\}\,.
\end{multline*}
\qed

\subsubsection{Refinement for finite regions}
\label{sec:refinements}

Note that in the proof of Prop.~\ref{prop:ILED} neither of the currents used for the high or the low spherical harmonic regime requires the use of Hardy inequalities for the boundary integrals in the asymptotic region; indeed in both cases the zeroth order terms $\phi^2$ can be estimated by the angular derivatives $\vert\nablab\phi\vert^2$, in the case of the current $J^{X_{\gamma,\alpha},1}$ for high angular frequencies by Poincar\'e's inequality Prop.~\ref{lemma:poincare}, and in the case of the current $J^{(\alpha)}$ for low angular frequencies as a result of the commutation with $\Omega_i$ in \eqref{currentalpha}.
Therefore we can in fact state a refinement of Prop.~\ref{prop:ILED} for finite regions, i.e.~an integrated local energy estimate on bounded domains in terms of the flux through the past boundary of that domain, that will be relevant in Section \ref{sec:iid}.

Let
\begin{gather}
  {}^{R}\mathcal{P}_{\tau_1}^{\tau_2}\doteq\mathcal{R}_{r_0,R}^\infty(2\tau_1+R^\ast,2\tau_2+R^\ast)\cap\{r\leq R\}\\
  \begin{split}
  {}^R\mathcal{D}\!\!\!\backslash\:{}_{\tau_1}^{\tau_2}&\doteq\Bigl\{(\us,\vs):\tau_1\leq \us\leq\tau_2\,,\vs-\us\geq R^\ast\,,\vs\leq\tau_2+R^\ast\Bigr\}\\
  &=\mathcal{R}_{r_0,R}^\infty(2\tau_1+R^\ast,2\tau_2+R^\ast,\tau_2+\frac{1}{2}(R^\ast-r_0^\ast),\tau_2+R^\ast)\setminus{}^{R}\mathcal{P}_{\tau_1}^{\tau_2}
  \label{def:s:Dslash}
  \end{split}
\end{gather}
and denote by $\Stt{\tau_1}{\tau_2}$ the past boundary of $\dP{R}{\tau_1}{\tau_2}\cup{}^R\mathcal{D}\!\!\!\backslash\:{}_{\tau_1}^{\tau_2}$ (see also figure \ref{fig:Sigmafinite}):
\begin{equation}
  \label{def:s:Sigmakomma}
  \begin{split}
  \Stt{\tau_1}{\tau_2}\doteq &\partial^{-}\Bigl(\dP{R}{\tau_1}{\tau_2}\cup{}^R\mathcal{D}\!\!\!\backslash\:{}_{\tau_1}^{\tau_2}\Bigr)\\
  =&\Bigl\{(\us,\vs):\ \vs=\tau_1+\frac{1}{2}(R^\ast+r_0^\ast)\,,\us\geq\tau_1+\frac{1}{2}(R^\ast-r_0^\ast)\Bigr\}\\&\cup\Bigl\{(\us,\vs):\ \us+\vs=2\tau_1+R^\ast\,,r_0^\ast\leq\vs-\us\leq R^\ast\Bigr\}\\&\cup\Bigl\{(\us,\vs):\ \us=\tau_1\,,R^\ast+\tau_1\leq \vs\leq R^\ast+\tau_2\Bigr\}
  \end{split}
\end{equation}
\begin{figure}[bt]
  \begin{center}
    \input{Sigmafinite.pstex_t}
    \caption{The past boundary $\Stt{\tau_1}{\tau_2}$ of $\dP{R}{\tau_1}{\tau_2}\cup{}^R\mathcal{D}\!\!\!\backslash\:{}_{\tau_1}^{\tau_2}$.}
    \label{fig:Sigmafinite}
  \end{center}
\end{figure}

\begin{prop}[integrated local energy decay on finite regions]\label{prop:rILED}
  Let $\phi$ be a solution of the wave equation $\Box_g\phi=0$, and $R>\rh$. Then
  there exists a constant $C(n,m,R)$, such that
  \begin{multline}\label{eq:rILED}
    \int_{\dP{R}{\tau_1}{\tau_2}}\Biggl\{\bigl(\frac{\partial\phi}{\partial\rs}\bigr)^2+\Bigl(\frac{\partial\phi}{\partial t}\Bigr)^2+\bigl(\cf\bigr)\normsph{\nablab\phi}^2\Biggr\}\dm{g}\\
    \leq C(n,m,R)\int_\Stt{\tau_1}{\tau_2}\Bigl(J^T(\phi)+J^T(T\cdot\phi),n\Bigr)
  \end{multline}
 for any $\tau_2>\tau_1$.
\end{prop}
\noindent In view of the remarks above the proof of Prop.~\ref{prop:rILED} is of course identical to the proof of Prop.~\ref{prop:ILED} given in Section \ref{sec:pfILED} by replacing the unbounded domain $\mathcal{R}_{r_0,r_1}^\infty(2\tau_1+R^\ast)$ by the bounded domain $\dP{R}{\tau_1}{\tau_2}\cup{}^R\mathcal{D}\!\!\!\backslash\:{}_{\tau_1}^{\tau_2}$.

However, this estimate does not include the zeroth order term, which we have covered seperately in Prop.~\ref{prop:zeroth}.
\begin{prop}[Refinement for zeroth order terms on timelike boundaries]\label{prop:zeroth:r}
  Let $\phi$ be solution of the wave equation \eqref{eq:wave}, and $R>\sqrt[n-2]{8nm}$. 
  Then there is a constant $C(n,m,R)$ such that
  \begin{multline}
    \label{eq:prop:zeroth:r}
    \int_{2\tau^\prime+R^\ast}^{2\tau+R^\ast}\ud t\int_\Sn\dm{\gn}\,\phi^2\vert_{r=R}\leq\\
    \leq C(n,m,R)\biggl\{\int_{2\tau^\prime+R^\ast}^{2\tau+R^\ast}\ud t\int_\Sn\dm{\gn}\,\Bigl\{\sqb{\ddrs{\phi}}+\sqv{\nablab\phi}\Bigr\}\bigr\vert_{r=R}\\
    +\int_\Stt{\tau^\prime}{\tau}\ned{T}{\phi}+\int_\Sn\!\!\!\!\dm{\gn}r^{n-2}\phi^2\vert_{(\sprime{\tau},R^\ast+\tau)}\biggr\}
  \end{multline}
for all $\tau^\prime<\tau$.
\end{prop}
\noindent The proof remains the same as for Prop.~\ref{prop:zeroth} on page \pageref{pg:pfzero} with the exception that Prop.~\ref{prop:fboundary:JX1} is used in place of Prop.~\ref{prop:boundary:JX1}.
\begin{proof}
More precisely, we have from the energy identity for $J^{X,1}$ on ${}^R\mathcal{D}\!\!\!\backslash\:{}_{\tau^\prime}^{\tau}$ that
\begin{multline}
\int_{R^\ast+2\taup}^{R^\ast+2\tau}\ud t\int_\Sn\dm{\gn} r^{n-1}\times\frac{n-1}{4R^2}\Bigl[\frac{1}{2}-(n-1)\frac{2m}{R^{n-2}}\Bigr]\bigl(1-\frac{2m}{R^{n-2}}\bigr)\phi^2\biggr\}\vert_{r=R}\displaybreak[1]\\
\leq\intT{\taup}{\tau}{R}{\frac{1}{2}\bigl(\cf\bigr)\normsph{\nablab\phi}^2+\frac{n-1}{2}\bigl(\cf\bigr)\bigl(\ddrs{\phi}\bigr)^2}\displaybreak[0]\\
+\int_\taup^\tau\ud\us\int_\Sn\dm{\gn}r^{n-1}\biggl\{\Bigl[\frac{1}{2}+\frac{n-1}{2}\bigl(\cf\bigr)\Bigr]\sqb{\ddus{\phi}}\\
+\frac{n-1}{4r^2}\Bigl[\frac{3}{2}-(n-1)\frac{2m}{r^{n-2}}\Bigr]\bigl(\cf\bigr)\phi^2\biggr\}\vert_{\vs=R^\ast+\tau}\displaybreak[0]\\
+\int_{R^\ast+\taup}^{R^\ast+\tau}\ud\vs\int_\Sn\dm{\gn}r^{n-1}\biggl\{\Bigl[\frac{1}{2}+\frac{n-1}{2}\bigl(\cf\bigr)\Bigr]\sqb{\ddvs{\phi}}\\
+\frac{n-1}{4r^2}\Bigl[\frac{3}{2}-(n-1)\frac{2m}{r^{n-2}}\Bigr]\bigl(\cf\bigr)\phi^2\biggr\}\vert_{\us=\taup}\,.
\end{multline}
On $\vs=R^\ast+\tau$, $\us=R^\ast+\tau-\rs$ is a function of $\rs$, hence one can change the variable of integration from $\us$ to $\rs$; furthermore for $R>\sqrt[n-2]{8nm}$ we note
\begin{gather*}
  (n-3)+n\frac{2m}{r^{n-2}}-(n-1)^2\bigl(\frac{2m}{r^{n-2}}\bigr)^2>\frac{3}{16}\\
  \frac{1}{2}-(n-1)\frac{2m}{r^{n-2}}>\frac{1}{4}\\
  \cf>\frac{3}{4}\,.
\end{gather*}
Hence \eqref{eq:prop:zeroth:r} indeed follows from Lemma \ref{lemma:hardyineqfinite} and the energy identity for $J^T$ on $\dP{R}{\tau_1}{\tau_2}\cup{}^R\mathcal{D}\!\!\!\backslash\:{}_{\tau_1}^{\tau_2}$.
\end{proof}

\section{The Decay Argument}
\label{sec:decay}

We will here prove energy decay of the solutions to the wave equation and higher order energy decay of their time derivatives in the interior based on the integrated local energy decay statements of Section \ref{sec:iled}, following the new physical-space approach to decay of \cite{DRNew}.

\begin{remark} Instead one could use the conformal Morawetz vectorfield \[Z=\us^2\dus+\vs^2\dvs\] to prove energy decay of solutions to the wave equation with a rate corresponding to the weights in $Z$; this is done in \cite{VSE}. Similary the use of the scaling vectorfield \[S=\vs\dvs+\us\dus\] should provide an alternative approach to prove higher order energy decay \cite{Lid}. Here however, we shall avoid the use of multipliers with weights in $t$.
\end{remark}

\subsection{Uniform Boundedness}
\label{sec:uniformboundedness}

A preliminary feature of the solutions to the wave equation \eqref{eq:wave} that is necessary to employ the decay mechanism of \cite{DRNew} is the uniform boundedness of their (nondegenerate) energy; this is a consequence of the conservation of the degenerate energy associated to the multiplier $T$, \emph{and the redshift effect} of Section \ref{sec:redshift}, which allows us to control the nondegenerate energy on the horizon.

Let $\Sigma$ be a (spherically symmetric) spacelike hypersurface in \M, $\sprime{\Sigma}=\Sigma\cap\{r\leq R\}$ and $\mathcal{N}$ the outgoing null hypersurface emerging from $\partial\sprime{\Sigma}$; moreover let \[\St{\tau}=\varphi_\tau\Bigl((\sprime{\Sigma}\cup\mathcal{N})\cap\mathcal{D}\Bigr)\] and
\begin{equation*}
  \Stp{\tau}=\St{\tau}\cap\{r\leq R\}\,,\qquad
  \Stt{\taup}{\tau}=\St{\taup}\cap\mathrm{J}^-(\Stp{\tau})\,.
\end{equation*}
\begin{figure}[tb]
  \begin{center}
    \input{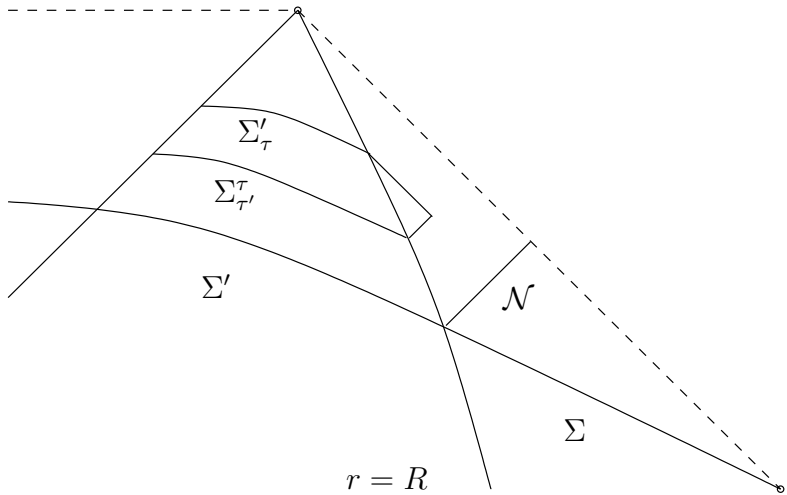}
    \caption{The construction of the surfaces $\Stp{\tau}$ from $\Sigma$.}
    \label{fig:Sigmaprime}
  \end{center}
\end{figure}

\begin{prop}[Uniform Boundedness]
  \label{prop:uniformboundedness}
  Let $\phi$ be a solution of the wave equation \eqref{eq:wave} with initial data on $\St{0}$, then there exists a constant $C(\St{0})$ such that
  \begin{equation}
    \label{eq:uniformboundedness}
    \int_\Stp{\tau}\ned{N}{\phi}\leq C\int_\Stt{0}{\tau}\ned{N}{\phi}\qquad(\tau>0)\,.
  \end{equation}
\end{prop}
\noindent\emph{Proof.}
One can proceed in analogy to the \emph{local observer's energy estimate} of \cite{DRC}; indeed, from the energy identity for $N$ on the domain $\mathcal{R}(\taup,\tau)=\cup_{\taup\leq\taub\leq\tau}\Stt{\taub}{\tau}$ it follows
\begin{equation}
  \label{eq:ub:eiN}
  \int_\Stp{\tau}\sned+\int_{\mathcal{R}(\taup,\tau)}K^N\leq\int_\Stt{\taup}{\tau}\sned
\end{equation}
since $(J^N,n_\mathcal{H})\geq 0$, and $(J^N,n_\mathcal{N})\geq 0$.
By Prop.~\ref{prop:redshift}, namely the redshift effect, $K^N$ is bounded from below by $\sned$ near the horizon, and from above by $(J^T,n)$ away from the horizon; since also the lapse of the foliation of $\mathcal{R}$ is bounded from above and below we conclude that there are constants $0<b<B$ only depending on $\Sigma$ and $N$ such that
\begin{multline}
  \int_\Stp{\tau}\sned+b\int_\taup^\tau\ud\taub\int_\Stt{\taub}{\tau}\sned\leq B\int_\taup^\tau\ud\taub\int_\Stt{\taub}{\tau}\sted+\int_\Stt{\taup}{\tau}\sned\leq\\
  \leq B(\tau-\taup)\int_\Stt{\taup}{\tau}\sted+\int_\Stt{\taup}{\tau}\sned
\end{multline}
where in the last step we have used the energy identity for $T$ on $\mathcal{R}(\taup,\taub)$ and $K^T=0$. Thus the desired energy bound follows from the elementary Lemma \ref{lemma:ub}.
\qed
\begin{lemma} \label{lemma:ub}
  Let $f:\mathbb{R}\to\mathbb{R}$ be a nonnegative function, $f\geq 0$, such that for all $t_1\leq t_2$ and two positive constants $0<c<C$
  \begin{equation*}
    f(t_2)+c\int_{t_1}^{t_2} f(t)\ud t\leq  C(t_2-t_1)+f(t_1)\,,
  \end{equation*}
  then
  \begin{equation*}
    f(t_2)\leq  f(t_1)+\frac{C}{c}\qquad(t_2\geq t_1)\,.
  \end{equation*}
\end{lemma}
\begin{proof} See e.g.~\cite{VST}.\end{proof}

\subsection{Energy Decay}
\label{sec:ed}

In this Section we prove quadratic decay of the nondegenerate energy.

Let
\begin{equation}
  \label{def:St:ed}
  \St{\tau_0}\doteq\partial^-\mathcal{R}_{r_0,R}^\infty(t_0)\qquad\tau_0=\frac{1}{2}(t_0-R^\ast)
\end{equation}
with $R>\sqrt[n-2]{8nm}$, $t_0>0$ and $r_0\doteq r_0^{(N)}$ according to Prop.~\ref{prop:localredshift}.

\begin{prop}[Energy decay]\label{prop:ed}
Let $\phi$ be a solution of the wave equation \eqref{eq:wave} with initial data on $\Sigma_{\tau_0}$ satisfying
\begin{equation}
  \label{eq:eda}
  D\doteq\int_{\tau_0+R^\ast}^\infty\int_\Sn\dm{\gn}\sum_{k=0}^1 r^2\sqb{\ddvs{\rn\partial_t^k\phi}}+\int_\St{\tau_0}\nedCT{2}<\infty\,,
\end{equation}
then there exists a constant $C(n,m)$ such that
\begin{equation}
  \label{eq:ed}
  \int_\St{\tau}\ned{N}{\phi}\leq\frac{C\,D}{\tau^2}\qquad(\tau>\tau_0)\,.
\end{equation}
\end{prop}

The proof is based on a \emph{weighted energy inequality}, derived from the energy identity for the current \eqref{eq:jr} on the domain
\begin{equation}
  \label{eq:dwein}
  {}^R\mathcal{D}_{\tau_1}^{\tau_2}
  =\bigl\{(\us,\vs):\tau_1\leq\us\leq\tau_2,\vs-\us\geq  R^\ast\bigr\}\,.
\end{equation}
\paragraph{Weighted energy identity.} Consider the current
\begin{equation}
\label{eq:jr}
\Jr_\mu(\phi)=T_{\mu\nu}(\psi)V^\nu
\end{equation}
where
\begin{gather}
\psi=r^\frac{n-1}{2}\phi\\
V=r^q\dvs\,,\quad q=p+1-n\,,\quad p\in\{1,2\}\,.
\end{gather}
This may also be viewed as the current to the multiplier vectorfield $r^p\dvs$, modified by the following terms:
\begin{multline*}
\Jr_\mu(\phi)=T_{\mu\nu}(\phi)\,r^p\bigl(\dvs\bigr)^\nu
+\bigl(\frac{n-1}{2}\bigr)^2\,r^{p-2}\bigl(\cf\bigr)(\partial_\mu r)\phi^2\\
+\frac{1}{2}\frac{n-1}{2}r^{p-1}(\partial_\mu r)\ddvs{\phi^2}
+\frac{1}{2}\frac{n-1}{2}r^{p-1}\bigl(\cf\bigr)\bigl(\partial_\mu\,\phi^2\bigr)\\
-\frac{1}{2}\bigl(\frac{n-1}{2}\bigr)^2\,r^{p-2}\bigl(\cf\bigr)\bigl(\dvs\bigr)_\mu\,\phi^2
-\frac{1}{2}\frac{n-1}{2}\bigl(\dvs\bigr)_\mu\,r^{p-1}\ddrs{\phi^2}
\end{multline*}
If $\Box_g\phi=0$ then we calculate
\begin{equation}
\label{eq:boxpsi}
\begin{split}
\Box_g\psi &= -\bigl(\cf\bigr)^{-1}\partial_{\us}\partial_{\vs}\psi+\frac{n-1}{r}\ddrs{\psi}+\frac{1}{r^2}\laplacessph\psi\\
&= \frac{n-1}{2}\bigl(\frac{n-3}{2}+\frac{n-1}{2}\frac{2m}{r^{n-2}}\bigr)\frac{1}{r^2}\psi+\frac{n-1}{r}\drs\psi\,.
\end{split}
\end{equation}
So the wave equation for $\phi$ \[\Box_g\phi=0\] is equivalent to the following equation for $\psi$:
\begin{multline}
\label{eq:wavepsi}
-\partial_{\us}\partial_{\vs}\psi+\bigl(\cf\bigr)\frac{1}{r^2}\laplacessph\psi\\
-\frac{n-1}{2}\bigl(\frac{n-3}{2}+\frac{n-1}{2}\frac{2m}{r^{n-2}}\bigr)\frac{1}{r^2}\bigl(\cf\bigr)\psi=0\,.
\end{multline}
Now,
\begin{equation}
\Kr(\phi)=\nabla^\mu \Jr_\mu(\phi)=\Box_g(\psi)\,V\cdot\psi+K^V(\psi)\,,
\end{equation}
where \[K^V(\psi)=\deformt{V}^{\mu\nu}\,T_{\mu\nu}(\psi)\,.\]
Since
\begin{gather}
\deformt{V}_{\us\us}=2\,q\,r^{q-1}\bigl(\cf\bigr)^2\notag\\
\deformt{V}_{\vs\vs}=0\notag\\
\deformt{V}_{\us\vs}=-\bigl(\cf\bigr)\,r^{q-1}\Bigl[q+(n-q-2)\frac{2m}{r^{n-2}}\Bigr]\\
\deformt{V}_{aA}=0\notag\\
\deformt{V}_{AB}=r^{q-1}\bigl(\cf\bigr)\,g_{AB}\notag
\end{gather}
we find
\begin{multline}\label{KVr}
\Kr\cdot r^{n-1}=\frac{n-1}{4}\Bigl(\frac{n-3}{2}+\frac{n-1}{2}\frac{2m}{r^{n-2}}\Bigr)\frac{r^p}{r^2}\ddvs{\psi^2}+\frac{p}{2}r^{p-1}\bigl(\ddvs{\psi}\bigr)^2\\
+\frac{1}{2}r^{p-1}\Bigl[(2-p)+(p-n)\frac{2m}{r^{n-2}}\Bigr]\normsph{\nablab\psi}^2\,.
\end{multline}
One may integrate the first term by parts to obtain:
\begin{multline}
\label{eq:wein:intparts}
\int_{\us+R^\ast}^\infty\ud\vs\,\Kr\cdot r^{n-1}=\frac{n-1}{4}\Bigl(\frac{n-3}{2}+\frac{n-1}{2}\frac{2m}{r^{n-2}}\Bigr)\frac{r^p}{r^2}\psi^2\Bigr\vert_{\us+R^\ast}^\infty\displaybreak[0]\\
+\int_{\us+R^\ast}^\infty\ud\vs\biggl\{\Bigl[\frac{n-1}{4}(2-p)\frac{n-3}{2}+\frac{n-1}{2}(n-p)\frac{2m}{r^{n-2}}\Bigr]\frac{r^p}{r^3}\bigl(\cf\bigr)\psi^2\\
+\frac{p}{2}r^{p-1}\bigl(\ddvs{\psi}\bigr)^2
+\frac{1}{2}r^{p-1}\Bigl[(2-p)+(p-n)\frac{2m}{r^{n-2}}\Bigr]\normsph{\nablab\psi}^2\biggr\}
\end{multline}
We can now write down the energy identity for the current $\Jr$ (see also Appendix \ref{sec:integration}):
\begin{equation*}
\int_{{}^R\mathcal{D}_{\tau_1}^{\tau_2}}\Kr\dm{g}=\int_{\partial{}^R\mathcal{D}_{\tau_1}^{\tau_2}}{}^\ast \Jr
\end{equation*}
Dropping the positive zeroth order terms, we obtain:
\begin{multline}
\label{eq:wein}
\int_{\tau_2+R^\ast}^\infty\ud\vs\int_\Sn\dm{\gn}\,\frac{1}{2}r^p\bigl(\ddvs{\psi}\bigr)^2\vert_{\us=\tau_2}\\
+\int_{\tau_1}^{\tau_2}\ud\us\int_{\us+R^\ast}^\infty\ud\vs\int_{\Sn}\dm{\gn}\times\\\times\biggl\{\frac{p}{2}r^{p-1}\bigl(\ddvs{\psi}\bigr)^2+\frac{1}{2}r^{p-1}\Bigl[(2-p)+(p-n)\frac{2m}{r^{n-2}}\Bigr]\normsph{\nablab\psi}^2\biggr\}\\
+\int_{\tau_1}^{\tau_2}\ud\us\int_\Sn\dm{\gn}\frac{1}{2}r^p\normsph{\nablab\psi}^2\vert_{\vs\to\infty}\leq\displaybreak[0]\\
\leq\bigl(1-\frac{2m}{R^{n-2}}\bigr)^{-1}\Biggl\{
\int_{\tau_1+R^\ast}^\infty\ud\vs\int_\Sn\dm{\gn}\frac{1}{2}r^p\bigl(\ddvs{\psi}\bigr)^2\vert_{\us=\tau_1}\\
+\int_{2\tau_1+R^\ast}^{2\tau_2+R^\ast}\ud t\int_\Sn\dm{\gn}\Bigl[\frac{1}{4}r^p\bigl(\ddvs{\psi}\bigr)^2+\frac{1}{4}r^p\normsph{\nablab\psi}^2\\+\frac{n-1}{4}\frac{1}{2}r^p\bigl(\frac{n-3}{2}+\frac{n-1}{2}\frac{2m}{R^{n-2}}\bigr)\frac{1}{r^2}\psi^2\Bigr]\vert_{r=R}
\Biggr\}
\end{multline}
Note that the powers of $r$ that appear in the bulk term are $1$ less than those that appear in the boundary terms. This allows for a hierarchy of inequalities \eqref{eq:wein} for different values of $p$, the so called \emph{$p$-hierarchy}.

\ \par\noindent\emph{Proof of Prop.~\ref{prop:ed}:}
In a first step the decay of the solutions at future null infinity will be deduced from the weighted energy inequality, and in a second step the continuation to the event horizon will be inferred from the redshift effect.

\noindent\textbf{Step 1.}
The $p$-hierarchy consists of two steps which exploits \eqref{eq:wein} first with $p=2$, then with $p=1$; but in a zeroth step we need to obtain control on the angular derivatives from \eqref{eq:wein} with $p=1$:

Since \[1-(n-1)\frac{2m}{r^{n-2}}>\frac{1}{2}\qquad(r>R)\] we have from the weighted energy inequality for $p=1$ on the domain ${}^{r_0^\prime}\mathcal{D}_{\tau^\prime}^{\tau}$ for $\tau>\tau^\prime\geq\tau_0\doteq\frac{1}{2}(t_0-R^\ast)$,
\begin{multline}\label{eq:edph:zero}
  \int_{\tau^\prime}^{\tau}\ud\us\int_{\us+{r_0^\prime}^\ast}^\infty\ud\vs\int_\Sn\dm{\gn}\frac{1}{4}\normsph{\nablab\psi}^2\leq\displaybreak[1]\\
  \\\leq\bigl(1-\frac{2m}{R^{n-2}}\bigr)^{-1}\int_{\tau^\prime+{r_0^\prime}^\ast}^\infty\ud\vs\int_\Sn\dm{\gn}\frac{1}{2}r\bigl(\ddvs{\psi}\bigr)^2\Bigr\vert_{\us=\tau^\prime}\displaybreak[0]\\
  +C(n,m)\bigl(1-\frac{2m}{R^{n-2}}\bigr)^{-1}\int_\St{\tau_0}\Bigl(J^T(\phi)+J^T(T\cdot\phi),n\Bigr)\,;
\end{multline}
here, we have estimated the boundary integrals as follows:

Choose $r_0^\prime\in(R^\ast,R^\ast+1)$ such that
\begin{multline*}
\int_{R^\ast}^{R^\ast+1}\!\!\!\!\ud\rs\int_{t_0+(\rs-R^\ast)}^\infty\!\!\!\!\!\ud t\int_\Sn\dm{\gn}\,\bigl(\cf\bigr)r^{n-1}\times\\\times\Bigl\{\frac{1}{r^n}\bigl(\ddrs{\phi}\bigr)^2+\frac{1}{r^{n+1}}\bigl(\ddt{\phi}\bigr)^2+\frac{1}{r^3}\bigl(\cf\bigr)\normsph{\nablab\phi}^2\Bigr\}\\
=\int_{t_0+({r_0^\prime}^\ast-R^\ast)}^\infty\!\!\!\!\ud t\int_\Sn\dm{\gn}\bigl(1-\frac{2m}{{r_0^\prime}^{n-2}}\bigr){r_0^\prime}^{n-1}\times\\\times\Bigl\{\frac{1}{{r_0^\prime}^n}\bigl(\ddrs{\phi}\bigr)^2+\frac{1}{{r_0^\prime}^{n+1}}\bigl(\ddt{\phi}\bigr)^2+\frac{1}{{r_0^\prime}^3}\bigl(1-\frac{2m}{{r_0^\prime}^{n-2}}\bigr)\normsph{\nablab\phi}^2\Bigr\}
\end{multline*}
then
\begin{multline*}
\int_{2\tau^\prime+{r_0^\prime}^\ast}^{2\tau+{r_0^\prime}^\ast}\ud t\int_\Sn\dm{\gn}\biggl[\frac{1}{4}r^p\bigl(\ddvs{\psi}\bigr)^2+\frac{1}{4}r^p\normsph{\nablab\psi}^2\\
+\frac{n-1}{4}\frac{1}{2}r^p\Bigl(\frac{n-3}{2}+\frac{n-1}{2}\frac{2m}{r^{n-2}}\Bigr)\frac{1}{r^2}\psi^2\biggr]\vert_{r={r_0^\prime}}\leq\displaybreak[1]\\
\leq\int_{2\tau^\prime+{r_0^\prime}^\ast}^{2\tau+{r_0^\prime}^\ast}\ud t\int_\Sn\dm{\gn}\,{r_0^\prime}^{p-2}\biggl[\frac{1}{2}\bigl(\frac{n-1}{2}\bigr)^2\phi^2+\frac{1}{2}{r_0^\prime}^2\bigl(\ddvs{\phi}\bigr)^2\\+\frac{1}{4}{r_0^\prime}^2\normsph{\nablab\phi}^2+\frac{n-1}{4}\frac{1}{2}\Bigl(\frac{n-3}{2}+\frac{n-1}{2}\frac{2m}{R^{n-2}}\Bigr)\phi^2\biggr]\Bigr\vert_{r={r_0^\prime}}\,r^{n-1}\leq\displaybreak[0]\\
\leq C(n,m)\int_\St{\tau_0}\Bigl(J^T(\phi)+J^T(T\cdot\phi),n\Bigr)
\end{multline*}
because by \eqref{boundaryintegralineq}
\begin{multline*}
\int_{{r_0^\prime}^\ast+2\tau^\prime}^{{r_0^\prime}^\ast+2\tau}\ud t\int_\Sn\dm{\gn}\,r^{n-1}\biggl[\frac{1}{4}\bigl(\ddvs{\phi}\bigr)^2+\frac{n-1}{(4{r_0^\prime})^2}\bigl(\cf\bigr)\phi^2\biggr]\Bigr\vert_{r={r_0^\prime}}\leq\displaybreak[1]\\
\leq\int_{{r_0^\prime}^\ast+2\tau^\prime}^{{r_0^\prime}^\ast+2\tau}\ud t\int_\Sn\dm{\gn}\,r^{n-1}\times\\\times\biggl[\frac{1}{2}\bigl(\cf\bigr)\normsph{\nablab\phi}^2+\frac{n-1}{2}\bigl(\cf\bigr)\bigl(\ddrs{\phi}\bigr)^2\biggr]\Bigr\vert_{r={r_0^\prime}}\displaybreak[0]\\+C(n,m)\int_\St{\tau_0}\Bigl(J^T(\phi),n\Bigr)
\end{multline*}
and by Prop.~\ref{prop:ILED} (and the choice of ${r_0^\prime}$):
\begin{multline*}
\int_{t_0+({r_0^\prime}^\ast-R^\ast)}^\infty\ud t\int_\Sn\dm{\gn}\,r^{n-1}\times\\\times\biggl[\frac{1}{{r_0^\prime}^n}\bigl(\ddrs{\phi}\bigr)^2+\frac{1}{{r_0^\prime}^3}\bigl(1-\frac{2m}{{r_0^\prime}^{n-2}}\bigr)\normsph{\nablab\phi}^2\biggr]\Bigr\vert_{r=r_0^\prime}\leq\displaybreak[0]\\
\leq C(n,m)\int_\St{\tau_0}\Bigl(J^T(\phi)+J^T(T\cdot\phi),n\Bigr)\,.
\end{multline*}
Note that for the use of \eqref{boundaryintegralineq} that with our choice of $R$
\begin{equation*}
  (n-3)+n\frac{2m}{r^{n-2}}-(n-1)\bigl(\frac{2m}{r^{n-2}}\bigr)^2>0 \qquad (r>R)\,.
\end{equation*}

\noindent\framebox[1.2\width]{$p=2$:} For $p=2$ \eqref{eq:wein} reads
\begin{multline*}
  \int_{\tau^\prime}^{\tau}\ud\us\int_{\us+{r_0^\prime}^\ast}^\infty\ud\vs\int_{\Sn}\dm{\gn}\biggl[r\bigl(\ddvs{\psi}\bigr)^2-\frac{1}{2}r(n-2)\frac{2m}{r^{n-2}}\normsph{\nablab\psi}^2\biggr]\leq\displaybreak[1]\\
  \leq\bigl(1-\frac{2m}{R^{n-2}}\bigr)^{-1}\int_{\tau^\prime+{r_0^\prime}^\ast}^\infty\ud\vs\int_\Sn\dm{\gn}\frac{1}{2}r^2\bigl(\ddvs{\psi}\bigr)^2\vert_{\us=\tau^\prime}\displaybreak[0]\\
  +C(n,m)\bigl(1-\frac{2m}{R^{n-2}}\bigr)^{-1}\int_\St{\tau_0}\Bigl(J^T(\phi)+J^T(T\cdot\phi),n\Bigr)\,.
\end{multline*}
Thus, with the previous estimate \eqref{eq:edph:zero},
\begin{multline}\label{eq:edph:two}
  \int_{\tau^\prime}^{\tau}\ud\us\int_{\us+{r_0^\prime}^\ast}^\infty\ud\vs\int_\Sn\dm{\gn}r\bigl(\ddvs{\psi}\bigr)^2\leq\displaybreak[0]\\
  \leq
  C(n,m)\bigl(1-\frac{2m}{R^{n-2}}\bigr)^{-1}\Biggl\{\int_{\tau^\prime+{r_0^\prime}^\ast}^\infty\ud\vs\int_\Sn\dm{\gn}\frac{1}{2}r^2\bigl(\ddvs{\psi}\bigr)^2\vert_{\us=\tau^\prime}\\+\int_\St{\tau_0}\Bigl(J^T(\phi)+J^T(T\cdot\phi),n\Bigr)\Biggr\}\,.
\end{multline}
Let us define \[\tau_{j+1}=2\tau_j\quad(j\in\mathbb{N}_0)\qquad\tau_0=\frac{1}{2}(t_0-R^\ast)\,,\] then there is a sequence $(\tau_j^\prime)_{j\in\mathbb{N}_0}$ with $\tau_j^\prime\in(\tau_j,\tau_{j+1})\ (j\in\mathbb{N}_0)$ such that
\begin{multline*}
\int_{\tau_j^\prime+{r_0^\prime}^\ast}^\infty\ud\vs\int_\Sn\dm{\gn}\,r\bigl(\ddvs{\psi}\bigr)^2\vert_{\us=\tau_j^\prime}\leq\displaybreak[0]\\
\leq\frac{1}{\tau_j}C(n,m)\Biggl[\int_{\tau_j+{r_0^\prime}^\ast}^\infty\ud\vs\int_\Sn\dm{\gn}\,r^2\bigl(\ddvs{\psi}\bigr)^2\vert_{\us=\tau_j}\\+\int_\St{\tau_0}\Bigl(J^T(\phi)+J^T(T\cdot\phi),n\Bigr)\Biggr]
\end{multline*}
and again by \eqref{eq:wein}
\begin{multline*}
  \int_{\tau_j+{r_0^\prime}^\ast}^\infty\ud\vs\int_\Sn\dm{\gn}\frac{1}{2}r^2\bigl(\ddvs{\psi}\bigr)^2\vert_{\us=\tau_j}\leq\displaybreak[1]\\
  \leq C(n,m)\Biggl[\int_{\tau_0+{r_0^\prime}^\ast}^\infty\ud\vs\int_\Sn\dm{\gn}\,r^2\bigl(\ddvs{\psi}\bigr)^2\vert_{\us=\tau_0}\displaybreak[0]\\
  +\int_\St{\tau_0}\Bigl(J^T(\phi)+J^T(T\cdot\phi),n\Bigr)\Biggr]\,.
\end{multline*}
Since $\frac{1}{\tau_j}\leq\frac{1}{\tau_j^\prime}\frac{\tau_{j+1}}{\tau_j}=\frac{2}{\tau_j^\prime}$ we have
\begin{multline}\label{eq:edph:two:f}
\int_{\tau_j^\prime+{r_0^\prime}^\ast}^\infty\ud\vs\int_\Sn\dm{\gn}\,r\bigl(\ddvs{\psi}\bigr)^2\vert_{\us=\tau_j^\prime}\leq\displaybreak[1]\\
\leq\frac{C(n,m)}{\tau_j^\prime}\Biggl[\int_{\tau_0+{r_0^\prime}^\ast}^\infty\ud\vs\int_\Sn\dm{\gn}\,r^2\bigl(\ddvs{\psi}\bigr)^2\vert_{\us=\tau_0}\displaybreak[0]\\
+\int_\St{\tau_0}\Bigl(J^T(\phi)+J^T(T\cdot\phi),n\Bigr)\Biggr]\,.
\end{multline}

\noindent\framebox[1.2\width]{$p=1$:} In order to deal with the timelike boundary integrals analogously to the above choose \[{r_j^{\prime\prime}}^\ast\in({r_0^\prime}^\ast,{r_0^\prime}^\ast+1)\] such that
\begin{multline*}
\int_{2\tau_j^\prime+{r_j^{\prime\prime}}^\ast}^{2\tau_{j+1}^\prime+{r_j^{\prime\prime}}^\ast}\ud t\int_\Sn\dm{\gn}\,r^{n-1}\times\\\times\Biggl[\frac{1}{r^n}\bigl(\ddrs{\phi}\bigr)^2+\frac{1}{r^{n+1}}\bigl(\ddt{\phi}\bigr)^2+\frac{1}{r^3}\bigl(\cf\bigr)\normsph{\nablab\phi}^2\Biggr]\vert_{r=r_j^{\prime\prime}}\leq\displaybreak[0]\\
\leq C(n,m)\int_{\Sigma_{\tau_j^\prime}}\Bigl(J^T(\phi)+J^T(T\cdot\phi),n\Bigr)\,.
\end{multline*}
Then, proceeding as before,
\begin{multline}\label{eq:edph:one:b}
\int_{2\tau_j^\prime+{r_j^{\prime\prime}}^\ast}^{2\tau_{j+1}^\prime+{r_j^{\prime\prime}}^\ast}\ud t\int_\Sn\dm{\gn}\times\\\times\biggl[\frac{1}{4}r\bigl(\ddvs{\psi}\bigr)^2+\frac{1}{4}r\normsph{\nablab\psi}^2+\frac{n-1}{4}\frac{1}{2}r\Bigl(\frac{n-3}{2}+\frac{n-1}{2}\frac{2m}{r^{n-2}}\Bigr)\frac{1}{r^2}\psi^2\biggr]\vert_{r=r_j^{\prime\prime}}\leq\displaybreak[0]\\
\leq C(n,m)\int_{\Sigma_{\tau_j^\prime}}\Bigl(J^T(\phi)+J^T(T\cdot\phi),n\Bigr)\,.
\end{multline}
Now apply \eqref{eq:wein} to the region ${}^{r_j^{\prime\prime}}\mathcal{D}_{\tau_j^\prime}^{\tau_{j+1}^\prime}$ to obtain:
\begin{multline*}
\int_{\tau_j^\prime}^{\tau_{j+1}^\prime}\ud\us\int_{\us+{r_j^{\prime\prime}}^\ast}^\infty\ud\vs\int_\Sn\dm{\gn}\times\\\times\biggl[\frac{1}{2}\bigl(\ddvs{\psi}\bigr)^2+\frac{1}{4}\normsph{\nablab\psi}^2\biggr]\leq\displaybreak[1]\\
\leq\bigl(1-\frac{2m}{R^{n-2}}\bigr)^{-1}\int_{\tau_j^\prime+{r_j^{\prime\prime}}^\ast}^\infty\ud\vs\int_\Sn\dm{\gn}\frac{1}{2}r\bigl(\ddvs{\psi}\bigr)^2\vert_{\us=\tau_j^\prime}\displaybreak[0]\\
+C(n,m)\bigl(1-\frac{2m}{R^{n-2}}\bigr)^{-1}\int_{\Sigma_{\tau_j^\prime}}\Bigl(J^T(\phi)+J^T(T\cdot\phi),n\Bigr)
\end{multline*}
By virtue of the result \eqref{eq:edph:two:f} from the case $p=2$, this yields
\begin{multline}\label{eq:edph:one}
\int_{\tau_j^\prime}^{\tau_{j+1}^\prime}\ud\us\int_{\us+{r_j^{\prime\prime}}^\ast}^\infty\ud\vs\int_\Sn\dm{\gn}\biggl[\frac{1}{2}\bigl(\ddvs{\psi}\bigr)^2+\frac{1}{4}\normsph{\nablab\psi}^2\biggr]\leq\displaybreak[1]\\
\leq\frac{C(n,m)}{\tau_j^\prime}\Biggl[\int_{\tau_0+{r_0^\prime}^\ast}^\infty\ud\vs\int_\Sn\dm{\gn}\,r^2\bigl(\ddvs{\psi}\bigr)^2\vert_{\us=\tau_0}
+\int_\St{\tau_0}\Bigl(J^T(\phi)+J^T(T\cdot\phi),n\Bigr)\Biggr]\displaybreak[0]\\
+C(n,m)\int_{\Sigma_{\tau_j^\prime}}\Bigl(J^T(\phi)+J^T(T\cdot\phi),n\Bigr)\,.
\end{multline}

\noindent\textbf{Step 2.}
Our aim is to prove decay for the \emph{non-degenerate} energy. Let us first find an estimate for
\begin{multline*}
\int_{\tau_j^\prime}^{\tau_{j+1}^\prime}\ud\tau\int_{\Sigma_\tau}\Bigl(J^N(\phi),n\Bigr)=\\
=\int_{\tau_j^\prime}^{\tau_{j+1}^\prime}\ud\tau\int_{\Sigma_\tau\cap\{r\leq r_j^{\prime\prime}\}}\Bigl(J^N(\phi),n\Bigr)
+\int_{\tau_j^\prime}^{\tau_{j+1}^\prime}\ud\tau\int_{\Sigma_\tau\cap\{r\geq r_j^{\prime\prime}\}}\Bigl(J^T(\phi),n\Bigr)\,.
\end{multline*}
The estimate of the first term is exactly the content of Cor.~\ref{cor:nILED}, and for the second term
\begin{multline*}
\int_{\tau_j^\prime}^{\tau_{j+1}^\prime}\ud\tau\int_{\Sigma_\tau^\prime\cap\{r\geq r_j^{\prime\prime}\}}\Bigl(J^T(\phi),n\Bigr)=\displaybreak[0]\\
=\int_{\tau_j^\prime}^{\tau_{j+1}^\prime}\!\!\!\ud\us\int_{\us+{r_j^{\prime\prime}}^\ast}^\infty\!\!\!\!\!\ud\vs\int_\Sn\dm{\gn}\,r^{n-1}\biggl[\frac{1}{2}\bigl(\ddvs{\phi}\bigr)^2+\frac{1}{2}\bigl(\cf\bigr)\normsph{\nablab\phi}^2\biggr]
\end{multline*}
we can use \eqref{eq:edph:one} once we have turned it into an estimate for the derivatives of $\phi$.
Note that
\begin{multline*}
\int_{\us+{r_0^\prime}^\ast}^\infty\ud\vs\,\bigl(\ddvs{\psi}\bigr)^2=\displaybreak[0]\\
=\int_{\us+{r_0^\prime}^\ast}^\infty\ud\vs\biggl[\frac{n-1}{2}\bigl(\cf\bigr)r^\frac{n-3}{2}\dvs\Bigl(r^\frac{n-1}{2}\phi^2\Bigr)+r^{n-1}\bigl(\ddvs{\phi}\bigr)^2\biggr]\displaybreak[1]\\
=-\frac{n-1}{2}\bigl(\cf\bigr)\,r^{n-2}\,\phi^2\vert_{\vs=\us+{r_0^\prime}^\ast}\\
+\int_{\us+{r_0^\prime}^\ast}^\infty\ud\vs\Biggl\{-\frac{n-1}{2}\bigl(\cf\bigr)\Bigl[(n-2)\frac{2m}{r^n}+\frac{n-3}{2}\bigl(\cf\bigr)\frac{1}{r^2}\Bigr]\phi^2+\bigl(\ddvs{\phi}\bigr)^2\biggr\}\,r^{n-1}
\end{multline*}
and by Lemma \ref{lemma:hardy}
\begin{multline*}
\int_{\us+{r_0^\prime}^\ast}^\infty\ud\vs\frac{1}{r^2}\phi^2\,r^{n-1}\leq\displaybreak[0]\\
\leq C(n,m)\int_{\us+{r_0^\prime}^\ast}^\infty\ud\vs\,\bigl(\ddvs{\psi}\bigr)^2+C(n,m)\,r^{n-1}\phi^2\vert_{(\us,\vs=\us+{r_0^\prime}^\ast)}\,.
\end{multline*}
Thus
\begin{equation*}
\int_{\us+{r_0^\prime}^\ast}^\infty\ud\vs\Bigl(\ddvs{\phi}\Bigr)^2\,r^{n-1}
\leq C(n,m)\biggl[\phi^2\vert_{(\us,\us+{r_0^\prime}^\ast)}+\int_{\us+{r_0^\prime}^\ast}^\infty\ud\vs\bigl(\ddvs{\psi}\bigr)^2\biggr]\,,
\end{equation*}
and finally in view of \eqref{eq:edph:one:b}
\begin{multline}
\int_{\tau_j^\prime}^{\tau_{j+1}^\prime}\ud\us\int_{\us+{r_j^{\prime\prime}}^\ast}^\infty\ud\vs\int_\Sn\dm{\gn}\bigl(\ddvs{\phi}\bigr)^2\,r^{n-1}\leq\displaybreak[1]\\
\leq C(n,m)\int_{\tau_j^\prime}^{\tau_{j+1}^\prime}\ud\us\int_{\us+{r_j^{\prime\prime}}^\ast}^\infty\ud\vs\int_\Sn\dm{\gn}\bigl(\ddvs{\psi}\bigr)^2\displaybreak[0]\\
+C(n,m)\int_{\Sigma_{\tau_j^\prime}}\Bigl(J^T(\phi)+J^T(T\cdot\phi),n\Bigr)\,.
\end{multline}
Therefore, putting the estimates for the two terms back together,
\begin{multline}\label{eq:edph:int}
\int_{\tau_j^\prime}^{\tau_{j+1}^\prime}\ud\tau\int_{\Sigma_\tau}\Bigl(J^N(\phi),n\Bigr)\leq\displaybreak[1]\\
\leq C(n,m)\int_{\tau_j^\prime}^{\tau_{j+1}^\prime}\ud\us\int_{\us+{r_j^{\prime\prime}}^\ast}^\infty\ud\vs\int_\Sn\dm{\gn}\biggl\{\bigl(\ddvs{\psi}\bigr)^2+\normsph{\nablab\psi}^2\biggr\}\displaybreak[0]\\+C(n,m)\int_{\Sigma_{\tau_j^\prime}}\Bigl(J^N(\phi)+J^T(T\cdot\phi),n\Bigr)\leq\displaybreak[1]\\
\leq\frac{C(n,m)}{\tau_j^\prime}\Biggl[\int_{\tau_0+{r_0^\prime}^\ast}^\infty\ud\vs\int_\Sn\dm{\gn}\,r^2\bigl(\ddvs{\psi}\bigr)^2\vert_{\us=\tau_0}+\int_\St{\tau_0}\Bigl(J^T(\phi)+J^T(T\cdot\phi),n\Bigr)\Biggr]\displaybreak[0]\\
+C(n,m)\int_{\Sigma_{\tau_j^\prime}}\Bigl(J^N(\phi)+J^T(T\cdot\phi),n\Bigr)
\end{multline}
where we have now used \eqref{eq:edph:one}.
The same inequality holds for $\tau_{j+2}^\prime$ in place of $\tau_{j+1}^\prime$, by adding the inequalities corresponding to the intervals $[\tau_j^\prime,\tau_{j+1}^\prime]$ and $[\tau_{j+1}^\prime,\tau_{j+2}^\prime]$ and using Prop.~\ref{prop:uniformboundedness} for the last term. So there is a sequence \[(\dprime{\tau}_j)_{j\in\mathbb{N}}\qquad \dprime{\tau}_j\in\bigl(\tau_j^\prime,\tau_{j+2}^\prime\bigr)\] such that
\[\int_{\tau_j^\prime}^{\tau_{j+2}^\prime}\ud\tau\int_{\Sigma_\tau}\Bigl(J^N(\phi),n\Bigr)\geq\tau_{j+1}\int_{\Sigma_{\dprime{\tau}_j}}\Bigl(J^N(\phi),n\Bigr)\]
and since $\frac{1}{\tau_{j+1}}\leq\frac{1}{\dprime{\tau}_j}\frac{\tau_{j+3}}{\tau_{j+1}}=\frac{4}{\dprime{\tau}_j}$ we have
\begin{equation}
\int_{\Sigma_{\dprime{\tau}_j}}\Bigl(J^N(\phi),n\Bigr)\leq\frac{4}{\dprime{\tau}_j}\int_{\tau_j}^{\tau_{j+2}^\prime}\ud\tau\int_{\Sigma_\tau^\prime}\Bigl(J^N(\phi),n\Bigr)\,.
\end{equation}
Now for any given $\tau>\tau_0$ we may choose \[j^\ast=\max\{j\in\mathbb{N}:\dprime{\tau}_j\leq\tau\}\] so that by \eqref{eq:uniformboundedness}
\[\int_{\Sigma_\tau}\Bigl(J^N(\phi),n\Bigr)\leq C\int_{\Sigma_{\dprime{\tau}_{j^\ast}}}\Bigl(J^N(\phi),n\Bigr)\]
with $\frac{\tau}{\dprime{\tau}_{j^\ast}}\leq\frac{\dprime{\tau}_{j^\ast+1}}{\dprime{\tau}_{j^\ast}}\leq 2^4$.
In particular we may estimate the last integral in \eqref{eq:edph:int}
\begin{equation*}
\int_{\Sigma_{\tau_j^\prime}}\Bigl(J^N(\phi)+J^T(T\cdot\phi),n\Bigr)
\leq\frac{C}{\sprime{\tau}_j}\int_{\tau_{j-1}^\prime}^{\tau_{j+1}^\prime}\ud\tau\int_{\Sigma_\tau}\Bigl(J^N(\phi)+J^N(T\cdot\phi),n\Bigr)
\end{equation*}
to see that in fact we have
\begin{multline}
\int_{\tau_j^\prime}^{\tau_{j+2}^\prime}\ud\tau\int_{\Sigma_\tau}\Bigl(J^N(\phi),n\Bigr)\leq\displaybreak[0]\\
\leq \frac{C(n,m)}{\tau_j^\prime}\Biggl[\int_{\tau_0+{r_0^\prime}^\ast}^\infty\ud\vs\int_\Sn\dm{\gn}\Bigl\{r^2\bigl(\frac{\partial r^\frac{n-1}{2}\phi}{\partial\vs}\bigr)^2+r^2\bigl(\frac{\partial r^\frac{n-1}{2}\ddt{\phi}}{\partial\vs}\bigr)\Bigr\}\\
+\int_\St{\tau_0}\Bigl(J^N(\phi)+J^N(T\cdot\phi)+J^T(T^2\phi),n\Bigr)\biggr]\,.
\end{multline}
Again with the sequence $(\dprime{\tau_j})_{j\in\mathbb{N}}$
\[\int_{\Sigma_{\dprime{\tau}_j}}\Bigl(J^N(\phi),n\Bigr)\leq\frac{1}{\tau_{j+1}}\int_{\tau_j^\prime}^{\tau_{j+2}^\prime}\ud\tau\int_{\Sigma_\tau}\Bigl(J^N(\phi),n\Bigr)\]
and since $\frac{1}{\tau_{j+1}}\frac{1}{\tau_j^\prime}\leq\frac{2^5}{{\dprime{\tau}_j}^2}$ we obtain by virtue of Prop.~\ref{prop:uniformboundedness} our final result:
\begin{multline}\label{EST}
\int_{\Sigma_\tau}\Bigl(J^N(\phi),n\Bigr)\leq\displaybreak[1]\\
\leq\frac{C(n,m)}{\tau^2}\Biggl[\int_{\tau_0+R^\ast}^\infty\ud\vs\int_\Sn\dm{\gn}\Bigl\{r^2\bigl(\ddvs{r^\frac{n-1}{2}\phi}\bigr)^2+r^2\bigl(\ddvs{r^\frac{n-1}{2}\ddt{\phi}}\bigr)^2\Bigr\}\displaybreak[0]\\
+\int_\St{\tau_0}\Bigl(J^N(\phi)+J^N(T\cdot\phi)+J^T(T^2\cdot\phi),n\Bigr)\Biggr]
\end{multline}
\qed

\subsection{Improved Interior Decay of the first order Energy}
\label{sec:iid}

In this Section we prove an energy estimate for the first order energy which improves the decay rate as compared to Prop.~\ref{prop:ed} in a bounded radial region.
\begin{remark}
The argument largely depends on the asymptotic properties of the spacetime, and is similar and slightly easier in Minkowski space, see \cite{VST}.
\end{remark}

\begin{prop}[Improved interior first order energy decay]
\label{prop:iid}
Let $0<\delta<\frac{1}{2}$, $R>\sqrt[n-2]{\frac{8nm}{\delta}}$, and let $\phi$ be a solution of the wave equation \eqref{eq:wave} with initial data on $\St{\tau_1}\ (\tau_1>0)$ satisfying
\begin{multline}
  D\doteq\intU{R}{\tau_1}{\sum_{k=0}^1r^{4-\delta}\sqb{\ddvs{(T^k\cdot\chi)}}\\+\sum_{k=0}^4r^2\sqb{\ddvs{(T^k\cdot\psi)}}+\sum_{k=0}^3\sumn r^2\sqb{\ddvs{T^k\Omega_i\psi}}}\\
  +\int_\St{\tau_1}\Bigl(\sum_{k=0}^5J^N(T^k\cdot\phi)+\sum_{k=0}^4\sumn J^N(T^k\Omega_i\phi),n\Bigr)<\infty\,.
\end{multline}
Then there exists a constant $C(n,m,\delta)$ such that
\begin{equation}
  \label{eq:iid}
  \int_\Stp{\tau}\ned{N}{\tphi}\leq\frac{C(n,m,\delta)}{\tau^{4-2\delta}}\,D\qquad(\tau>\tau_1)
\end{equation}
where $\Stp{\tau}=\St{\tau}\cap\{r\leq R\}$.
\end{prop}

In addition to the weighted energy identity arising from the multiplier $r^p\dvs$ that was used to prove Prop.~\ref{prop:ed} we will here also use a commutation with $\dvs$ to obtain the energy decay for $\ddt{\phi}$ of Prop.~\ref{prop:iid}.

\paragraph{Weighted energy and commutation.}
Consider the current
\begin{equation}
  \label{eq:s:vcurrent}
  \Jv_\mu(\phi)\doteq T_{\mu\nu}(\chi)V^\nu\,,
\end{equation}
where now
\begin{gather}
\label{eq:s:chi}
\chi=\partial_\vs\psi=\ddvs{(r^\frac{n-1}{2}\phi)}\qquad V=r^q\dvs\\ q=p-(n-1)\notag\qquad 2<p<4\qquad\delta=4-p\,.
\end{gather}
\begin{notation} To make the dependence on $p$ explicit, we denote by
  \begin{equation}
    \label{def:s:Kv}
    \Kv{\phi}{p}\doteq\nabla^\mu\Jv_\mu(\phi)\,.    
  \end{equation}
\end{notation}
The error terms for $\stackrel{\scriptscriptstyle{v}}{K}$ arise from the fact that $\chi$ is not a solution of \eqref{eq:wave}; here, similarly to \eqref{eq:boxpsi}, we find:
\begin{multline}
  \label{eq:boxchi}
  \Box_g\chi=-\frac{n-1}{2r^3}\Biggl\{(n-3)+n\frac{2m}{r^{n-2}}-(n-1)^2\sqb{\frac{2m}{r^{n-2}}}\Biggr\}\psi\\
  +\frac{n-1}{4r^2}\Bigl[(n-3)+(n-1)\frac{2m}{r^{n-2}}\Bigr]\chi+\frac{1}{r}\Bigl[2-n\frac{2m}{r^{n-2}}\Bigr]\laplacesphs\psi
  +\frac{n-1}{r}\ddrs{\chi}    
\end{multline}
Hence
\begin{align}
  \Kv{\phi}{p}=&\Box(\chi)\,V\cdot\chi+K^V(\chi)\notag\\
  =&\frac{1}{2}p\,r^{q-1}\sqb{\ddvs{\chi}}
  +\frac{1}{2}\Bigl[(2-p)-(n-p)\frac{2m}{r^{n-2}}\Bigr]\,r^{q-1}\sqv{\nablab\chi}\\
  &-\frac{n-1}{2}r^{q-3}\Biggl[(n-3)+n\frac{2m}{r^{n-2}}-(n-1)^2\sqb{\frac{2m}{r^{n-2}}}\Biggr]\psi\dddvs{\psi}\\
  &+\frac{n-1}{8}r^{q-2}\Bigl[(n-3)+(n-1)\frac{2m}{r^{n-2}}\Bigr]\ddvs{\chi^2}\\
  &+r^{q-1}\Bigl[2-n\frac{2m}{r^{n-2}}\Bigr]\bigl(\laplacesphs\psi\bigr)\Bigl(\ddvs{\chi}\Bigr)
\end{align}
which is not positive definite. However, we have
\begin{multline}
  \label{eq:s:Kvcontrols}
  \frac{1}{4}p\,r^{p-1}\sqb{\ddvs{\chi}}\leq\,\Kv{\phi}{p}\cdot r^{n-1}+\frac{1}{2}\Bigl[(p-2)+(n-p)\frac{2m}{r^{n-2}}\Bigr]r^{p-1}\sqv{\nablab\chi}\\
  +\frac{(n-1)^2(n-2)^2}{2}r^{(p-2)-1}\frac{1}{r^2}\psi^2+\frac{4}{p}r^{(p-2)-1}r^2\sqb{\laplacesphs\psi}\\
  -\frac{n-1}{8}r^{p-2}\Bigl[(n-3)+(n-1)\frac{2m}{r^{n-2}}\Bigr]\ddvs{\chi^2}\,,
\end{multline}
where we have used that
\begin{equation}
  n-2>n-3+n\frac{2m}{r^{n-2}}-(n-1)^2\sqb{\frac{2m}{r^{n-2}}}\geq n-3
\end{equation}
(is decreasing) on $r>\sqrt[n-2]{4nm}$.
The key insight here is that we are able to control all other terms on the right hand side of \eqref{eq:s:Kvcontrols} by the current $\Jr$ of Section \ref{sec:ed} with $p-2$ in the role of $p$, i.e.
\begin{multline}
  \label{eq:cwein:pre}
  \int_{\tau_1}^{\tau_2}\!\!\!\!\ud\us\int_{\us+R^\ast}^\infty\!\!\!\!\ud\vs\int_\Sn\dm{\gn}\,r^{p-1}\sqb{\ddvs{\chi}}\leq\\
  \leq C(n,m,\delta,p)\int_{\dD{R}{\tau_1}{\tau_2}}\Bigl\{\Kv{\phi}{p}+\Krp{\phi}{p-2}+\sumn\Krp{\Omega_i\phi}{p-2}\Bigr\}\\
  +C(n,m,\delta,p)\int_{2\tau_1+R^\ast}^{2\tau_2+R^\ast}\!\!\!\!\ud t\int_\Sn\dm{\gn}\biggl\{\psi^2+\sqb{\ddvs{\psi}}+\sumn\sqb{\Omega_i\psi}+\sumn\sqv{\nablab\Omega_i\psi}\biggr\}\vert_{r=R}\,.
\end{multline}
Indeed, the first term $\vert\nablab\partial_\vs\psi\vert^2$ can be integrated by parts twice (such that we can absorb the resulting $\partial_\vs\chi$ term in the left hand side):
\begin{multline}
  \int_{\tau_1}^{\tau_2}\!\!\!\!\ud\us\int_{\us+R^\ast}^\infty\!\!\!\!\ud\vs\int_\Sn\dm{\gn}\,r^{p-1}\sqv{\nablab\chi}=\\
  =-\int_{\tau_1}^{\tau_2}\!\!\!\!\ud\us\int_{\us+R^\ast}^\infty\!\!\!\!\ud\vs\int_\Sn\dm{\gn}\,r^{p-1}\laplacesphs\partial_{\vs}\psi\cdot\partial_{\vs}\psi\\
  =-\int_{\tau_1}^{\tau_2}\!\!\!\!\ud\us\int_\Sn\dm{\gn}\,r^{p-1}\,\laplacesphs\psi\,\ddvs{\psi}\vert_{\us+R^\ast}^\infty\\
  +\int_{\tau_1}^{\tau_2}\!\!\!\!\ud\us\int_{\us+R^\ast}^\infty\!\!\!\!\ud\vs\int_\Sn\dm{\gn}\,\biggl\{(p-1)r^{p-2}\bigl(\cf\bigr)(\laplacesphs\psi)(\ddvs{\psi})\\
  +r^{p-1}(\laplacesphs\psi)\ddvs{\chi}+r^{p-1}\frac{2}{r}\bigl(\cf\bigr)(\laplacesphs\psi)\ddvs{\psi}\biggr\}\leq\\
  \leq\int_{\tau_1}^{\tau_2}\!\!\!\!\ud\us\int_\Sn\dm{\gn}\,r^{p-1}\,(\laplacesphs\psi)\,(\ddvs{\psi})\vert_{\vs=\us+R^\ast}\\
  +\int_{\tau_1}^{\tau_2}\!\!\!\!\ud\us\int_{\us+R^\ast}^\infty\!\!\!\!\ud\vs\int_\Sn\dm{\gn}\,\biggl\{\Bigl(p-1+\frac{n-2}{p}+2\Bigr)r^{(p-2)-1}\sqb{\laplacesphs\psi}r^2\\
  +(p-1+2)r^{(p-2)-1}\sqb{\ddvs{\psi}}+\frac{1}{2}\frac{p}{4}\frac{2}{n-2}r^{p-1}\sqb{\ddvs{\chi}}\biggr\}
\end{multline}
The second term in \eqref{eq:s:Kvcontrols} is controlled by the Hardy inequality
\begin{multline}
  \frac{1}{2}\int_{\us+R^\ast}^\infty \ud\vs\,r^{(p-2)-1}\frac{1}{r^2}\psi^2\leq
  \frac{1}{4-p}\frac{1}{R^{4-p}}\frac{1}{1-\frac{2m}{R^{n-2}}}\psi^2\vert_{(\us,\us+R^\ast)}\\+\frac{2}{(4-p)^2(1-\frac{2m}{R^{n-2}})^2}\int_{\us+R^\ast}^\infty\ud\vs\,r^{(p-2)-1}\sqb{\ddvs{\psi}}\,,
\end{multline}
and the third term simply by the following commutation with $\Omega_i$:
\begin{lemma}\label{lemma:s:laplacesq}
For each $n\geq 3$, it holds
\begin{equation}
  \label{eq:s:laplacesq}
  r^2\sqb{\laplacesphs\psi}\leq C(n)\sum_{j=1}^\frac{n(n-1)}{2}\sqv{\nablab r^\frac{n-1}{2}\Omega_j\phi}\,.
\end{equation}
\end{lemma}
The last term in \eqref{eq:s:Kvcontrols} we can rearrange as follows:
\begin{multline}
  \label{eq:cwein:intparts}
  -\frac{n-1}{8}r^{p-2}\Bigl[n-3+(n-1)\frac{2m}{r^{n-2}}\Bigr]\ddvs{\chi^2}=\\
  =-\dvs\biggl\{\frac{n-1}{8}r^{p-2}\Bigl[(n-3)+(n-1)\mrn\Bigr]\sqb{\ddvs{\psi}}\biggr\}\\
  +\frac{n-1}{8}r^{(p-2)-1}\Bigl[(p-2)(n-3)+(n-1)\Bigl((p-2)+(n-2)\Bigr)\mrn\Bigr]\bigl(\cf\bigr)\sqb{\ddvs{\psi}}
\end{multline}
Therefore (see also Appendix \ref{sec:integration})
\begin{multline}
  \intDloc{\tau_1}{\tau_2}{R}\frac{1}{8}p\,r^{p-1}\sqb{\ddvs{\chi}}\leq\\
  \leq\frac{1}{2}\frac{1}{1-\frac{2m}{R^{n-2}}}\intD{R}\Kv{\phi}{p}\dm{g}\\
  +C(n,p,\delta,R)\int_{\tau_1}^{\tau_2}\!\!\!\!\ud\us\int_\Sn\dm{\gn}\biggl\{r^{p-2}\sqb{\laplacesphs\psi}r^2+r^{p-2}\sqb{\ddvs{\psi}}+\psi^2\biggr\}\vert_{\vs=\us+R^\ast}\\
  +C(p,n,\delta,R)\intDloc{\tau_1}{\tau_2}{R}\\\times\biggl\{r^{(p-2)-1}\sumn\sqv{\nablab r^\frac{n-1}{2}\Omega_i\phi}+r^{(p-2)-1}\sqb{\ddvs{\psi}}\biggr\}\\
  -\frac{n-1}{8}\intDloc{\tau_1}{\tau_2}{R}r^{p-2}\Bigl[n+3+(n-1)\mrn\Bigr]\ddvs{\chi^2}\,.
\end{multline}
Now, recall \eqref{KVr}, and note that
\begin{equation}
  \delta-\bigl(n-(2-\delta)\bigr)\mrn>\frac{\delta}{2}\qquad(r>\sqrt[n-2]{\frac{4nm}{\delta}})\,,
\end{equation}
to see that
\begin{multline}
  r^{(p-2)-1}\sumn\sqv{\nablab r^\frac{n-1}{2}\Omega_i\phi}\leq\frac{4}{\delta}\sumn\Krp{\Omega_i\phi}{p-2}r^{n-1}\\
  -\frac{4}{\delta}\sumn\frac{n-1}{4}\Bigl[\frac{n-3}{2}+\frac{n-1}{2}\mrn\Bigr]r^{(p-2)-2}\ddvs{\sqb{\rn\Omega_i\phi}}\,.
\end{multline}
So
\begin{multline}
  \intDloc{\tau_1}{\tau_2}{R}p\,r^{p-1}\sqb{\ddvs{\chi}}\leq\\
  \leq C(n,m,\delta,p)\intD{R}\biggl\{\Kv{\phi}{p}+\Krp{\phi}{p-2}+\sumn\Krp{\Omega_i\phi}{p-2}\biggr\}\\
  -C(n,m,\delta,p)\intDloc{\tau_1}{\tau_2}{R}\frac{n-1}{8}\Bigl[n-3+(n-1)\mrn\Bigr]\times\\\times\biggl\{\frac{r^{p-2}}{r^2}\ddvs{\psi^2}+\frac{r^{p-2}}{r^2}\sumn\ddvs{\bigl(\Omega_i\psi\bigr)^2}+\frac{r^p}{r^2}\ddvs{\chi^2}\biggr\}\\
  +C(n,m,\delta,p)\int_{2\tau_1+R^\ast}^{2\tau_2+R^\ast}\!\!\!\!\ud t\int_\Sn\dm{\gn}\biggl\{\sumn\sqv{\nablab\Omega_i\phi}+\sqb{\ddvs{\psi}}+\psi^2\biggr\}\vert_{r=R}
\end{multline}
which upon integrating by parts yields \eqref{eq:cwein:pre}; note that the $\partial_\vs\psi^2$ and $\partial_\vs(\Omega_i\psi)^2$ terms generate boundary terms at infinity and zeroth order bulk terms \emph{with the right sign} by \eqref{eq:wein:intparts} while the $\partial_\vs\chi^2$ is reduced to a $(\partial_\vs\psi)^2$ term by \eqref{eq:cwein:intparts}.

By virtue of Stokes' theorem \eqref{eq:appendix:stokes} and in view of \eqref{eq:appendix:boundary} we conclude
\begin{multline}
  \label{eq:cwein}
  \intU{R}{\tau_2}{r^p\sqb{\ddvs{\chi}}+r^{p-2}\sqb{\ddvs{\psi}}+\sumn r^{p-2}\sqb{\ddvs{\Omega_i\psi}}}\\
  +\intDloc{\tau_1}{\tau_2}{R}r^{p-1}\sqb{\ddvs{\chi}}\leq\\
  \leq C(n,m,\delta,p)\int_{\tau_1+R^\ast}^\infty\!\!\!\!\ud\vs\int_\Sn\!\!\dm{\gn}\times\\\times\biggl\{r^p\sqb{\ddvs{\chi}}+r^{p-2}\sqb{\ddvs{\psi}}+\sumn r^{p-2}\sqb{\ddvs{\Omega_i\psi}}\biggr\}\Bigr\vert_{\us=\tau_1}\\
  +C(n,m,\delta,p)\int_{2\tau_1+R^\ast}^{2\tau_2+R^\ast}\!\!\!\!\ud t\int_\Sn\!\!\dm{\gn}\biggl\{\psi^2+\sqb{\ddvs{\psi}}+\sqb{\dddvs{\psi}}\\+\sumn\Bigl[\sqb{\Omega_i\psi}+\sqb{\ddvs{\Omega_i\psi}}\Bigr]+\sqv{\nablab\chi}+\sqv{\nablab\psi}+\sumn\sqv{\nablab\Omega_i\psi}\biggr\}\Bigr\vert_{r=R}\,.
\end{multline}

\noindent\emph{Proof of Prop.~\ref{prop:iid}.}
We shall use this \emph{weighted energy inequality} for $\chi$ to proceed in a hierarchy of four steps.\par
\noindent\framebox[1.1\width]{$p=4-\delta$:}
Let $\tau_1>0$, and $\tau_{j+1}=2\tau_j\quad(j\in\mathbb{N})$.
In a first step we use \eqref{eq:cwein} with $p=4-\delta$ and \eqref{eq:wein} with $p=2$ as an estimate for the spacetime integral of $\pvs\chi$, $\pvs\psi$, and $\pvs(\Omega_j\psi)$ on $\dD{R}{\tau_j}{\tau_{j+1}}$, and in a second step as an estimate for the corresponding integral on the future boundary of $\dD{R}{\tau_1}{\tau_j}$:
\begin{multline}
  \label{eq:pfour}
  \intDloc{\tau_j}{\tau_{j+1}}{R}\biggl\{r^{3-\delta}\sqb{\ddvs{\chi}}+r\sqb{\ddvs{\psi}}+\sumn r\sqb{\ddvs{\Omega_i\psi}}\biggr\}\leq\displaybreak[1]\\
  \leq C(n,m,\delta)\intU{R}{\tau_j}{r^{4-\delta}\sqb{\ddvs{\chi}}+r^2\sqb{\ddvs{\psi}}+\sumn r^2\sqb{\ddvs{\Omega_j\psi}}}\displaybreak[0]\\
  +C(n,m,\delta)\intT{\tau_j}{\tau_{j+1}}{R}{\psi^2+\sqb{\ddvs{\psi}}+\sqb{\dddvs{\psi}}\\+\sqv{\nablab\psi}+\sqv{\nablab\ddvs{\psi}}+\sumn\Bigl[\sqb{\Omega_i\psi}+\sqb{\ddvs{\Omega_i\psi}}+\sqv{\nablab{\Omega_i\psi}}\Bigr]}\leq\displaybreak[1]\\
  \leq C(n,m,\delta)\intU{R}{\tau_1}{r^{4-\delta}\sqb{\ddvs{\chi}}+r^2\sqb{\ddvs{\psi}}+\sumn r^2\sqb{\ddvs{\Omega_i\psi}}}\displaybreak[0]\\
  +C(n,m,\delta)\intT{\tau_1}{\tau_{j+1}}{R}{\psi^2+\sqb{\ddvs{\psi}}+\sqb{\dddvs{\psi}}\\+\sqv{\nablab\psi}+\sqv{\nablab\ddvs{\psi}}+\sumn\Bigl[\sqb{\Omega_i\psi}+\sqb{\ddvs{\Omega_i\psi}}+\sqv{\nablab{\Omega_i\psi}}\Bigr]}
\end{multline}
Thus by the mean value theorem of integration we obtain a sequence $\tau_j^\prime\in(\tau_j,\tau_{j+1})\quad(j\in\mathbb{N})$ such that the corresponding integral from the left hand side on $\us=\tau_j^\prime$ is bounded by $\tau_j^{-1}\times$ the right hand side of \eqref{eq:pfour}.

\framebox[1.1\width]{$p=3-\delta$:}
Next we shall use \eqref{eq:cwein} with $p=3-\delta$ on $\dD{R_j}{\taujp{2j-1}}{\taujp{2j+1}}$, (with $R_j^\ast\in(R^\ast,R^\ast+1)\quad(j\in\mathbb{N})$ chosen appropriately below). However, the quantity we are actually interested in is not $\partial_\vs\chi$, but rather
\begin{multline}
  \sqb{\ddvs{\rn T\cdot\phi}}=\sqb{\ddvs{T\cdot\rn\phi}}=\sqb{\frac{1}{2}\dddvs{\psi}+\frac{1}{2}\frac{\partial^2\psi}{\partial\us\partial\vs}}=\\
  \stackrel{\eqref{eq:wavepsi}}{=}\sqb{\frac{1}{2}\dddvs{\psi}+\frac{1}{2}\bigl(\cf\bigr)\laplacesphs\psi-\frac{1}{2}\frac{n-1}{2}\bigl(\frac{n-3}{2}+\frac{n-1}{2}\mrn\bigr)\frac{1}{r^2}\bigl(\cf\bigr)\psi}\\
  \leq\sqb{\dddvs{\psi}}+\sqb{\laplacesphs\psi}+\frac{n-1}{4}2(n-2)\frac{1}{r^4}\psi^2\,.
\end{multline}
Using the simple Hardy inequality
\begin{multline}
  \frac{1}{2}\int_{\us+R^\ast}^\infty\ud\vs\,r^{2-\delta}\frac{1}{r^4}\psi^2\leq\\
  \leq\frac{1}{1-\frac{2m}{R^{n-2}}}\frac{1}{r^{1+\delta}}\psi^2(\us,\us+R^\ast)
    +\frac{2}{\bigl(1-\frac{2m}{R^{n-2}}\bigr)^2}\int_{\us+R^\ast}^\infty\ud\vs\frac{1}{r^\delta}\sqb{\ddvs{\psi}}
\end{multline}
and again the commutation introduced in Lemma \ref{lemma:s:laplacesq} we obtain
\begin{multline}
  \intDloc{\taujp{2j-1}}{\taujp{2j+1}}{R_j}{r^{2-\delta}\sqb{\rn T\cdot\phi}}\leq\displaybreak[1]\\
  \leq\intDloc{\taujp{2j-1}}{\taujp{2j+1}}{R_j}\biggl\{r^{2-\delta}\sqb{\ddvs{\chi}}\\+C(n)r^{-\delta}\sumn\sqv{\nablab\Omega_i\psi}+\frac{(n-1)(n-2)}{2}\frac{2}{\bigl(1-\frac{2m}{R^{n-2}}\bigr)^2}r^{-\delta}\sqb{\ddvs{\psi}}\biggr\}\\
  +\frac{1}{1-\frac{2m}{R^{n-2}}}\intT{\taujp{2j-1}}{\taujp{2j+1}}{R_j}{\frac{1}{r^{1+\delta}}\psi^2}\leq\displaybreak[1]\\
  \leq C(n,m,\delta)\intDloc{\taujp{2j-1}}{\taujp{2j+1}}{R_j}\biggl\{r^{2-\delta}\sqb{\ddvs{\chi}}\\+\Krp{\phi}{1-\delta}r^{n-1}+\sumn \Krp{\Omega_i\phi}{1-\delta}r^{n-1}\biggr\}\displaybreak[0]\\
  +C(n,m,\delta)\intT{\taujp{2j-1}}{\taujp{2j+1}}{R_j}{\psi^2+\sumn\sqb{\Omega_i\psi}}
\end{multline}
where in the last step we have again used \eqref{eq:wein:intparts}.
Furthermore, by now applying \eqref{eq:cwein} with $p=3-\delta$,
\begin{multline}
  \intDloc{\taujp{2j-1}}{\taujp{2j+1}}{R_j}{r^{2-\delta}\sqb{\rn T\cdot\phi}}\leq\displaybreak[1]\\
  \leq C(n,m,\delta)\intU{R_j}{\taujp{2j-1}}{r^{3-\delta}\sqb{\ddvs{\chi}}\\+r\sqb{\ddvs{\psi}}+\sumn r\sqb{\ddvs{\Omega_i\psi}}}\\
  +C(n,m,\delta)\intT{\taujp{2j-1}}{\taujp{2j+1}}{R_j}{\psi^2+\sqb{\ddvs{\psi}}+\sqb{\dddvs{\psi}}\\+\sqv{\nablab\psi}+\sqv{\nablab\ddvs{\psi}}+\sumn\Bigl[\sqb{\Omega_i\psi}+\sqb{\ddvs{\Omega_i\psi}}+\sqv{\nablab\Omega_i\psi}\Bigr]}\,,
\end{multline}
we obtain a sequence $\taujdp{j}\in(\taujp{2j-1},\taujp{2j+1})\quad(j\in\mathbb{N})$ such that in view of the previous step:
\begin{multline}
  \intU{R_j}{\taujdp{j}}{r^{2-\delta}\sqb{\ddvs{(\rn T\cdot\phi)}}}\leq\displaybreak[1]\\
  \leq\frac{C(n,m,\delta)}{\tau_{2j}\,\tau_{2j-1}}\intU{R}{\tau_1}{r^{4-\delta}\sqb{\ddvs{\chi}}\\+r^2\sqb{\ddvs{\psi}}+r^2\sumn\sqb{\ddvs{\Omega_i\psi}}}\displaybreak[0]\\
  +\frac{C(n,m,\delta)}{\tau_{2j}\,\tau_{2j-1}}\intT{\tau_1}{\tau_{2j+1}}{R}{\psi^2+\sqb{\ddvs{\psi}}+\sqb{\dddvs{\psi}}\\+\sqv{\nablab\psi}+\sqv{\nablab\ddvs{\psi}}+\sumn\Bigl[\sqb{\Omega_i\psi}+\sqb{\ddvs{\Omega_i\psi}}+\sqv{\nablab\Omega_i\psi}}\displaybreak[0]\\
  +C(n,m,\delta)\intT{\taujp{2j-1}}{\taujp{2j+1}}{R_j}{\psi^2+\sqb{\ddvs{\psi}}+\sqb{\dddvs{\psi}}\\+\sqv{\nablab\psi}+\sqv{\nablab\ddvs{\psi}}+\sumn\Bigl[\sqb{\Omega_i\psi}+\sqb{\ddvs{\Omega_i\psi}}+\sqv{\nablab\Omega_i\psi}}
\end{multline}
Now, by writing out the derivatives of $\psi=\rn\phi$, and using \eqref{eq:wavepsi}, we calculate that
\begin{multline}
  \psi^2+\sqb{\ddvs{\psi}}+\sqb{\dddvs{\psi}}\\+\sqv{\nablab\psi}+\sqv{\nablab\ddvs{\psi}}+\sumn\Bigl[\sqb{\Omega_i\psi}+\sqb{\ddvs{\Omega_i\psi}}+\sqv{\nablab\Omega_i\psi}\Bigr]\leq\displaybreak[0]\\
  \leq C(R)\biggl\{\phi^2+\sqb{\ddvs{\phi}}+\sqb{\ddus{\phi}}\\+\sqb{\ddvs{T\cdot\phi}}+\sqv{\nablab\phi}+\sumn\Bigl[\sqv{\nablab\Omega_i\phi}+\sqb{\ddvs{\Omega_i\phi}}\Bigr]\biggr\}\vert_{r=R}\,;
\end{multline}
by applying Prop.~\ref{prop:ILED} first to the domain $\dD{r_1}{\tau_1}{\tau_{2j+1}}\subset\mathcal{R}_{r_0,r_1}^\infty(2\tau_1+r_1^\ast)$ where $r_1>\sqrt[n-2]{\frac{4nm}{\delta}}$ to fix the radius $R$ and then to the domain $\dD{r(\rs=R^\ast+1)}{\taujp{2j-1}}{\taujp{2j+1}}\setminus\dD{R}{\taujp{2j-1}}{\taujp{2j+1}}\subset\mathcal{R}_{r_0,R}^\infty(2\taujp{2j-1}+R^\ast)$ to fix the radii $R_j\ (j\in\mathbb{N})$ by using the mean value theorem for the integration in $\rs$ this yields (see also Appendix \ref{ref:dyadic})
\begin{multline}
  \intU{R_j}{\taujdp{j}}{r^{2-\delta}\sqb{\ddvs{\rn T\cdot\phi}}}\leq\displaybreak[1]\\
  \leq\frac{C(n,m,\delta)}{(\taujdp{j})^2}\Biggl\{\intU{R}{\tau_1}{r^{4-\delta}\sqb{\ddvs{\chi}}\\+r^2\sqb{\ddvs{\psi}}+r^2\sumn\sqb{\ddvs{\Omega_i\psi}}}\\+\int_{\St{\tau_1}}\Bigl(J^T(\phi)+J^T(T\cdot\phi)+J^T(T^2\cdot\phi)+\sumn\bigl[J^T(\Omega_i\phi)+J^T(T\cdot\Omega_i\phi)\bigr],n\Bigr)\Biggr\}\displaybreak[0]\\
  +C(n,m,\delta)\int_\St{\tau_{2j-1}}\Bigl(J^T(\phi)+J^T(T\cdot\phi)+J^T(T^2\cdot\phi)\\+\sumn\bigl[J^T(\Omega_i\phi)+J^T(T\cdot\Omega_i\phi)\bigr],n\Bigr)\,.
\end{multline}
Therefore, by Prop.~\ref{prop:ed}:
\begin{multline}\label{eq:edac}
    \intU{R_j}{\taujdp{j}}{r^{2-\delta}\sqb{\ddvs{\rn T\cdot\phi}}}\leq\displaybreak[1]\\
    \leq\frac{C(n,m,\delta)}{(\taujdp{j})^2}\Biggl\{\intU{R}{\tau_1}{r^{4-\delta}\sqb{\ddvs{\chi}}\\+\sum_{k=0}^3 r^2\sqb{\ddvs{T^k\cdot\psi}}+\sum_{k=0}^2\sumn r^2\sqb{\ddvs{T^k\Omega_i\psi}}}\displaybreak[0]\\+\int_\St{\tau_1}\Bigl(\sum_{k=0}^4 J^N(T^k\cdot\phi)+\sum_{k=0}^3 J^N(T^k\Omega_i\phi),n\Bigr)\Biggr\}
\end{multline}
\begin{remark}
This statement should be compared to the assumptions of Prop.~\ref{prop:ed} \eqref{eq:eda}, from which all that one can deduce with \eqref{eq:wein} is
\begin{equation}
  \intU{R}{\tau}{r^2\sqb{\ddvs{(\rn T\cdot\phi)}}}<\infty\qquad(\tau>\tau_0)\,.
\end{equation}
\end{remark}

We shall now proceed along the lines of the proof of Prop.~\ref{prop:ed} in Section \ref{sec:ed}, just that we have \eqref{eq:edac} as a starting point for the solution $T\cdot\phi$ of \eqref{eq:wave}, (and \eqref{eq:ed}); however, as opposed to Prop.~\ref{prop:ed} the hierarchy does not descend from $p=2$ but $p<2$, which introduces a degeneracy in the last step, and requires the refinement of Prop.~\ref{prop:ILED} to Prop.~\ref{prop:rILED}, and Prop.~\ref{prop:zeroth} to Prop.~\ref{prop:zeroth:r}, see Section \ref{sec:refinements}.\par

\begin{lemma}[Pointwise decay under special assumptions]\label{lemma:pointwisedecay:special}
Let $\phi$ be a solution of the wave equation \eqref{eq:wave}, with initial data on $\Sigma_{\tau_1}\ (\tau_1>0)$ satisfying
\begin{multline*}
    D\doteq\intU{R}{\tau_1}{r^{4-\delta}\sqb{\ddvs{\chi}}\\+\sum_{k=0}^3 r^2\sqb{\ddvs{T^k\psi}}+\sum_{k=0}^2\sumn r^2\sqb{\ddvs{T^k\Omega_i\psi}}}\displaybreak[0]\\+\int_\St{\tau_1}\Bigl(\sum_{k=0}^4 J^N(T^k\cdot\phi)+\sum_{k=0}^3 J^N(T^k\Omega_i\phi),n\Bigr)<\infty
\end{multline*}
for some $\delta>0$ and 
\begin{equation}
  \label{eq:pointwisedecay:special:a}
  \int_{\taup+R^\ast}^\infty\ud\vs\int_\Sn\!\!\!\!\dm{\gn}\times r^{2-\delta}\sqb{\ddvs{\tpsi}}\vert_{\us=\taup}\leq\frac{C(n,m,\delta)\,D}{\taup^2}
  \tag{$\ast$}
\end{equation}
for some $\taup>\tau_1$.
Then there is a constant $C(n,m,R)$ such that
\begin{equation*}
  \int_\Sn\dm{\gn}\,r^{n-1-\frac{\delta}{2}}\bigl(\tphi\bigr)^2\vert_{(\us=\taup,\vs=R^\ast+\tau)}\leq\frac{C(n,m,\delta,R)\,D}{\taup^2}
\end{equation*}
for all $\tau>\tau^\prime$.
\end{lemma}
\begin{remark}
  Note the gain in powers of $r$ in comparison to the boundary term arising in Prop.~\ref{prop:zeroth:r}.
\end{remark}
\begin{proof}
First, integrating from infinity,
\begin{equation*}
  (\tphi)(\taup,R^\ast+\taup)=-\int_{\taup+R^\ast}^\infty\ddvs{(\tphi)}\ud\vs
\end{equation*}
and then by Cauchy's inequality, and the Sobolev inequality on the sphere,
\begin{multline*}
  (\tphi)^2(\taup,R+\taup)\leq\int_{R^\ast+\taup}^\infty\frac{1}{r^{n-1}}\ud\vs\times\int_{R+\taup}^\infty\bigl(\ddvs{(\tphi)}\Bigr)^2r^{n-1}\ud\vs\displaybreak[0]\\
  \leq\frac{1}{2}\bigl(1-\frac{2m}{r^{n-2}\vert_{(\us=\taup,\vs=R^\ast+\taup)}}\bigr)^{-1}\frac{1}{n-2}\frac{1}{r^{n-2}}\times\\\times C(m,n)\int_{R^\ast+\taup}^\infty\ud\vs\int_\Sn\dm{\gn}\,r^{n-1}\biggl\{\Bigl(\ddvs{(\tphi)}\Bigr)^2+\sumn\Bigl(\Omega_i\ddvs{(\tphi)}\Bigr)^2\biggr\}\,.
\end{multline*}
Therefore, by Prop.~\ref{prop:ed},
\begin{multline}\label{eq:pointwisedecay:special:pf:step1}
  \bigl(r^{n-2}(\tphi\bigr)^2)(\taup,R^\ast+\taup)\leq\frac{C(n,m)}{1-\frac{2m}{R^{n-2}}}\int_{\Sigma_{\taup}}\Bigl(J^T(\tphi)+\sumn J^T(\Omega_i\tphi),n\Bigr)\\\leq\frac{C(n,m,R)}{\taup^2}\,D\,.  
  \tag{$\ast\ast$}
\end{multline}
Now
\begin{multline*}
  r^{n-1}\int_\Sn\dm{\gn}\,(\tphi)^2\,(\taup,R^\ast+\tau)=\displaybreak[0]\\
  =\int_\Sn\dm{\gn}\bigl(r^{n-1}(\tphi)^2\bigr)(\taup,R^\ast+\taup)+\int_{R^\ast+\taup}^{R^\ast+\tau}\!\!\!\!\ud\vs\int_\Sn\dm{\gn}2\tpsi\ddvs{\tpsi}\leq\displaybreak[0]\\
  \leq R^{n-1}\int_\Sn\dm{\gn}(\tphi)^2(\taup,R^\ast+\taup)\\+2r^\frac{\delta}{2}\vert_{\begin{subarray}{l}(\us=\taup,\\\ \vs=R^\ast+\tau)\end{subarray}}\sqrt{\int_{R^\ast+\taup}^\infty\!\!\!\!\ud\vs\int_\Sn\dm{\gn}\frac{1}{r^2}(\tphi)^2 r^{n-1}}\times\\\times\sqrt{\int_{R^\ast+\taup}^\infty\!\!\!\!\ud\vs\,r^{2-\delta}\Bigl(\ddvs{\tpsi}\Bigr)^2}\,,
\end{multline*}
which proves the pointwise estimate of the Lemma in view of the Hardy inequality of Lemma \ref{lemma:hardy}, Prop.~\ref{prop:ed}, the assumption \eqref{eq:pointwisedecay:special:a} and \eqref{eq:pointwisedecay:special:pf:step1}.
\end{proof}

\noindent\framebox[1.1\width]{$p=2-\delta$:}
By the weighted energy inequality with $p=1-\delta$ and $\rn T\cdot\phi$ in the role of $\psi$, see \eqref{eq:wein:intparts} in particular,
\begin{multline}
  \intDloc{\idpr{\tau}{2j-1}}{\idpr{\tau}{2j+1}}{\ipr{R}{j}}{r^{1-\delta}\sqb{\ddvs{T\cdot\psi}}}\leq\displaybreak[0]\\
  \leq C(n,m)\intDloc{\idpr{\tau}{2j-1}}{\idpr{\tau}{2j+1}}{\ipr{R}{j}}{\Krp{T\cdot\phi}{2-\delta}}\\
  +C(n,m)\intT{\idpr{\tau}{2j-1}}{\idpr{\tau}{2j+1}}{\ipr{R}{j}}{\sqb{T\cdot\psi}}\leq\displaybreak[0]\\
  \leq C(n,m)\intU{\ipr{R}{j}}{\idpr{\tau}{2j-1}}{r^{2-\delta}\sqb{\ddvs{T\cdot\psi}}}\\
  +C(n,m)\intT{\idpr{\tau}{2j-1}}{\idpr{\tau}{2j+1}}{\ipr{R}{j}}{\sqb{T\cdot\psi}+\sqb{\ddvs{\tpsi}}+\sqv{\nablab\tpsi}}
\end{multline}
where we choose $\ipr{R}{j}^\ast\in(R^\ast+1,R^\ast+2)$ such that Prop.~\ref{prop:ILED} applied to the domain $\dD{r(\rs=R^\ast+2)}{\idpr{\tau}{2j-1}}{\idpr{\tau}{2j+1}}\setminus\dD{r(\rs=R^\ast+1)}{\idpr{\tau}{2j-1}}{\idpr{\tau}{2j+1}}$ yields an estimate for the integral on the timelike boundary above in terms of the first and second order energies on $\St{\idpr{\tau}{2j-1}}$ which in turn decays by Prop.~\ref{prop:ed}. Therefore there exists a sequence $\itpr{\tau}{j}\in(\idpr{\tau}{2j-1},\idpr{\tau}{2j+1})\ (j\in\mathbb{N})$ such that
\begin{multline}
  \label{eq:iid:ptwo}
  \intU{\ipr{R}{j}}{\itpr{\tau}{j}}{r^{1-\delta}\sqb{\ddvs{T\cdot\psi}}}\leq\displaybreak[1]\\
  \leq\frac{C(n,m,\delta)}{(\itpr{\tau}{j})^3}\Biggl\{
  \intU{R}{\tau_1}{r^{4-\delta}\sqb{\ddvs{\chi}}\\+\sum_{k=0}^3r^2\sqb{\ddvs{(T^k\cdot\psi)}}+\sum_{k=0}^2\sumn r^2\sqb{\ddvs{T^k\Omega_i\psi}}}\displaybreak[0]\\
  +\int_\St{\tau_1}\Bigl(\sum_{k=0}^4J^N(T^k\cdot\phi)+\sum_{k=0}^3\sumn J^N(T^k\Omega_i\phi),n\Bigr)\Biggr\}\,.
\end{multline}
\noindent\framebox[1.1\width]{$p=1-\delta$:}
Since, by integrating by parts,
\begin{multline}
  \int_{\us+R^\ast}^\infty\ud\vs\frac{1}{r^\delta}\sqb{\ddvs{\psi}}=\\
  =\int_{\us+R^\ast}^\infty\ud\vs\frac{1}{r^\delta}\Bigl\{\frac{n-1}{2r}\rn\bigl(\cf\bigr)\ddvs{\bigl(\rn\phi^2\bigr)}+r^{n-1}\sqb{\ddvs{\phi}}\Bigr\}\displaybreak[0]\\
  =\frac{1}{r^\delta}\frac{n-1}{2r}\bigl(\cf\bigr)\psi^2\vert_{\us+R^\ast}^\infty
  +\int_{\us+R^\ast}^\infty\ud\vs\biggl\{\frac{\delta}{r^{1+\delta}}\frac{n-1}{2r}\bigl(\cf\bigr)^2\psi^2\\
  +\frac{1}{r^\delta}\frac{n-1}{2r^2}\bigl(\cf\bigr)\psi^2\Bigl[(n-2)+\bigl(\cf\bigr)\frac{n-3}{2}\Bigr]
  +\frac{1}{r^\delta}\sqb{\ddvs{\phi}}r^{n-1}\biggr\}
\end{multline}
we have by \eqref{KVr} that also (with $\idpr{R}{j}^\ast\in(R^\ast+2,R^\ast+3)$),
\begin{multline}
  \intDloc{\itpr{\tau}{2j-1}}{\itpr{\tau}{2j+1}}{\idpr{R}{j}}{\frac{1}{r^\delta}\Bigl\{\sqb{\ddvs{\tphi}}+\sqv{\nablab\tphi}\Bigr\}r^{n-1}}\leq\displaybreak[0]\\
  \leq C(n,m)\biggl\{\intDloc{\itpr{\tau}{2j-1}}{\itpr{\tau}{2j+1}}{\idpr{R}{j}}\Krp{\tphi}{1-\delta}\cdot r^{n-1}\\
  +\intT{\itpr{\tau}{2j-1}}{\itpr{\tau}{2j+1}}{\idpr{R}{j}}{\sqb{\tpsi}}\biggr\}\,.
\end{multline}
By virtue of Stokes theorem \eqref{eq:appendix:stokes}, \eqref{eq:appendix:boundary} and our previous result \eqref{eq:iid:ptwo} we obtain
\begin{multline}
  \label{eq:iid:pone:pre}
  \intDloc{\itpr{\tau}{2j-1}}{\itpr{\tau}{2j+1}}{\idpr{R}{j}}{\frac{1}{r^\delta}\Bigl(J^T(\tphi),\dvs\Bigr)r^{n-1}}\leq\displaybreak[1]\\
  \leq C(n,m)\biggl\{\intU{\idpr{R}{j}}{\itpr{\tau}{2j-1}}{r^{1-\delta}\sqb{\ddvs{(\tpsi)}}}\displaybreak[0]\\
  +\intT{\itpr{\tau}{2j-1}}{\itpr{\tau}{2j+1}}{\idpr{R}{j}}{\sqb{\ddvs{(\tpsi)}}+\sqv{\nablab\tpsi}+\sqb{T\cdot\psi}}\biggr\}\displaybreak[1]\\  
  \leq\frac{C(n,m,\delta)}{(\itpr{\tau}{j})^3}\Biggl\{
  \intU{R}{\tau_1}{r^{4-\delta}\sqb{\ddvs{\chi}}\\+\sum_{k=0}^3r^2\sqb{\ddvs{(T^k\cdot\psi)}}+\sum_{k=0}^2\sumn r^2\sqb{\ddvs{T^k\Omega_i\psi}}}\displaybreak[0]\\
  +\int_\St{\tau_1}\Bigl(\sum_{k=0}^4J^N(T^k\cdot\phi)+\sum_{k=0}^3\sumn J^N(T^k\Omega_i\phi),n\Bigr)\Biggr\}\displaybreak[0]\\
  +C(n,m)\biggl\{\int_\Stt{\idpr{\tau}{2(2j-1)-1}}{\itpr{\tau}{2j+1}}\Bigl(J^T(\tphi)+J^T(T^2\cdot\phi),n\Bigr)\\+\int_\Sn\!\!\!\!\dm{\gn}r^{n-2}\sqb{\tphi}\vert_{\begin{subarray}{l}(\us=\idpr{\tau}{2(2j-1)-1},\\\ \vs=\idpr{R}{j}^\ast+\itpr{\tau}{2j+1})\end{subarray}}\biggr\}\,,
\end{multline}
where in the last inequality we have used Prop.~\ref{prop:zeroth:r}, and then chosen $\idpr{R}{j}\ (j\in\mathbb{N})$ suitably by Prop.~\ref{prop:rILED};
furthermore the inequality still holds if we add the integral of the nondegenerate energy on $\dP{\idpr{R}{j}}{\itpr{\tau}{2j-1}}{\itpr{\tau}{2j+1}}$ on the left hand side and replace $J^T$ by $J^N$ in the first term of the integral on $\Stt{\idpr{\tau}{2(2j-1)-1}}{\itpr{\tau}{2j+1}}$ on the right hand side.
The last two terms on the right hand side of \eqref{eq:iid:pone:pre} in fact decay with almost the same rate as the first;
for first note here that we could have used Prop.~\ref{prop:zeroth} and Cor.~\ref{cor:nILED} instead, and then employ Prop.~\ref{prop:ed} to obtain in any case that
\begin{multline}
  \int_{\itpr{\tau}{2j-1}}^{\itpr{\tau}{2j+1}}\ud\tau\int_\St{\tau}{\frac{1}{r^\delta}\ned{N}{\tphi}}\leq\displaybreak[1]\\
  \leq\frac{C(n,m,\delta)}{(\itpr{\tau}{2j-1})^2}\Biggl\{
  \intU{R}{\tau_1}{r^{4-\delta}\sqb{\ddvs{\chi}}\\+\sum_{k=0}^3r^2\sqb{\ddvs{(T^k\cdot\psi)}}+\sum_{k=0}^2\sumn r^2\sqb{\ddvs{T^k\Omega_i\psi}}}\displaybreak[0]\\
  +\int_\St{\tau_1}\Bigl(\sum_{k=0}^4J^N(T^k\cdot\phi)+\sum_{k=0}^3\sumn J^N(T^k\Omega_i\phi),n\Bigr)\Biggr\}\,.
\end{multline}
It then follows that there exists a sequence $\iqpr{\tau}{j}\in(\itpr{\tau}{2j-1},\itpr{\tau}{2j+1})$ such that
\begin{multline}
  \label{eq:iid:pone:pre:first}
  \int_\Stt{\iqpr{\tau}{j}}{\itpr{\tau}{2(j+2)+1}}\ned{N}{\tphi}\leq r^\delta\vert_{\begin{subarray}{l}(\us=\iqpr{\tau}{j},\\\ \vs=\idpr{R}{j+2}^\ast+\itpr{\tau}{2(j+2)+1})\end{subarray}}\int_\St{\iqpr{\tau}{j}}\frac{1}{r^\delta}\ned{N}{\tphi}\leq\displaybreak[1]\\
  \leq\frac{C(n,m,\delta)}{(\iqpr{\tau}{j})^{3-\delta}}\Biggl\{
  \intU{R}{\tau_1}{r^{4-\delta}\sqb{\ddvs{\chi}}\\+\sum_{k=0}^3r^2\sqb{\ddvs{(T^k\cdot\psi)}}+\sum_{k=0}^2\sumn r^2\sqb{\ddvs{T^k\Omega_i\psi}}}\displaybreak[0]\\
  +\int_\St{\tau_1}\Bigl(\sum_{k=0}^4J^N(T^k\cdot\phi)+\sum_{k=0}^3\sumn J^N(T^k\Omega_i\phi),n\Bigr)\Biggr\}
\end{multline}
because $\iqpr{\tau}{j}(\itpr{\tau}{2(j+2)+1}-\iqpr{\tau}{j})^{-1}\leq 1$.
And second the assumptions of Lemma \ref{lemma:pointwisedecay:special} are satisfied in view of \eqref{eq:edac} on $\us=\idpr{\tau}{j}\ (j\in\mathbb{N})$ which yields
\begin{multline}
  \label{eq:iid:pone:pre:second}
  \int_\Sn\dm{\gn}r^{n-2}\sqb{\tphi}\vert_{(\us=\idpr{\tau}{2(2j-1)-1},\vs=\idpr{R}{j}^\ast+\itpr{\tau}{2j+1})}\leq\displaybreak[1]\\
  \leq\frac{C(n,m,\delta)}{(\itpr{\tau}{2j-1})^{3-\frac{\delta}{2}}}\Biggl\{\intU{R}{\tau_1}{r^{4-\delta}\sqb{\ddvs{\chi}}\\+\sum_{k=0}^3 r^2\sqb{\ddvs{T^k\cdot\psi}}+\sum_{k=0}^2\sumn r^2\sqb{\ddvs{T^k\Omega_i\psi}}}\displaybreak[0]\\+\int_\St{\tau_1}\Bigl(\sum_{k=0}^4 J^N(T^k\cdot\phi)+\sum_{k=0}^3 J^N(T^k\Omega_i\phi),n\Bigr)\Biggr\}
\end{multline}
because also $\itpr{\tau}{2j-1}(\itpr{\tau}{2j+1}-\idpr{\tau}{2(2j-1)-1})^{-1}\leq C$.
We shall now return to \eqref{eq:iid:pone:pre} (and its extension that includes the nondegenerate energy on $\dP{\idpr{R}{j}}{\itpr{\tau}{2j-1}}{\itpr{\tau}{2j+1}}$) to find that, after inserting \eqref{eq:iid:pone:pre:first} and using Prop.~\ref{prop:uniformboundedness},
\begin{multline}
  \int_\Stt{\idpr{\tau}{2(2j-1)-1}}{\itpr{\tau}{2j+1}}\Bigl(J^N(\tphi)+J^T(T^2\cdot\phi),n\Bigr)
  \leq\int_\Stt{\itpr{\tau}{(2j-1)-1}}{\itpr{\tau}{2j+1}}\Bigl(J^N(\tphi)+J^T(T^2\cdot\phi),n\Bigr)\leq\\
  \leq\int_\Stt{\iqpr{\tau}{j-2}}{\itpr{\tau}{2j+1}}\Bigl(J^N(\tphi)+J^T(T^2\cdot\phi),n\Bigr)\leq\displaybreak[1]\\
  \leq\frac{C(n,m,\delta)}{(\iqpr{\tau}{j-2})^{3-\delta}}\Biggl\{
  \intU{R}{\tau_1}{r^{4-\delta}\sqb{\ddvs{\chi}}+r^{4-\delta}\sqb{\ddvs{(T\cdot\chi)}}\\+\sum_{k=0}^4r^2\sqb{\ddvs{(T^k\cdot\psi)}}+\sum_{k=0}^3\sumn r^2\sqb{\ddvs{T^k\Omega_i\psi}}}\displaybreak[0]\\
  +\int_\St{\tau_1}\Bigl(\sum_{k=0}^5J^N(T^k\cdot\phi)+\sum_{k=0}^4\sumn J^N(T^k\Omega_i\phi),n\Bigr)\Biggr\}
\end{multline}
and using \eqref{eq:iid:pone:pre:second}, that there exists (another) sequence $\iqpr{\tau}{j}\in(\itpr{\tau}{2j-1},\itpr{\tau}{2j+1})\ (j\in\mathbb{N})$ such that
\begin{multline}
  \label{eq:iid:pone:dyadic}
  \int_\St{\iqpr{\tau}{j}}\frac{1}{r^\delta}\ned{N}{\tphi}\leq\\
  \leq\frac{C(n,m,\delta)}{(\iqpr{\tau}{j})^{4-\delta}}\Biggl\{
  \intU{R}{\tau_1}{\sum_{k=0}^1r^{4-\delta}\sqb{\ddvs{(T^k\cdot\chi)}}\\+\sum_{k=0}^4r^2\sqb{\ddvs{(T^k\cdot\psi)}}+\sum_{k=0}^3\sumn r^2\sqb{\ddvs{T^k\Omega_i\psi}}}\displaybreak[0]\\
  +\int_\St{\tau_1}\Bigl(\sum_{k=0}^5J^N(T^k\cdot\phi)+\sum_{k=0}^4\sumn J^N(T^k\Omega_i\phi),n\Bigr)\Biggr\}\,.
\end{multline}
So for any $\tau>\tau_1$ we can choose $j\in\mathbb{N}$ such that $\tau\in(\itpr{\tau}{2j-1},\itpr{\tau}{2j+1})$ to obtain finally by Prop.~\ref{prop:uniformboundedness} that
\begin{multline}
  \label{eq:iid:pone:interior}
  \int_{\St{\tau}\cap\{r\leq R\}}\ned{N}{\tphi}\leq\displaybreak[0]\\
  \leq\int_\Stt{\iqpr{\tau}{j-1}}{\itpr{\tau}{2j+1}}\ned{N}{\tphi}\leq r^\delta\vert_{\begin{subarray}{l}(\us=\iqpr{\tau}{j-1},\\\ \vs=R^\ast+\itpr{\tau}{2j+1})\end{subarray}}\int_\St{\iqpr{\tau}{j-1}}\frac{1}{r^\delta}\ned{N}{\tphi}\displaybreak[1]\\
  \leq\frac{C(n,m,\delta)}{\tau^{4-2\delta}}\Biggl\{
  \intU{R}{\tau_1}{\sum_{k=0}^1r^{4-\delta}\sqb{\ddvs{(T^k\cdot\chi)}}\\+\sum_{k=0}^4r^2\sqb{\ddvs{(T^k\cdot\psi)}}+\sum_{k=0}^3\sumn r^2\sqb{\ddvs{T^k\Omega_i\psi}}}\displaybreak[0]\\
  +\int_\St{\tau_1}\Bigl(\sum_{k=0}^5J^N(T^k\cdot\phi)+\sum_{k=0}^4\sumn J^N(T^k\Omega_i\phi),n\Bigr)\Biggr\}\,.
\end{multline}
\qed

\begin{remark}
\label{remark:iid:extension}
Note that for the removal of the restriction to dyadic sequences in the last step of the proof, \eqref{eq:iid:pone:dyadic} - \eqref{eq:iid:pone:interior}, we could have equally obtained a decay estimate for the energy flux through $\St{\tau}\cap\{\rs\leq R^\ast+\tau^k\}$ (with $k\in\mathbb{N}$) by replacing $\Stt{\iqpr{\tau}{j-1}}{\itpr{\tau}{2j+1}}$ by $\Stt{\iqpr{\tau}{j-1}}{\iqpr{\tau}{j-1}+\tau^k}$ in the first estimate in \eqref{eq:iid:pone:interior}; if $\delta>0$ for a chosen $k\in\mathbb{N}$ is restricted to $\delta<(1+k)^{-1}$ we then still obtain a decay rate of $\tau^{4-(1+k)\delta}$ for the energy flux through $\St{\tau}\cap\{\rs\leq R^\ast+\tau^k\}$.
\end{remark}

\section{Pointwise Bounds}
\label{sec:pointwise}

In this Section we first prove pointwise estimates on $\lvert\phi\rvert$ and $\lvert\partial_t\phi\rvert$ separately based on the energy decay results Prop.~\ref{prop:ed} and Prop.~\ref{prop:iid} in Section \ref{sec:decay}. Then we give the interpolation argument to improve the pointwise decay on $\lvert\phi\rvert$. As we shall see in view of the nondegenerate energy estimates of Section \ref{sec:decay} we may restrict ourselves in the first place to a radial region away from the horizon.

\begin{prop}[Pointwise decay]
\label{prop:pointwisedecay}
\begin{itemize}
\item[(i)] Let $\phi$ be a solution of the wave equation \eqref{eq:wave}, with initial data on $\St{\tau_0}$ such that
\begin{equation}
  D=\int_{\tau_0+R^\ast}^\infty\int_\Sn\dm{\gn}\sum_{k=0}^{[\frac{n}{2}]+1} r^2\sqb{\ddvs{T^k\cdot\psi}}+\int_\St{\tau_0}\nedCT{[\frac{n}{2}]+2}<\infty\,.
\end{equation}
Then there is a constant $C(n,m)$ such that for $r_0<r<R$,
\begin{equation}
  \label{eq:pointwisedecay}
  \vert\phi(t,r)\vert\leq\frac{C(n,m)\,\sqrt{D}}{\tau}\qquad(\tau=\frac{1}{2}(t-R^\ast)>\tau_0)\,.
\end{equation}
\item[(ii)] If moreover, the initial data satisfies
\begin{multline}
  D=\intU{R}{\tau_0}{\sum_{k=0}^{[\frac{n}{2}]+1}r^{4-\delta}\sqb{\ddvs{(T^k\cdot\chi)}}\\+\sum_{k=0}^{[\frac{n}{2}]+4}r^2\sqb{\ddvs{(T^k\cdot\psi)}}+\sum_{k=0}^{[\frac{n}{2}]+3}\sumn r^2\sqb{\ddvs{T^k\Omega_i\psi}}}\\
  +\int_\St{\tau_0}\Bigl(\sum_{k=0}^{[\frac{n}{2}]+5}J^N(T^k\cdot\phi)+\sum_{k=0}^{[\frac{n}{2}]+4}\sumn J^N(T^k\Omega_i\phi),n\Bigr)<\infty
\end{multline}
for some $0<\delta<\frac{1}{4}$, and $R>\sqrt[n-2]{\frac{8nm}{\delta}}$, then there is a constant $C(n,m)$ such that for $r_0<r<R$,
\begin{equation}
  \label{eq:pointwisedecayt}
  \lvert\partial_t\phi(t,r)\rvert\leq\frac{C(n,m)\,\sqrt{D}}{\tau^{2-2\delta}}\qquad(\tau=\frac{1}{2}(t-R^\ast)>\tau_0)\,.
\end{equation}
\end{itemize}
\end{prop}

The pointwise bounds are obtained from the energy estimates of Section \ref{sec:decay} using Sobolev inequalties and elliptic estimates; the former provide the link betweeen pointwise and integral quantities, and the latter allow for the expression of these integral quantities in terms of higher order energies.

\paragraph{Sobolev embedding.} By the extension theorem applied to the Sobolev embedding $\mathrm{H}^s(\mathbb{R}^n)\subset\mathrm{L}^\infty(\mathbb{R}^n)\ (s>\frac{n}{2})$ we have, for $r_0<\overline{r}<R$,
\begin{equation}
  \label{eq:sobolev}
  \vert\phi(\overline{t},\overline{r})\vert^2\leq C(n)\int_{r_0^\ast}^{R^\ast}\ud\rs\int_\Sn\dm{\gn}\Biggl\{\phi^2+\sum_{\vert\alpha\vert\geq 1}^{\vert\alpha\vert\leq[\frac{n}{2}]+1}\sqv{\overline{\nabla}^\alpha\phi}\Biggr\}r^{n-1}\Bigr\vert_{t=\overline{t}}
\end{equation}
where $\overline{\nabla}$ denote the tangential derivatives to the hypersurface $\overline{\Sigma}_{t}$, and $\alpha$ denotes a multindex of order $n$.

\paragraph{Elliptic estimates.} Note that for any solution $\phi$ of the wave equation
\begin{equation}
  T^2\cdot\phi=\frac{\partial^2\phi}{\partial\rs^2}+\bigl(\cf\bigr)\frac{n-1}{r}\ddrs{\phi}+\bigl(\cf\bigr)\laplacesph\phi\doteq L\cdot\phi
\end{equation}
where the operator
\begin{equation}
  L=\bigl(\cf\bigr)\overline{g}^{ij}\overline{\nabla}_i\partial_j
\end{equation}
is clearly elliptic, (here $\overline{g}_t=g\vert_{\overline{\Sigma}_t}$ denotes the restriction of $g$ to the spacelike hypersurfaces $\overline{\Sigma}_t$, a Riemannian metric on $\overline{\Sigma}_t$, and $i,j=1,\ldots,n$).
In view of the standard higher order interior elliptic regularity estimate,
\begin{equation}
  \lVert\phi\rVert_{\mathrm{H}^{m+2}(\widehat{\Sigma}_t)}\leq C\Bigl(\lVert L\cdot\phi\rVert_{\mathrm{H}^m(\widehat{\Sigma}_t)}+\lVert\phi\rVert_{\mathrm{L}^2(\widehat{\Sigma}_t)}\Bigr)\qquad\widehat{\Sigma}_t\doteq\overline{\Sigma}_t\cap\{r_0<r<R\}\,,
\end{equation}
we conclude with \eqref{eq:sobolev} that in the case where $[\frac{n}{2}]+1$ is even,
\begin{equation}
  \lvert\phi\rvert^2\leq C(n,m)\int_{r_0^\ast}^{R^\ast}\ud\rs\int_\Sn\dm{\gn}\sum_{l=0}^{[\frac{n}{2}]+1}\sqb{T^l\cdot\phi}r^{n-1}\,;
\end{equation}
in general we have:
\begin{lemma}[Pointwise estimate in terms of higher order energies]
\label{lemma:pointwisedecay}
Let $\phi$ be a solution of the wave equation \eqref{eq:wave}, and $n\geq 3$. Then there exists a constant $C(n,m)$ such that for all $r_0<r<R$:
\begin{equation}
  \label{eq:lemma:pointwisedecay}
  \lvert\phi(t,r)\rvert^2\leq C(n,m)\biggl[\lVert\phi\rVert^2_{\mathrm{L}^2(\widehat{\Sigma}_t)}+\int_{\widehat{\Sigma}_t}\sum_{l=0}^{[\frac{n}{2}]}\ned{T}{T^l\cdot\phi}\biggr]
\end{equation}
\end{lemma}

\begin{proof}[Proof of Prop.~\ref{prop:pointwisedecay}]
In view of the Lemma \ref{lemma:pointwisedecay} and the energy decay estimates of Section \ref{sec:decay} it remains to control the zeroth order term $\lVert\phi\rVert_{\mathrm{L}^2(\widehat{\Sigma}_t)}$; we multiply the integrand by $(\frac{R}{r})^2\geq 1$ and extend the integral to $\us=\tau=\frac{1}{2}(t-R^\ast)$, $\vs\geq\frac{1}{2}(t+R^\ast)$.
\begin{itemize}
\item[(i)] By Lemma \ref{lemma:hardy} we can then estimate $\lVert\phi\rVert^2_{\mathrm{L}^2(\widehat{\Sigma}_t)}$ by the energy flux through $\St{\tau=\frac{1}{2}(t-R^\ast)}$, and apply Prop.~\ref{prop:ed} to the higher order energies of Lemma \ref{lemma:pointwisedecay}.
\item[(ii)] Here we extend the integral only to $\tau+R^\ast\leq\vs\leq\tau+R^\ast+\tau^3$ and apply Lemma \ref{lemma:hardyineqfinite} to obtain
  \begin{multline}
    \label{eq:pointwisedecay:zeroth}
    \int_{r_0^\ast}^{R^\ast}\ud\rs\int_\Sn\dm{\gn}(\partial_t\phi)^2 r^{n-1}\leq
    C(n,m)\,R^2\,\int_{\St{\tau}\cap\{r^\ast\leq R^\ast+\tau^3\}}\ned{T}{\partial_t\phi}\\+C(n,m)\frac{R^2}{r}\int_{\Sn}r^{n-1}(\partial_t\phi)^2\vert_{(\us=\tau,\vs=\tau+R^\ast+\tau^3)}\,.
  \end{multline}
  As in the proof of Lemma \ref{lemma:pointwisedecay:special} we obtain by integrating from infinity and Cauchy's inequality that
  \begin{multline}
    r^{n-2}(\partial_t\phi)^2(\tau,\tau+R^\ast)\leq\frac{C(n,m)}{1-\frac{2m}{R^{n-2}}}\int_\St{\tau}\Bigl(J^T(\partial_t\phi)+\sumn J^T(\Omega_i\partial_t\phi),n\Bigr)
  \end{multline}
  which decays by Prop.~\ref{prop:ed} with a rate $\tau^{-2}$.
  Moreover, as in the proof of Lemma \ref{lemma:pointwisedecay:special},
  \begin{multline}
    \int_\Sn\dm{\gn}r^{n-1}(\partial_t\phi)^2\vert_{(\us=\tau,\vs=\tau+R^\ast+\tau^3)}=\\=\int_\Sn\dm{\gn}r^{n-1}(\partial_t\phi)^2\vert_{(\us=\tau,\vs=\tau+R^\ast)}+\int_{\tau+R^\ast}^{\tau+R^\ast+\tau^3}\!\!\!\!\!\!\!\!\ud\vs 2\,\partial_t\psi\,\ddvs{\partial_t\psi}\vert_{\us=\tau}
  \end{multline}
  and
  \begin{multline}
    \int_{\tau+R^\ast}^{\tau+R^\ast+\tau^3}\!\!\!\!\!\!\!\!\ud\vs \partial_t\psi\,\ddvs{\partial_t\psi}\vert_{\us=\tau}\leq\displaybreak[0]\\
    \leq\sqrt{\int_{\tau+R^\ast}^\infty\int_\Sn\dm{\gn}\frac{1}{r^2}(\partial_t\phi)^2r^{n-1}}\times\sqrt{\int_{\tau+R^\ast}^\infty\int_\Sn\dm{\gn}r^2\sqb{\ddvs{\rn\partial_t\phi}}}\,,
  \end{multline}
  the first factor decaying with a rate $\tau^{-1}$ by Lemma \ref{lemma:hardy} and Prop.~\ref{prop:ed}, and the second factor bounded by the weighted energy inequality for $\rn\partial_t\phi$ in place of $\psi$ with $p=2$. Therefore
  \begin{equation}
    \int_\Sn r^{n-1}(\partial_t\phi)^2\vert_{(\us=\tau,\vs=\tau+R^\ast+\tau^3)}\leq\frac{C(n,m)}{1-\frac{2m}{R^{n-2}}}\frac{D}{\tau}\,.
  \end{equation}
  By virtue of Prop.~\ref{prop:iid}, compare in particular Remark \ref{remark:iid:extension} on page \pageref{remark:iid:extension}, the first term on the right hand side of \eqref{eq:pointwisedecay:zeroth} decays with a rate of $\tau^{4-4\delta}$, and this is matched by the second term in view of the prefactor $r^{-1}=(R^\ast+\tau^3)^{-1}$, which is the result of our choice of powers of $\tau$ in the extension of the integral \eqref{remark:iid:extension}. Lemma \ref{lemma:pointwisedecay} applied to the solution $\partial_t\phi$ of \eqref{eq:wave} then yields the pointwise decay result \eqref{eq:pointwisedecayt} after having applied Prop.~\ref{prop:iid} to the higher order energies on the right hand side of \eqref{eq:lemma:pointwisedecay}.\qedhere
\end{itemize}
\end{proof}

\paragraph{Interpolation.}
We shall now interpolate between the results Prop.~\ref{prop:pointwisedecay} (i) and (ii) to improve the pointwise estimate for $\lvert\phi\rvert$.
Our argument can in some sense be compared to the proof of improved decay in \cite{Lid}.
The basic observation underlying this argument is that for $r_0<r<R$ and $t_1>t_0$
\begin{multline}
  \label{eq:interpolate}
  r^{n-2}\phi^2(r,t_1)=r^{n-2}\phi^2(r,t_0)+\int_{t_0}^{t_1}2\,\phi(t,r)\,\ddt{\phi}(t,r)\,r^{n-2}\ud t\\
  \leq r^{n-2}\phi^2(r,t_0)+\frac{1}{t_0^{1-2\delta}}\int_{t_0}^{t_1}\phi^2(t,r)\,r^{n-2}\ud t+t_0^{1-2\delta}\int_{t_0}^{t_1}\sqb{\ddt{\phi}}(t,r)\,r^{n-2}\ud t\,.
\end{multline}
Moreover, as a consequence of Lemma \ref{lemma:jump}, 
\begin{equation}
  \label{eq:jump:st}
  r^{n-2}\phi^2(t,r)\leq R^{n-2}\phi^2(t,R)+\bigl(1-\frac{2m}{r_0^{n-2}}\bigr)^{-1}\int_{\rs}^{R^\ast}\sqb{\ddrs{\phi}}r^{n-1}\ud\rs\,,
\end{equation}
we obtain an estimate for the timelike integrals in terms of the corresponding integrals at $r=R$ and spacetime integrals, using the Sobolev inequality on the sphere:
\begin{multline}
  \label{eq:jump:timelike}
  \int_{t_0}^{t_1}r^{n-2}\phi^2(t,r)\ud t\leq\int_{t_0}^{t_1}\ud t\int_\Sn\dm{\gn}\biggl\{R^{n-2}\phi^2(t,R)+\sumn R^{n-2}\sqb{\Omega_i\phi}(t,R)\biggr\}\\
  +\bigl(1-\frac{2m}{r_0^{n-2}}\bigr)^{-1}\int_{t_0}^{t_1}\ud t\int_{\rs}^{R^\ast}\ud\rs\int_\Sn\dm{\gn}r^{n-1}\biggl\{\sqb{\ddrs{\phi}}+\sumn\sqb{\ddrs{\Omega_i\phi}}\biggr\}
\end{multline}

\begin{lemma}
\label{lemma:jump}
Let $a<b\in\mathbb{R}$ and $\phi\in\mathrm{C}^1([a,b])$ then
\begin{equation}
  a^{n-2}\phi^2(a)\leq b^{n-2}\phi^2(b)+\int_a^b\Bigl(\frac{\ud\phi}{\ud x}\Bigr)^2 x^{n-1}\ud x
\end{equation}
for all $n\geq 3$.
\end{lemma}
\begin{proof}
Since, by integration by parts,
\begin{multline*}
  \int_a^b2\phi(x)\frac{\ud\phi}{\ud x}(x)x^{n-2}\ud x=2\phi^2(x)x^{n-2}\vert_a^b\\
  -\int_a^b2\phi(x)\frac{\ud\phi}{\ud x}(x)x^{n-2}\ud x-\int_a^b2\phi^2(x)(n-2)x^{n-3}\ud x\,,
\end{multline*}
it clearly follows, with Cauchy's inequality,
\begin{multline*}
  a^{n-2}\phi^2(a)\leq b^{n-2}\phi^2(b)+\int_a^b\Bigl(\frac{\ud\phi}{\ud x}\Bigr)^2x^{n-1}\ud x\\
  +\bigl[1-(n-2)\bigr]\int_a^b\frac{1}{x^2}\phi^2(x)x^{n-1}\ud x\,.\qedhere
\end{multline*}
\end{proof}

\begin{prop}[Improved interior pointwise decay]\label{prop:iipd}
Let $\phi$ be a solution of the wave equation \eqref{eq:wave}, with initial data on $\St{\tau_0}\ (\tau_0>1)$ satisfying 
\begin{multline}
  D\doteq\intU{R}{\tau_0}{\sum_{k=0}^2r^{4-\delta}\sqb{\ddvs{(T^k\cdot\chi)}}+\sum_{k=0}^2\sumn r^{4-\delta}\sqb{\ddvs{(T^k\cdot\Omega_i\chi)}}\\+\sum_{k=0}^5r^2\sqb{\ddvs{(T^k\cdot\psi)}}+\sum_{k=0}^5\sumn r^2\sqb{\ddvs{T^k\Omega_i\psi}}+\sum_{k=0}^4\sum_{i,j=1}^\frac{n(n-1)}{2} r^2\sqb{\ddvs{T^k\Omega_i\Omega_j\psi}}}\\
  +\int_\St{\tau_0}\Bigl(\sum_{k=0}^6J^N(T^k\cdot\phi)+\sum_{k=0}^6\sumn J^N(T^k\Omega_i\phi)+\sum_{k=0}^5\sum_{i,j}^\frac{n(n-1)}{2} J^N(T^k\Omega_i\Omega_j\phi),n\Bigr)<\infty\,.
\end{multline}
for some $0<\delta<\frac{1}{4}$, where $R>\sqrt[n-2]{\frac{8nm}{\delta}}$, $n\geq 3$.
Then there exists a constant $C(n,m,\delta)$ such that for $\rh<r_0<r<R$,
\begin{equation}
  \label{eq:iipd}
  r^\frac{n-2}{2}\lvert\phi\rvert(t,r)\leq\frac{C(n,m,\delta)\,D}{t^{\frac{3}{2}-\delta}}\,.
\end{equation}
\end{prop}

\begin{proof}
Let $\bar{t}_0=2(\tau_0+\tau_0)+R^\ast$ and $\bar{t}_1=\bar{t}_0+2\tau_0$ then by \eqref{eq:jump:timelike}, Prop.~\ref{prop:zeroth} and Prop.~\ref{prop:ILED}
\begin{multline}
  \label{eq:pf:iipd:a}
  \int_{\bar{t}_0}^{\bar{t}_1}\phi^2(t,r)r^{n-2}\ud t\leq C(n,m)\int_\St{2\tau_0}\Bigl(J^T(\phi)+J^T(T\cdot\phi)\\+\sumn J^T(\Omega_i\phi)+\sumn J^T(T\cdot\Omega_i\phi),n\Bigr)\,;
\end{multline}
hence by Prop.~\ref{prop:ed} there exists $t_0^\prime\in(\bar{t}_0,\bar{t}_1)$ such that
\begin{equation}
  \label{eq:iipd:ind}
  r^{n-2}\phi^2(t_0^\prime,r)\leq \frac{C(n,m)\,D}{\bar{t}_0^3}\,.
\end{equation}
Now set $\tau_0^\prime=\frac{1}{2}(t_0^\prime-R^\ast)$ and $\tau_j^\prime=2\tau_{j-1}^\prime\ (j\in\mathbb{N})$, and $t_j^\prime=2\tau_j^\prime+R^\ast\ (j\in\mathbb{N})$; note that $t_{j+1}^\prime-t_j^\prime=\frac{1}{2}(t_j^\prime-R^\ast)$.
Now consider \eqref{eq:interpolate} with $t_1=t^\prime_{j+1}$, $t_0=t^\prime_j$; since by \eqref{eq:jump:timelike}, together with Prop.~\ref{prop:ILED} and Prop.~\ref{prop:zeroth},
\begin{multline}
  \label{eq:pf:iipd:b}
  \int_{t_j^\prime}^{t_{j+1}^\prime}r^{n-2}\phi^2(t,r)\ud t\leq C(n,m)\int_\St{\tau_j^\prime}\Bigl(J^T(\phi)+J^T(T\cdot\phi)\\+\sumn J^T(\Omega_i\phi)+\sumn J^T(T\cdot\Omega_i\phi),n\Bigr)\,,
\end{multline}
and by Prop.~\ref{prop:rILED} and Prop.~\ref{prop:zeroth:r},
\begin{multline}
  \label{eq:pf:iipd:c}
  \int_{t_j^\prime}^{t_{j+1}^\prime}r^{n-2}(\partial_t\phi)^2(t,r)\ud t\leq
  C(n,m)\biggl\{\int_{\St{\tau_j^\prime}\cap\{\rs\leq R^\ast+(\tau_j^\prime)^3\}}\Bigl(J^T(\partial_t\phi)+J^T(T\cdot\partial_t\phi)\\+\sumn J^T(\Omega_i\partial_t\phi)+\sumn J^T(T\cdot\Omega_i\partial_t\phi),n\Bigr)\\
  +\int_\Sn\dm{\gn}\Bigl[r^{n-2}(\partial_t\phi)^2+\sumn r^{n-2}(\Omega_i\partial_t\phi)^2\Bigr]\vert_{(\us=\tau_j^\prime,\vs=R^\ast+\tau_j^\prime+(\tau_j^\prime)^3)}\biggr\}\,,
\end{multline}
which decays with the rate $\tau^{4-4\delta}$ as is shown in the proof of Prop.~\ref{prop:pointwisedecay} (ii), we obtain
\begin{multline}
  \label{eq:iipd:it}
  r^{n-2}\phi^2(r,t_{j+1}^\prime)\leq\\\leq r^{n-2}\phi^2(r,t_j^\prime)
  +\frac{C(n,m)}{(t_j^\prime)^{1-2\delta}}\frac{D}{(\tau_j^\prime)^2}+C(n,m,\delta)(t_j^\prime)^{1-2\delta}\frac{D}{(\tau_j^\prime)^{4-4\delta}}\leq\\
  \leq r^{n-2}\phi^2(r,t_j^\prime)+\frac{C(n,m,\delta)\,D}{(t_j^\prime)^{3-2\delta}}\,.
\end{multline}
In fact, by induction on $j\in\mathbb{N}$ using \eqref{eq:iipd:ind} for $j=0$, we have shown
\begin{equation}
  \label{eq:iipd:d}
  r^{n-2}\phi^2(r,t_j^\prime)\leq\frac{C(n,m,\delta)\,D}{(t_j^\prime)^{3-2\delta}}\qquad(j\in\mathbb{N}\cup\{0\})\,.
\end{equation}
Finally for any $t\geq t_0^\prime$ we may choose $j\in\mathbb{N}\cup\{0\}$ such that $t\in(t_j^\prime,t_{j+1}^\prime)$ and conclude the proof by applying \eqref{eq:iipd:d} and \eqref{eq:iipd:it} which holds with $t$ in place of $t_{j+1}^\prime$.
\end{proof}

\paragraph{Extension to the horizon.}
Note that for $\rh\leq r<r_0$ the same interpolation \eqref{eq:interpolate} by integration along lines of constant radius $r<r_0$ can be carried out.
However, on the right hand sides of \eqref{eq:jump:st} and \eqref{eq:jump:timelike} a new term results from the integration on $\vs=\frac{1}{2}(t_0+r_0^\ast)$ from the radius $r<r_0$ to $r=r_0$; but we infer from the explicit construction \eqref{Y} that the resulting integrand
\begin{equation}
  \Bigl(\frac{2}{\cf}\ddus{\phi}\Bigr)^2\leq T[\phi](Y,Y)\leq\Bigl(J^N[\phi],N\Bigr)
\end{equation}
is controlled by Cor.~\ref{cor:nILED} and the proof of Prop.~\ref{prop:iipd} above extends to that of Thm.~\ref{thm:pd} by replacing $J^T$ by $J^N$ on the right hand sides of \eqref{eq:pf:iipd:a}, \eqref{eq:pf:iipd:b} and \eqref{eq:pf:iipd:c}.

\appendix

\section{Notation}
\label{sec:notation}
\subsection*{Contraction} We sum over repeated indices. Also we use interchangeably
\begin{equation}
  g(V,N) \doteq (V,N) \doteq V_\mu N^\mu \,.
\end{equation}

\subsection*{Integration} 
Let $\mathcal{D}$ in \M\ be a domain bounded by two homologous hypersurfaces, $\Sigma_1$ and $\Sigma_2$ being its past and future boundary respectively. We then write $\int_\St{1}(J,n)$ for the boundary terms on $\Sigma_1$ arising from a general current $J$ in the expression $\int_{\partial\mathcal{D}}{}^\ast J$. If $\mathcal{S}\subset\Sigma_1$ is spacelike, then $(J,n)=g(J,n)$ is in fact the inner product of $J$ with the timelike normal $n$ to $\Sigma_1$; e.g.~on constant $t$-slices $\overline{\Sigma}_t$ (see Section \ref{sec:geometry}) we have $n=(\cf)^{-\frac{1}{2}}\dt$. If $\mathcal{U}\subset\Sigma_1$ is an outgoing null segment then $\int_\mathcal{U}(J,n)$ denotes an integral of the form $\int\ud v\int_\mathrm{S}\dm{\gamma}g(J,\frac{\partial}{\partial v})$; e.g.~on the outgoing null segments of the hypersurfaces $\Sigma_\tau$ (see Section \ref{sec:iled}) used throughout this paper we have
\begin{equation}
  \label{eq:not:intnull}
  \int_{\St{\tau}\cap\{r\geq R\}}(J,n)\doteq\int_{\tau+R^\ast}^\infty\ud\vs\int_\Sn\dm{\gn}\,r^{n-1}\,\Bigl(J,\dvs\Bigr)\,.
\end{equation}
The volume form is usually omitted,
\begin{equation*}
  \int_\mathcal{D}f\doteq\int_\mathcal{D}f\,\dm{g}\qquad(\mathcal{D}\subset\mathcal{M})\,.
\end{equation*}

\section{Formulas for Reference}
In this appendix we summarize a few formulas for reference.

\subsection*{The wave equation}
The d'Alembert operator in \eqref{eq:wave} can we written out in any coordinate system according to
\begin{equation}
  \Box_g\phi=(g^{-1})^{\mu\nu}\nabla_\mu\partial_\nu\phi
\end{equation}
where $\nabla$ denotes the covariant derivative of the Levi-Civita connection of $g$.

\subsection*{Components of the energy momentum tensor}
\label{ref:nd}
The components of the energy momentum tensor \[T_{\mu\nu}(\phi)=\partial_\mu\phi\,\partial_\nu\phi-\frac{1}{2}g_{\mu\nu}\,\partial^\alpha\phi\,\partial_\alpha\phi\] tangential to \Q\ are given in $(\us,\vs)$-coordinates by
\begin{subequations}
\begin{gather}
\label{eq:ndem}
T_{\us\us}=\bigl(\frac{\partial\phi}{\partial\us}\bigr)^2\\
T_{\vs\vs}=\bigl(\frac{\partial\phi}{\partial\vs}\bigr)^2\\
T_{\us\vs}=\bigl(1-\frac{2m}{r^{n-2}}\bigr)\normsph{\nablab\phi}^2\,.
\end{gather}
\end{subequations}
We also refer to \eqref{eq:ndem} as the \emph{null decomposition} of the energy momentum tensor.
Note here that
\begin{equation*}
\partial^\alpha\phi\,\partial_\alpha\phi=-\frac{1}{1-\frac{2m}{r^{n-2}}}\bigl(\frac{\partial\phi}{\partial\us}\bigr)\bigl(\frac{\partial\phi}{\partial\vs}\bigr)+\normsph{\nablab\phi}^2
\end{equation*}
and
\begin{equation*}
\frac{1}{r^2}\gn^{\phantom{-}AB}T_{AB}=\normsph{\nablab\phi}^2-\frac{1}{2}(n-1)\partial^\alpha\phi\,\partial_\alpha\phi\,.
\end{equation*}

\subsection*{Integration}
\label{sec:integration}
A typical domain of integration that we use is
\begin{equation}
  \label{eq:appendix:D}
  \dD{R}{\tau_1}{\tau_2}=\Bigl\{(\us,\vs):\tau_1\leq\us\leq\tau_2\,,\vs-\us\geq R^\ast\Bigr\}\,.
\end{equation}
In local coordinates we have, by calculating the volume form from \eqref{eq:mefc}, that
\begin{equation}
  \label{eq:int:vol}
  \intD{R}\dm{g}=\int_{\tau_1}^{\tau_2}\!\!\!\!\ud\us\int_{\us+R^\ast}^\infty\!\!\!\!\ud\vs\int_\Sn\!\!\!\!\dm{\gn}\,2\bigl(\cf\bigr)r^{n-1}\,.
\end{equation}
For a general current $J$ the energy identity on this domain reads
\begin{equation}
\label{eq:appendix:stokes}
\intD{R}K^X\dm{g}=\intpD{R}{}^\ast J\,,
\end{equation}
where the right hand side is given more explicitly by
\begin{equation}
  \begin{split}
    \intpD{R}{}^\ast J =&-\int_{R^\ast+\tau_2}^\infty\!\!\!\!\ud\vs\int_\Sn\dm{\gn}\,r^{n-1}\,g(J,\dvs)\vert_{\us=\tau_2} \\
    &-\int_{\tau_1}^{\tau_2}\!\!\!\!\ud\us\int_\Sn\dm{\gn}\,r^{n-1}g(J,\dus)\vert_{\vs\to\infty}\\
    &+\int_{R^\ast+\tau_1}^\infty\!\!\!\!\ud\vs\int_\Sn\dm{\gn}\,r^{n-1}\,g(J,\dvs)\vert_{\us=\tau_1}\\
    &-\int_{R^\ast+2\tau_1}^{R^\ast+2\tau_2}\ud t\int_\Sn\,r^{n-1}g(J,\drs)\vert_{r=R}\,.
  \end{split}
  \label{eq:appendix:boundary}
\end{equation}

\subsection*{Radial functions}
\label{ref:rrs}

In this appendix we summarize some statements on the relation between $r$ and
\begin{equation}
  \rs=\int_{\rphp}^r \frac{1}{1-\frac{2m}{r^{n-2}}}\ud r
\end{equation}
The proofs are omitted here, but can be found in \cite{VST}.

\begin{prop}\label{prop:limitrsr}
For all $n\geq 3$, \[\lim_{\frac{r}{\rh}\to\infty}\frac{\rs}{r}=1\,.\]
\end{prop}

\noindent While this fact concerns the region $\rs\geq 0$ and is essentially due to $\lim_{x\to\infty}\frac{\log x}{x}=0$, the next concerns $\rs\leq 0$ and is similarly due to $\lim_{x\to 0}x\log x=0$.
\begin{prop}\label{prop:limitminusrs}
For all $n\geq 3$,\[\lim_{\frac{r}{\rh}\to 1}\bigl(1-\frac{2m}{r^{n-2}}\bigr)(-\rs)=0\,.\]
\end{prop}

\noindent In fact we have:
\begin{prop} \label{prop:minusrsupperbound}
For $\rs<0$, \[\Bigl(1-\frac{2m}{r^{n-2}}\Bigr)\leq\frac{\rhp}{(-\rs)}\,.\]
\end{prop}

\noindent This being an upper bound on $(-\rs)$ we will also need a lower bound:
\begin{prop} \label{prop:minusrslowerbound}
For $\rs\leq 0$, \[(-\rs)\geq\frac{\rhp}{n-2}\log\Bigl[\frac{\rphn-1}{\rphn+1}\frac{\frac{r}{\rh}+1}{\frac{r}{\rh}-1}\Bigr]\,.\]
\end{prop}

\subsection*{Dyadic sequences}
\label{ref:dyadic}
In our argument, Section \ref{sec:iid} in particular, we construct a hierarchy of \emph{dyadic sequences}, beginning with a sequence of real numbers $(\tau_j)_{j\in\mathbb{N}}$ where $\tau_1>0$ and $\tau_{j+1}=2\tau_{j}\ (j\in\mathbb{N})$. We then obtain (by the mean value theorem of integration) a sequence $(\tau_j^\prime)_{j\in\mathbb{N}}$ with $\tau_j^\prime$ in the interval $(\tau_j,\tau_{j+1})$ of length $\tau_j$ for all $j\in\mathbb{N}$. We then built up on these values another sequence $(\tau_j^{\prime\prime})_{j\in\mathbb{N}}$ which takes values (as selected by the mean value theorem) in the intervals $(\tau_{2j-1}^\prime,\tau_{2j+1}^\prime)\ni\tau_j^{\prime\prime}$; note that their length is at least $\tau_{2j+1}^\prime-\tau_{2j-1}^\prime\geq\tau_{2j+1}-\tau_{2j}=\tau_{2j}$. In the same fashion the sequence $(\tau_j^{\prime\prime\prime})_{j\in\mathbb{N}}$ is built upon $(\tau_j^{\prime\prime})_{j\in\mathbb{N}}$, etc.

\section{Boundary Integrals and Hardy Inequalities}
\label{sec:boundary}

In this appendix we prove appropriate Hardy inequalities that are needed in our argument to estimate boundary terms that typically arise in the energy identities.

\paragraph{$X$-type currents.}
Let $X=f(\rs)\drs$ and recall the modification \eqref{eq:firstmodifiedcurrent}.

\begin{prop}[Boundary terms near null infinity] \label{prop:boundary:JX1}
Let $f=\mathcal{O}(1)$, $f^\prime=\mathcal{O}(\frac{1}{r})$, and $f^{\prime\prime}=\mathcal{O}(\frac{1}{r^2})$, then there exists a constant $C(n,m)$ such that
  \begin{equation}
    \label{est:boundary:D}
    \int_{\partial{}^R\mathcal{D}_{\tau_1}^{\tau_2}\setminus\{r=R\}}{}^\ast J^{X,1}\leq C(n,m)\int_{\Sigma_{\tau_1}}\ned{T}{\phi}\,.
  \end{equation}
\end{prop}
\begin{proof}
For the boundary integrals on the null segments $\us=\tau_1,\tau_2$ we find
\begin{multline}
  \biggl\vert\int_{R^\ast+\tau_i}^\infty\!\!\!\!\ud\vs\int_\Sn\dm{\gn}g(J^{X,1},\dvs)\,r^{n-1}\biggr\vert\leq\\
  \leq C(n)\int_{R^\ast+\tau_i}^\infty\!\!\!\!\ud\vs\int_\Sn\dm{\gn}\,r^{n-1}\biggl\{\sqb{\ddvs{\phi}}+\sqv{\nablab\phi}\\
  +\Bigl[\frac{\vert f\vert}{r^2}+\frac{\vert f^\prime\vert}{r}+\vert f^\prime\vert^2+\vert f^{\prime\prime}\vert\Bigl]\phi^2\biggr\}
\end{multline}
and in view of the Hardy inequality Lemma \ref{lemma:hardy}
\begin{equation}
  \int_{R^\ast+\tau_i}^\infty\!\!\!\!\ud\vs\int_\Sn\dm{\gn}\,\frac{1}{r^2}\phi^2\,r^{n-1}\vert_{\us=\tau_i}\leq
  C(n,m)\int_\St{\tau_i}\ned{T}{\phi}\,;
\end{equation}
note that the corresponding zero order terms vanish at future null infinity, cf.~Remark \ref{rmk:aprioriasymptotics}.
Then \eqref{est:boundary:D} follows from the energy identity for $T$ on ${}^R\mathcal{D}_{\tau_1}^{\tau_2}$.
\end{proof}

\begin{lemma}[Hardy inequality]\label{lemma:hardy}
Let $\phi\in\mathrm{C}^1([a,\infty))$, $a>0$, with $\vert\phi(a)\vert<\infty$ and
\begin{equation}
\label{eq:hardy:ac}
\lim_{x\to\infty}x^\frac{n-2}{2}\phi(x)=0\,,
\end{equation}
then a constant $C(n)>0$ exists such that
\begin{equation}
\label{eq:hardy}
\int_a^\infty\frac{1}{x^2}\phi^2(x)\,x^{n-1}\ud x\leq C(n)\int_a^\infty\bigl(\frac{\ud \phi}{\ud x}\bigr)^2\,x^{n-1}\ud x\,.
\end{equation}
\end{lemma}
\begin{proof}
This is a consequence of the Cauchy-Schwarz inequality, after integration by parts
\[\int_a^\infty\frac{1}{x^2}\phi^2(x)\,x^{n-1}\ud x=\int_a^\infty \sprime{g}(x)\phi^2(x)\ud x\] with \[g(x)=\int_a^x y^{n-3}\ud y\,.\qedhere\]
\end{proof}

\begin{remark}\label{rmk:aprioriasymptotics}
The conditions of the Lemma on $\phi$ are in fact satisfied for any solution of the wave equation \eqref{eq:wave}. By a density argument we may assume without loss of generality that the initial data is compactly supported. Then for a fixed $\tau$, and $\vs$ large enough $\phi(\tau,\vs)=0$ and for $\us\geq\tau$ \[\phi(\us,\vs)=\int_\tau^{\us}\ddus{\phi}\ud\us\,.\]
Thus \[\phi(\us,\vs)\leq\biggl(\int_\tau^{\us}\bigl(\ddus{\phi}\bigr)^2\,r^{n-1}\ud\us\biggr)^\frac{1}{2}\biggl(\int_\tau^{\us}\frac{1}{r^{n-1}}\ud\us\biggr)^\frac{1}{2}\,.\]
On one hand \[\int_\tau^{\us}\int_\Sn\bigl(\ddus{\phi}\bigr)^2\,r^{n-1}\dm{\gn}\ud\us\leq\int_\St{\tau}\bigl(J^T(\phi),n\bigr)<\infty\,,\] whereas on the other hand
\begin{multline*}
\int_\tau^{\us}\frac{1}{r^{n-1}}\ud\us=\frac{1}{n-2}\int_\tau^{\us}\bigl(\cf\bigr)^{-1}\dus\bigl(\frac{1}{r^{n-2}}\bigr)\ud\us\\
\leq\frac{1}{n-2}\Bigl(1-\frac{2m}{R^{n-2}}\Bigr)^{-1}\biggl(1-\Bigl(\frac{r(\us,\vs)}{r(\tau,\vs)}\Bigr)^{n-2}\biggr)\,\frac{1}{r^{n-2}}
\end{multline*}
if we restrict $\us\geq\tau$ to $r(\us,\vs)\geq R$.
Hence \[\lim_{\vs\to\infty}r^\frac{n-2}{2}\phi=0\,.\]
\end{remark}

Instead of \eqref{eq:hardy} which requires \eqref{eq:hardy:ac} one can prove the corresponding Hardy inequality for finite intervals:
\begin{lemma}[Hardy inequality for finite intervals]
\label{lemma:hardyineqfinite}
Let $0<a<b$, and $\phi\in\mathrm{C}^1((a,b))$ then
\begin{multline}
  \label{eq:hardyineqfinite}
  \frac{1}{2}\int_a^b\frac{1}{x^2}\phi^2(x)x^{n-1}\ud x\leq\\
  \leq\frac{1}{n-2}b^{n-2}\phi^2(b)+2\sqb{\frac{2}{n-2}}\int_a^b\sqb{\frac{\ud\phi}{\ud x}}x^{n-1}\ud x\,.
\end{multline}
\end{lemma}
\begin{proof}
Let
\begin{equation*}
  g(x)=\int_a^x y^{n-3}\ud y=\frac{1}{n-2}y^{n-2}\vert^x_a
\end{equation*}
then by integration by parts and using Cachy's inequality
\begin{multline*}
  \int_a^b\frac{1}{x^2}\phi^2(x)x^{n-1}\ud x=g\,\phi^2\vert_a^b-\int_a^b g(x)2\phi(x)\frac{\ud \phi}{\ud x}\ud x\leq\\
  \leq g(b)\phi^2(b)+2\epsilon\int_a^b\frac{1}{x^2}\phi^2(x)x^{n-1}\ud x
  +\frac{1}{2\epsilon}\int_a^b\frac{g(x)^2}{x^{n-3}}\sqb{\frac{\ud\phi}{\ud x}}\ud x
\end{multline*}
where $\epsilon>0$; \eqref{eq:hardyineqfinite} follows for $\epsilon=\frac{1}{4}$ because
\begin{gather*}
  g(b)\leq\frac{1}{n-2}b^{n-2}\\
  \frac{g(x)^2}{x^{n-3}}\leq\frac{2}{n-2}\Bigl(1+\bigl({\frac{a}{x}}\bigr)^{2(n-2)}\Bigr)x^{n-1}\,.\qedhere
\end{gather*}
\end{proof}

Recall the domain \eqref{def:s:Dslash}; by using Lemma \ref{lemma:hardyineqfinite} instead of Lemma \ref{lemma:hardy} we can prove the following refinement of Prop.~\ref{prop:boundary:JX1} to bounded domains:
\begin{prop}[Boundary terms on bounded domains] \label{prop:fboundary:JX1}
Let $f=\mathcal{O}(1)$, $f^\prime=\mathcal{O}(\frac{1}{r})$, and $f^{\prime\prime}=\mathcal{O}(\frac{1}{r^2})$, then there exists a constant $C(n,m)$ such that
  \begin{multline}
    \label{est:fboundary:D}
    \int_{\partial\dDb{R}{\tau_1}{\tau_2}\setminus\{r=R\}}{}^\ast J^{X,1}\leq\\\leq C(n,m)\biggl\{\int_\Stt{\tau_1}{\tau_2}\ned{T}{\phi}+\int_\Sn\!\!\!\!\dm{\gn}r^{n-2}\phi^2\vert_{(\us=\tau_1,\vs=R^\ast+\tau_2)}\biggr\}\,.
  \end{multline}
\end{prop}

Recall the domain \eqref{def:Rinfty}.

\begin{prop}[Boundary terms near the event horizon] \label{prop:vboundary:JX1}
Let $f=\mathcal{O}(1)$, $f^\prime=\mathcal{O}(\frac{1}{\vert\rs\vert^4})$, and $f^{\prime\prime}=\mathcal{O}(\frac{1}{\vert\rs\vert^5})$, and \[\pi_l\phi=0\qquad(0\leq l<L)\,,\] for some $L\in\mathbb{N}$, then there exists a constant $C(n,m,L)$ such that
  \begin{equation}
    \label{est:boundary:R}
    \int_{\partial\mathcal{R}_{r_0,r_1}^\infty(t_0)}{}^\ast J^{X,1}\leq C(n,m,L)\int_{\Sigma_{\tau_0}}\ned{T}{\phi}\,.
  \end{equation}
where $\tau_0=\frac{1}{2}(t_0-r_1^\ast)$.
\end{prop}

The proof is given in Section \ref{sec:pfILED} in the special case $f=f_{\gamma,\alpha}$ using the following Lemma.

\begin{lemma}[Hardy inequality]\label{ineq:vhardy}
Let $a>0$, $\phi\in\mathrm{C}^1([a,\infty))$ with \[\lim_{x\to\infty}\vert\phi(x)\vert<\infty\,.\] Then
\begin{multline}
  \int_a^\infty\frac{1}{1+x^2}\phi^2(x)\ud x\leq\\
\leq 8\frac{1+a^2}{a^2}\int_a^\infty\bigl(\frac{\ud\phi}{\ud x}\bigr)^2\ud x+2\pi\int_a^{a+1}\Bigl\{\phi^2+\bigl(\frac{\ud\phi}{\ud x}\bigr)^2\Bigr\}\ud x\,.
\end{multline}
\end{lemma}

\begin{proof}
Let us first assume that $\phi(a)=0$. Define \[g(x)=-\int_x^\infty\frac{1}{1+y^2}\ud y\] then
\begin{multline*}
\int_a^\infty\frac{1}{1+x^2}\phi^2(x)\ud x=\int_a^\infty \sprime{g}(x)\phi^2(x)\ud x\\
=g(x)\phi^2(x)\vert_a^\infty-2\int_a^\infty g(x)\phi(x)\frac{\ud\phi}{\ud x}\ud x\\
\leq 2\Bigl(\int_a^\infty\frac{g(x)^2}{\sprime{g}(x)}\bigl(\frac{\ud\phi}{\ud x}\bigr)^2\ud x\Bigr)^\frac{1}{2}\Bigl(\int_a^\infty\sprime{g}(x)\phi^2(x)\ud x\Bigr)^\frac{1}{2}\,.
\end{multline*}
Since $\vert g(x)\vert\leq\frac{1}{x}$ we have \[\frac{g(x)^2}{\sprime{g}(x)}\leq\frac{1+x^2}{x^2}\leq\frac{1+a^2}{a^2}\] and therefore
\[\int_a^\infty\frac{1}{1+x^2}\phi^2(x)\ud x\leq 4\int_a^\infty\frac{g(x)^2}{\sprime{g}(x)}\bigl(\frac{\ud\phi}{\ud x}\bigr)^2\ud x\leq 4\frac{1+a^2}{a^2}\int_a^\infty\bigl(\frac{\ud\phi}{\ud x}\bigr)^2\ud x\,.\]
Without the assumption $\phi(a)=0$ this applied to the function $\phi(x)-\phi(a)$ yields
\begin{multline*}
  \int_a^\infty\frac{1}{1+x^2}\phi^2(x)\ud x\leq\\
\leq 2\int_a^\infty\frac{1}{1+x^2}\bigl(\phi(x)-\phi(a)\bigr)^2\ud x+2\int_a^\infty\frac{1}{1+x^2}\phi(a)^2\ud x\\
\leq 8\frac{1+a^2}{a^2}\int_a^\infty\bigl(\frac{\ud\phi}{\ud x}\bigr)^2\ud x +\pi\phi(a)^2\,.
\end{multline*}
We conclude the proof with the following pointwise bound:
On one hand for some $\sprime{a}\in(a,a+1)$ \[\int_a^{a+1}\phi(x)^2\ud x=\phi(\sprime{a})^2\] and on the other hand
\[\phi(\sprime{a})^2-\phi(a)^2=\int_a^{\sprime{a}}\frac{\ud}{\ud x}\phi(x)^2\ud x\leq\int_a^{\sprime{a}}\Bigl\{\phi(x)^2+\bigl(\frac{\ud\phi}{\ud x}\bigr)^2\Bigr\}\ud x\,.\]
Hence
\begin{multline*}
  \phi(a)^2\leq\int_a^{\sprime{a}}\Bigl\{\phi(x)^2+\bigl(\frac{\ud\phi}{\ud x}\bigr)^2\Bigr\}\ud x+\int_a^{a+1}\phi(x)^2\ud x\\
\leq 2\int_a^{a+1}\Bigl\{\phi(x)^2+\bigl(\frac{\ud \phi}{\ud x}\bigr)^2\Bigr\}\ud x\,.\qedhere
\end{multline*}
\end{proof}

\paragraph{Auxiliary currents.}
For auxiliary currents of the form
\begin{equation}
  \label{eq:auxcurrent}
  J^\text{aux}_\mu=\frac{1}{2}h(r)\partial_\mu(\phi^2)
\end{equation}
we have the same results.
\begin{prop} \label{prop:boundary:aux}
Let $h=\mathcal{O}(\frac{1}{r})$, then there exists a constants $C(n,m)$ such that
  \begin{equation}
    \label{est:boundary:aux}
    \int_{\partial{}^R\mathcal{D}_{\tau_1}^{\tau_2}\setminus\{r=R\}}{}^\ast J^\text{aux}\leq C(n,m)\int_{\Sigma_{\tau_1}}\ned{T}{\phi}\,,
  \end{equation}
and moreover for a constant $C(n,m)$ we have the refinement
  \begin{multline}
    \label{est:fboundary:aux}
    \int_{\partial\dDb{R}{\tau_1}{\tau_2}\setminus\{r=R\}}{}^\ast J^\text{aux}\leq\\\leq C(n,m)\biggl\{\int_\Stt{\tau_1}{\tau_2}\ned{T}{\phi}+\int_\Sn\!\!\!\!\dm{\gn}r^{n-2}\phi^2\vert_{(\tau_1,R^\ast+\tau_2)}\biggr\}\,.
  \end{multline}
\end{prop}
\begin{proof}
Note that here, in comparison to the proof of Prop.~\ref{prop:boundary:JX1},
\begin{equation*}
  \Bigl\vert g(J^\text{aux},\dvs)\Bigr\vert\leq h^2\phi^2+\sqb{\ddvs{\phi}}\,.\qedhere
\end{equation*}
\end{proof}

\begin{prop}
Let $h=\mathcal{O}(\frac{1}{\vert\rs\vert})$, then there exists a constant $C(n,m)$ such that
  \begin{equation}
    \int_{\partial\mathcal{R}_{r_0,r_1}^\infty(t_0)}{}^\ast J^\text{aux}\leq C(n,m)\int_{\Sigma_{\tau_0}}\ned{T}{\phi}\,.
  \end{equation}
where $\tau_0=\frac{1}{2}(t_0-r_1^\ast)$.
\end{prop}

\begin{remark} Note that in view of Prop.~\ref{prop:minusrsupperbound} the function $h=\frac{1}{r}\bigl(\cf\bigr)$ satisfies the assumption of the Proposition.\end{remark}

\end{document}